\documentclass[twocolumn,amsmath,trackchanges]{aastex702}

\usepackage{amsmath}
\usepackage{gensymb}
\usepackage{multirow}
\usepackage{makecell}
\usepackage{rotating}
\usepackage{threeparttable}

\shorttitle{The Galactic Stellar Extinction Database}
\shortauthors{Zhang et al.}

\begin{document}

\title{GSED: The Galactic Stellar Extinction Database}

\author[orcid=0009-0003-2888-6317]{Baisong Zhang}
\affiliation{School of Integrated Circuits, Dezhou Vocational and Technical College, Dezhou, Shandong 253034, China}
\affiliation{South-Western Institute for Astronomy Research, Yunnan University, Kunming, Yunnan 650091, China}
\email{}

\author[orcid=0000-0003-2472-4903]{Bingqiu Chen}
\affiliation{South-Western Institute for Astronomy Research, Yunnan University, Kunming, Yunnan 650091, China}
\email[show]{bchen@ynu.edu.cn}
\correspondingauthor{Bingqiu Chen}

\author[0000-0002-8669-5370]{Dongwei Fan}
\affiliation{National Astronomical Observatories, Chinese Academy of Sciences, Beijing 100101, China}
\affiliation{National Astronomical Data Center, Beijing 100101, China}
\email[show]{fandongwei@nao.cas.cn}
\correspondingauthor{Dongwei Fan}

\author[0000-0003-2471-2363]{Haibo Yuan}
\affiliation{School of Physics and Astronomy, Beijing Normal University,  Beijing 100875, China}
\affiliation{Institute for Frontiers in Astronomy and Astrophysics, Beijing Normal University, Beijing 102206, China}
\email{}

\author{Jian Cui}
\affiliation{School of Integrated Circuits, Dezhou Vocational and Technical College, Dezhou, Shandong 253034, China}
\email{}

\author[0009-0007-5623-2475]{Pinjian Chen}
\affiliation{National Astronomical Observatories, Chinese Academy of Sciences, Beijing 100101, China}
\affiliation{School of Astronomy and Space Science, University of the Chinese Academy of Sciences, Beijing 100049, China}
\email{}

\author[0000-0001-5737-6445]{Helong Guo}
\affiliation{Research Center of Astronomy, Qinghai University, Xining, Qinghai 810016, China}
\affiliation{Department of Physics and Astronomy, Qinghai University, Xining, Qinghai 810016, China}
\email{}

\author[0000-0003-2751-2172]{Lin Zhang}
\affiliation{South-Western Institute for Astronomy Research, Yunnan University, Kunming, Yunnan 650091, China}
\email{}

\author[0009-0001-4291-0147]{Baokun Sun}
\affiliation{Institute of Astrophysics, Chuxiong Normal College, Chuxiong, Yunnan 675000, China}
\email{}

\author[0009-0006-5847-9271]{Xingzhu Zou}
\affiliation{South-Western Institute for Astronomy Research, Yunnan University, Kunming, Yunnan 650091, China}
\email{}

\author[0000-0003-3618-9960]{Lunwei Zhang}
\affiliation{School of Mathematics, Physics and Statistics, Honghe University, Mengzi, Yunnan 661100, China}
\email{}

\author{Yanan Cao}
\affiliation{South-Western Institute for Astronomy Research, Yunnan University, Kunming, Yunnan 650091, China}
\email{}

\author[0009-0007-6842-8117]{Longfei Ding}
\affiliation{South-Western Institute for Astronomy Research, Yunnan University, Kunming, Yunnan 650091, China}
\email{}

\begin{abstract}
	Reliable extinction correction is essential for nearly all astrophysical studies within the Galaxy. We present the Galactic Stellar Extinction Database (GSED, \url{https://nadc.china-vo.org/data/gsed/}), a homogenised database that unifies six representative 3D extinction datasets under a common $E(B-V)$ and parallax-distance baseline. A six-layer multilayer perceptron is designed to correct the systematic differences in both extinction and distance across the heterogeneous input catalogues. Applying the trained models yields a catalogue of over 1.9 billion homogenised entries, which is built into a publicly accessible, real-time query service: a user supplies a coordinate and a search radius, the system retrieves the data, fits the distance--extinction relation, returns $E(B-V)$ together with $E(G_{\rm BP}-G_{\rm RP})$ and $A_V$, and allows the raw catalogue and the fitted curve to be downloaded. By delivering extinction as raw stellar measurements rather than voxelised map products and retaining the capacity to incorporate future datasets, GSED provides a flexible, traceable, and extensible new tool for Galactic extinction correction and dust-structure studies.
\end{abstract}


\section{Introduction}

Nearly a century ago, \citet{Trumpler1930PASP} established through studies of Galactic star clusters that interstellar dust is ubiquitous and dims and reddens background objects. Dust causes extinction and reddening in the ultraviolet, optical, and near-infrared through absorption and scattering, and it constitutes a foreground contamination that almost every relevant observation must confront \citep{Draine2003ARAA}. To recover the luminosity and colour of an object, and thereby to assess its intrinsic properties and physical parameters accurately, the extinction and reddening of dust must be removed. Whether a reliable extinction estimate can be obtained for an arbitrary object or line of sight therefore often determines the precision of the associated study directly.

Such extinction information has long come mainly from extinction maps that describe the spatial distribution of dust. Depending on whether distance information is included, extinction maps fall into two-dimensional (2D) and three-dimensional (3D) classes. 2D extinction maps \citep[e.g.][]{sfd1998ApJ, Schultheis1999AA, Gonzalez2012AA, Planck2014AA, Gontcharov2025RAA}, provide the cumulative total extinction along the line of sight and cannot distinguish dust at different distances, so they tend to overestimate the actual extinction of disc stars embedded within the dust layer \citep{Berry2012ApJ, chen2014mnrasanti}. 3D extinction maps add the crucial distance dimension and depict the 3D structure of interstellar dust more accurately, which makes them a focus of recent research, and we list only a few representative works here. Early 3D maps relied mainly on photometric or model-dependent distance estimates and thus achieved limited distance precision. \citet{marshall2006AA} construct a 3D extinction map of the inner Galactic disc ($\left| l \right|\leq100\degree$, $\left| b \right|\leq10\degree$), \citet{chen2013AA3dmap} and \citet{Schultheis2014AA} construct 3D maps of the Galactic bulge region, \citet{chen2014mnrasanti} construct a 3D map toward the anticentre ($140\degree<l<240\degree$, $-60\degree<b<40\degree$), and \citet{Hanson2016MNRAS} construct a 3D map near the Galactic plane ($0\degree<l<250\degree$, $\left| b \right|<4.5\degree$). After the release of the Gaia data \citep{gaia2018AA}, high-precision trigonometric parallaxes greatly improved the precision of stellar distances and gave rise to a series of 3D maps with higher distance accuracy. \citet{chen2019mnras}, hereafter Chen19, use Gaia DR2 and the Two Micron All Sky Survey (2MASS; \citealt{Skrutskie2006AJ}) photometry to cover the entire Galactic disc ($0\degree<l<360\degree$, $\left| b \right|<10\degree$); \citet{green2019ApJ}, hereafter Green19, use Gaia DR2, Pan-STARRS1 \citep{2016arXiv161205560C}, and 2MASS to construct a 3D map covering $\delta>-30\degree$; \citet{guo2021ApJ}, hereafter Guo21, use SkyMapper \citep{Wolf2018PASA}, Gaia DR2, and 2MASS to construct a 3D map of the southern sky; and \citet{Zucker2025ApJ}, hereafter Zucker25, use deep photometry from the Dark Energy Camera Plane Survey 2 (DECaPS2; \citealt{Saydjari2023ApJS}) to construct a map of the southern Galactic disc ($239\degree<l<6\degree$, $\left| b \right|<10\degree$). \citet{wang2025ApJS} provide an all-sky 3D map based on Gaia XP  \citep{zxy2023MNRAS, zxy2025Sci} and the Large Sky Area Multi-Object Fiber Spectroscopic Telescope (LAMOST; \citealt{Zhao2012RAA}) data. We summarize the detailed information in Table~\ref{extin_summary}.

\begin{table*}
	\centering
	\caption{Summary of the 3D dust extinction maps. The “Coverage” column specifies the sky region covered by each map, and the “Used in this work” column indicates whether the dataset is incorporated into our final database.}
	\label{extin_summary}
	\setlength{\tabcolsep}{5mm}{
		\begin{tabular}{ccc}
			\hline
			Source & Coverage  & Used in this work  \\
			\hline
			\citet{marshall2006AA}&$\left| l \right|\leq100\degree$, $\left| b \right|\leq10\degree$&No\\
			\citet{chen2013AA3dmap} &$\left| l \right|<10\degree$, $\left| b \right|<2\degree$&No\\
			\citet{Schultheis2014AA}&$\left| l \right|<10\degree$, $-10\degree<b<5\degree$&No\\
			\citet{chen2014mnrasanti}&$140\degree<l<240\degree$, $-60\degree<b<40\degree$&No\\
			\citet{Hanson2016MNRAS}&$0\degree<l<250\degree$, $\left| b \right|<4.5\degree$&No\\
			\citet{chen2019mnras} & $0\degree<l<360\degree$, $\left| b \right|<10\degree$ &Yes \\
			\citet{green2019ApJ} & $\delta>-30\degree$&Yes\\
			\citet{guo2021ApJ} &Southern sky&Yes\\
			\citet{sun2025RAA} &Southern sky&Yes\\
			\citet{Zucker2025ApJ} &$239\degree<l<6\degree$, $\left| b \right|<10\degree$&Yes\\
			\citet{zxy2023MNRAS}&All-sky&No\\
			\citet{zxy2025Sci}&All-sky&Yes\\
			\citet{wang2025ApJS}&All-sky&No\\
			\citet{Yu2026MNRAS}&-&Planned\\
			\hline
	\end{tabular}}
\end{table*}

With these maps in hand, objects within the Galaxy can be corrected using extinction-correction tools, among which Dustmaps \citep{2018dustmaps} and GALExtin \citep{2021GALExtin} are the most widely used. Dustmaps integrates a variety of extinction maps into a single query interface within a unified framework, supporting the 2D maps of \citet{sfd1998ApJ} and \citet{Planck2014AA} together with the 3D maps of \citet{marshall2006AA}, \citet{chen2014mnrasanti}, \citet{green2015a3dmap, green2018MNRAS, green2019ApJ}, \citet{Edenhofer2024AA}, and \citet{Zucker2025ApJ}, etc., from which users may select as needed and call through the common interface. GALExtin is an online tool designed specifically for extinction correction that integrates multiple maps through a web interface in an intuitive and convenient manner. Beyond these two comprehensive tools, some authors also provide dedicated query sites or programs for their own maps, such as \citet{guo2021ApJ}, who release extinction-correction tools for the southern and all-sky regions based on the Dustmaps framework, and \citet{wang2025ApJS}, who also provide an online site for querying their all-sky map.

Although these tools offer important computational interfaces, three limitations remain difficult to avoid in practice. The first is the fragmentation of coverage. As noted above, a single extinction map often covers only a specific region or a specific distance range, so the maps complement one another yet do not connect. With both Dustmaps and GALExtin, users must specify the map in advance and be familiar with its coverage boundaries, and once a target source lies at the edge of a map or beyond its coverage, obtaining a valid or reliable extinction estimate becomes difficult, which raises the threshold for use. The second is the lack of a common calibration baseline together with the presence of systematic differences. Different maps adopt different extinction tracers such as $E(B-V)$, $E(g-r)$, $A_{r}$, and $A_{V}$, different distance scales, and different inversion methods and priors, and therefore carry non-negligible systematic differences among one another. Directly stitching them together or converting between them with fixed extinction coefficients introduces clear biases and seams into the result, so they cannot simply be merged into a single homogeneous all-sky map. The third is the absence of raw information and insufficient flexibility. Existing tools return only the extinction value at one point and at one fixed resolution, without providing the raw stellar sample along that line of sight; the window used to fit the distance--extinction relation is fixed in advance, so a reliable estimate cannot be obtained where the data are sparse, while lowering the resolution in exchange for more data introduces bias, and users cannot redo the fit from the raw data themselves. In addition, Dustmaps requires the large volume of extinction data to be downloaded locally before use.

To address these shortcomings, we construct the Galactic Stellar Extinction Database (GSED). Unlike redrawing another extinction map, GSED stores the extinction and distance measurements of roughly $1.936$ billion individual stars. We adopt the accurate colour excess \(E(B-V)\) and the parallax distances from \citet[][hereafter Zhang25]{zxy2025Sci} as a common baseline, normalising the extinction systematics and distance scales of six representative datasets onto a single system. This integrates previously separate and disconnected data into a homogeneous, all-sky database whose components can speak to one another. Because the data are stored at the stellar level, GSED allows users to specify a query radius according to their needs in order to adjust the resolution of the extinction data dynamically, to obtain the raw stellar sample within the search region, and to obtain more accurate and traceable results in distance and extinction. Moreover, the systematic-correction and distance-correction framework of GSED can be reused for extinction datasets released in the future, giving the database the capacity to expand continuously, with its data volume and coverage growing as subsequent surveys advance.

\section{Data}

As the first version of the GSED database, we select six recently published and representative large catalogues of stellar distances and extinction. We note that we do not include every published work of this kind; the design of GSED allows new extinction data to be incorporated continuously, and other extinction catalogues, for example \citet{Yu2026MNRAS}, will be added in due course.

Chen19 uses the photometry and astrometry of Gaia DR2 together with near-infrared photometry and applies a random forest machine learning method to obtain the $E(G_{\rm BP}-G_{\rm RP})$ reddening of about 56 million stars; after removing stars with large parallax errors in Gaia DR2, the work provides extinction and distance information for more than 35 million stars covering the disc region $0\degree<l<360\degree$, $|b|<10\degree$. Green19 uses multi-band photometry from Gaia DR2, Pan-STARRS1, and 2MASS and applies a Bayesian method to infer stellar atmospheric parameters, distance, and extinction simultaneously, providing the reddening and distance of about 799 million stars covering $\delta>-30\degree$. Guo21 combines multi-band data from SkyMapper, Gaia DR2, and near-infrared photometry and derives reliable $A_{r}$ extinction and distances for about 19 million stars within roughly $14{,}000\,\mathrm{deg}^2$ of the southern sky through SED fitting. \citet{sun2025RAA}, hereafter Sun25, uses data including SkyMapper and Gaia and obtains the stellar parameters, $E(BP-RP)$ reddening, and distances of about 140 million southern-sky stars through the SPar method \citep{Sun2023AJ}. Zucker25 uses deep photometry including the Dark Energy Camera Plane Survey 2 (DECaPS2; \citealt{Saydjari2023ApJS}) together with Gaia DR3 parallaxes and applies the brutus stellar-parameter inference framework to give the $A_V$ extinction and distances of about 700 million stars in the southern Galactic disc ($239\degree<l<6\degree$, $|b|<10\degree$). Zhang25 uses Gaia XP low-resolution spectra and near-infrared photometry such as 2MASS to develop a data-driven forward model and measures the extinction curve $R(V)$, $E(B-V)$, stellar parameters, and distances of about 130 million high-quality stars covering the whole sky.

These six datasets differ in their extinction tracers, distance scales, and stellar-parameter spaces. To calibrate them onto a single system, we prepare MLP inputs for each dataset. The atmospheric parameters or intrinsic colour are the feature inputs required by the MLP model to learn the extinction systematics, as described in Section~3. Because the catalogues provide different raw information, we need to fill in the missing parameters.

Chen19 and Guo21 compute only the stellar extinction, adopt the Gaia DR2 parallax distance, and do not provide $T_{\rm eff}$ or [Fe/H], so the intrinsic colour must be derived from Gaia DR3 colours as a substitute input. We therefore cross-match Chen19 and Guo21 with Gaia DR3 \citep{gaia2023A&A} within a radius of $1\,\mathrm{arcsec}$ to obtain the Gaia DR3 parallaxes and the $G_{\rm BP}-G_{\rm RP}$ colour, hereafter $BP-RP$, and we remove entries with null parallax or $BP-RP$ colour. We also impose the following constraints on the parallaxes: \texttt{Plx > 0.02}, \texttt{e\_Plx > 0}, and \texttt{e\_Plx / Plx < 0.5}. The distance $d$ and distance error $d\_err$ of Chen19 and Guo21 follow directly from the parallaxes. From the Gaia DR3 $BP-RP$ colour and the extinction of Chen19 and Guo21, combined with the extinction law given by Chen19, we further obtain the intrinsic colour $(BP-RP)_0$ of each star. For Green19, we cross-match its data with Pan-STARRS1 within a radius of $1\,\mathrm{arcsec}$ to obtain the $g$- and $r$-band photometry, and after removing null values we derive the intrinsic colour $(g-r)_{\rm 0}$ using the extinction $E(g-r)$. For Sun25, and Zucker25, the catalogues already provide atmospheric parameters such as $T_{\rm eff}$ and [Fe/H] directly, which serve as input features of the MLP model without additional cross-matching.

\section{Correction of Extinction and Distance Systematics}

As described in Section~2, the six extinction datasets differ in their extinction tracers, distance scales, and stellar-parameter spaces and cannot be merged directly. To caribrate them into a single system, we select a reasonable reference baseline. 
We adopt the Zhang25 catalogue as the baseline for both extinction and distance. For extinction, they developed a data-driven model for Gaia BP/RP spectra, trained on LAMOST atmospheric parameters, Gaia DR3 parallaxes, and Bayestar19 reddening \citep{green2019ApJ}, alongside 2MASS and WISE photometry. Applying this model to all Gaia BP/RP sources, they derived stellar parameters, parallaxes, and extinctions for approximately 220 million stars, with the derived \(E(B-V)\) measurements having typical uncertainties of \(0.03\,\mathrm{mag}\). For distance, the catalogue provides a high-precision and self-consistent all-sky distance scale based on Gaia trigonometric parallaxes, processed with strict quality selection and a forward model. On this baseline, we design a unified six-layer MLP framework that takes the raw extinction and distance of each catalogue as the main inputs, supplemented by auxiliary features such as atmospheric parameters, intrinsic colour, or position, and uses the baseline $E(B-V)$ and the baseline distance as supervision values to learn the complex relations between each dataset and the baseline through a mapping in a high-dimensional feature space, thereby correcting the extinction and distance systematics in a unified manner.

\subsection{Correction of Extinction Systematics}

The systematic differences among extinction datasets arise mainly from two sources. The first is the different stellar intrinsic colours or atmospheric parameters adopted, which lead to differences in the extinction zero point. The second is the different extinction laws assumed and extinction coefficients used, since the extinction coefficient itself varies with stellar effective temperature and extinction \citep{Shen2022MNRAS, ZhangYuan2023ApJS}, so that the conversion between the same extinction tracers differs from one work to another. The extinction systematics therefore change with stellar spectral type and metallicity and also depend on the magnitude of the extinction itself, and they cannot be corrected by a single fixed scaling factor. Based on this understanding, we use the raw extinction together with the atmospheric parameters or intrinsic colour as the input features of the MLP, so that the model can learn how the extinction bias varies with stellar parameters. The specific choice of inputs is also constrained by the information each catalogue provides. For the catalogues that supply $T_{\rm eff}$ and [Fe/H], namely Sun25 and Zucker25, we use these two parameters directly; for the catalogues without an effective temperature, we adopt the stellar intrinsic colour as a proxy for spectral type, with Chen19 and Guo21 providing $(BP-RP)_0$ and Green19 providing $(g-r)_{\rm 0}$; and since Green19 additionally provides [Fe/H], we include it as well.

\begin{figure*}
	\centering
	\includegraphics[width=0.82\linewidth]{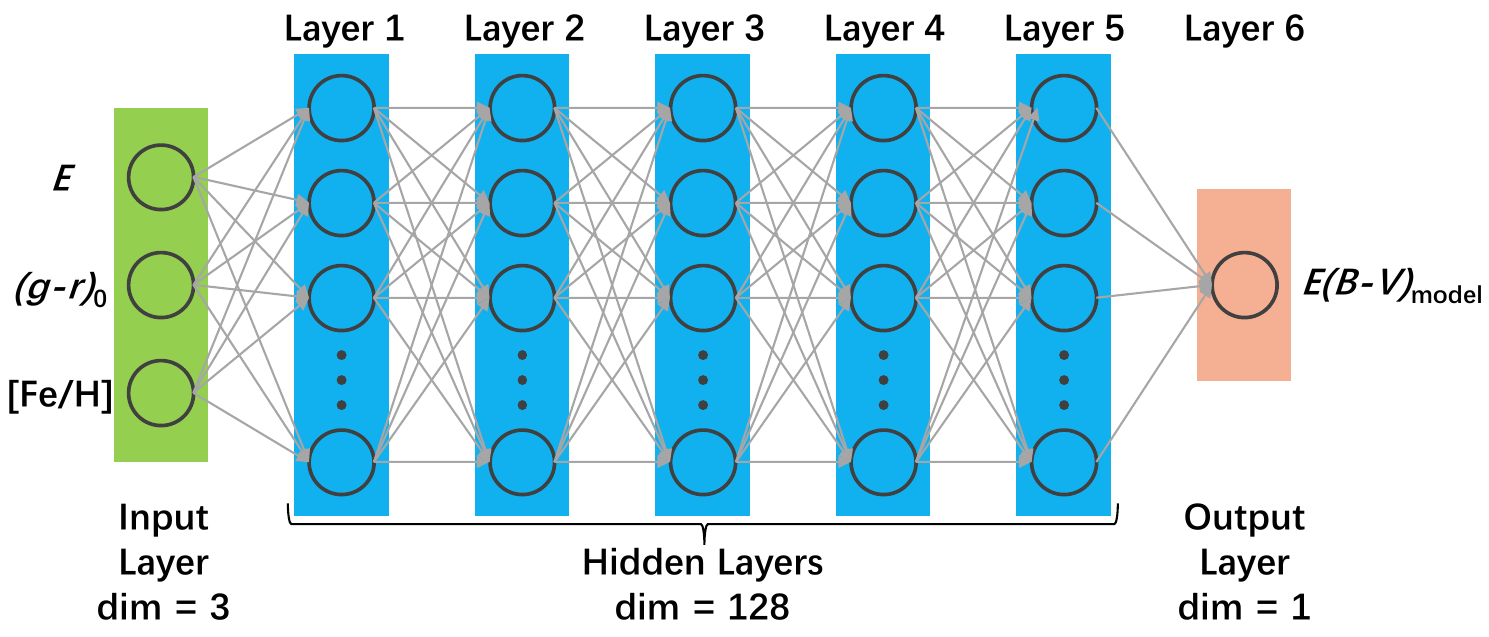}
	\caption{Network architecture of the extinction-systematics correction model. The input parameters differ slightly among works, and this figure takes the correction of the extinction systematics of \citet{green2019ApJ} as an example.}
	\label{fc_ebv}
\end{figure*}

Fig.~\ref{fc_ebv} shows the architecture of the deep neural network proposed in this work. It adopts a fully connected design and comprises an input layer, five hidden layers, and one output layer for a total of six layers, with a fixed hidden width of $128$ dimensions, ReLU activation, and no activation in the output layer so as to preserve the continuity and dynamic range of the regression. The core logic of the network is to learn the systematic difference between the input extinction and the target baseline extinction through a mapping in a high-dimensional feature space. During training, the colour excess $E(B-V)$ of the Zhang25 extinction catalogue serves as the supervision value, and the model is optimised with an L1 loss, namely the mean absolute error (MAE),
\begin{equation}
	\mathcal{L} = \frac{1}{N} \sum_{i=1}^{N} \left| E(B-V)_i^{\mathrm{true}} - E(B-V)_i^{\mathrm{model}} \right|,
\end{equation}
where $N$ is the total number of samples and $E(B-V)_i^{\mathrm{true}}$ and $E(B-V)_i^{\mathrm{model}}$ denote the target extinction and the model prediction for the $i$-th star. We use the MAE because its error term grows linearly and is insensitive to outliers, so it remains robust when the data contain many points with large scatter.

\begin{table}
	\centering
	\caption{Input parameters and loss results of the extinction-systematics correction models for the different catalogues.}
	\label{ebv_result}
	\begin{tabular}{ccc}
		\hline
		Source & Input Parameters  & Loss (MAE)  \\
		\hline
		Chen19 & $E(BP-RP)$, $(BP-RP)_{\rm 0}$&0.0729 \\
		Green19 & $E$, $(g-r)_{\rm 0}$, [Fe/H]& 0.0641 \\
		Guo21 &$A_r$, $(BP-RP)_{\rm 0}$&0.0443\\
		Sun25 &$E(BP-RP)$, $T_{\rm eff}$, [Fe/H]&0.0514\\
		Zucker25 &$A_V$, $T_{\rm eff}$, [Fe/H]&0.0403\\
		\hline
	\end{tabular}
\end{table}

Each of the six extinction datasets is cross-matched with the Wang et al. catalogue, and an MLP model of identical structure is trained separately. The input parameters, data selection, and detailed results of each model are given in Appendix~\ref{appendix_A}, and the overall performance is summarised here. Table~\ref{ebv_result} summarises the input parameters and the test-set performance of the extinction-systematics correction models, all of which use the same learning rate of $2.0\times10^{-3}$. The Chen19 and Guo21 models take the extinction and $(BP-RP)_{\rm 0}$ as inputs and achieve test-set losses of $0.0729$ and $0.0443$, respectively. The Green19 model uses the extinction, $(g-r)_{\rm 0}$, and [Fe/H] as inputs, achieving a loss of $0.0641$. Finally, the Sun25 and Zucker25 models take the extinction, $T_{\rm eff}$, and [Fe/H] as inputs and yield losses of $0.0514$ and $0.0403$, respectively.

This variation in performance can be primarily attributed to the differences in input features and photometric data. For instance, Green19, Chen19, and Sun25 provide broadband extinction or reddening in other passbands based on multi-band photometry, whose typical extinction uncertainty reaches \(0.05-0.06\,\mathrm{mag}\) \citep{Chen2015MNRAS}, introducing substantially larger scatter relative to the baseline. In contrast, the relatively small losses of Guo21 and Zucker25 (\(0.0443\) and \(0.0403\)) stem from distinct reasons. The Guo21 catalogue contains a relatively limited number of sources, so the model training remains effective. On the other hand, the Zucker25 catalogue covers a relatively narrow sky region (\(239^\circ < l < 6^\circ\), \(|b| < 10^\circ\)). This restricted spatial coverage limits the diversity of the extinction and stellar parameters within the sample, making the prediction task less complex compared to large sky area datasets, which consequently may yield a lower loss.

The figures in the appendix show that, for each dataset, the raw extinction carries a systematic difference that varies markedly with the atmospheric parameters and the extinction, and that after the MLP correction the residual between the corrected value and the baseline $E(B-V)_{\rm Zhang25}$ concentrates around zero and no longer shows any systematic trend with any parameter, indicating that the model removes the extinction systematics effectively.
Furthermore, in Appendix~\ref{appendix_C}, we test whether the input parameter [Fe/H] affects the results.

We note that the MLP essentially learns the average mapping between the input and the baseline, so its prediction represents the most probable baseline extinction under given input conditions rather than an exact reproduction of every data point. When the scatter between the raw extinction and the baseline is large and the correlation is weak, the prediction residual grows accordingly.

\subsection{Correction of Distance Systematics}

The differences in distance among catalogues arise mainly from two aspects. The first is the different distance-estimation methods, including photometric distances, trigonometric parallax distances, and their respective priors and processing pipelines, which may introduce a systematic zero-point offset. The second is the different sky coverage and detection depth of the surveys, so the distance systematics may vary with the line of sight. We therefore use the Galactic longitude $l$, the Galactic latitude $b$, and the raw distance $d$ as the input features of the MLP, so that the model can learn how the distance bias varies with position and distance simultaneously. The distance-correction model adopts the same six-layer MLP structure as the extinction-correction model and differs only in its input features and supervision value, taking $l$, $b$, and the raw distance $d$ as inputs and the distance $D_{\rm Zhang25}$ converted from parallax in the Zhang25 catalogue as the supervision value.

\begin{table}
	\centering
	\caption{Input parameters and loss results of the distance-systematics correction models for the different catalogues.}
	\label{d_result}
	\setlength{\tabcolsep}{4.5mm}{
		\begin{tabular}{ccc}
			\hline
			Source & Input Parameters  & Loss (MAE)  \\
			\hline
			Chen19 & $l$, $b$, $D_{\rm Chen19}$&0.2508 \\
			Green19 & $l$, $b$, $D_{\rm Green19}$&0.7883 \\
			Guo21 &$l$, $b$, $D_{\rm Guo21}$&0.0393 \\
			Sun25 &$l$, $b$, $D_{\rm Sun25}$&0.8336 \\
			Zucker25 &$l$, $b$, $D_{\rm Zucker25}$&0.8498 \\
			\hline
	\end{tabular}}
\end{table}

Five of the datasets, with Zhang25 itself serving as the distance baseline and requiring no calibration, are each cross-matched with the Zhang25 catalogue, and a MLP model of identical structure is trained separately. The input parameters, data selection and detailed results of each model are given in Appendix~\ref{appendix_B}, and the overall performance is summarised here. Table~\ref{d_result} summarises the input parameters and the test-set performance of the distance-correction models, all of which take the Galactic longitude $l$, the Galactic latitude $b$, and the distance $D$ as inputs and use the same learning rate of $2.0\times10^{-3}$. The differences in the distance-correction performance among the models are closely tied to the characteristics of the data themselves. The Guo21 loss is the smallest at $0.0393$, for three reasons: its total sample is small and its data quality is fairly uniform; limited by the sky coverage of the early SkyMapper data release, it contains few stars in the low-latitude, high-extinction, and high-density regions of the disc; and its stars are generally bright, so the Gaia trigonometric parallaxes are measured with high precision and the raw distances deviate little from the baseline.

The losses of Chen19, Green19, Sun25, and Zucker25 are relatively large, ranging from $0.25$ to $0.85$, for two reasons. First, a large fraction of the stars in these four datasets lie in the high-density regions of the disc, where the scatter of the Gaia trigonometric parallaxes increases markedly \citep{Lindegren2020AA} and the raw distances are inherently uncertain. Second, Green19, Sun25, and Zucker25 all contain a considerable proportion of distant stars whose distances rely mainly on photometric parallaxes rather than Gaia trigonometric parallaxes and therefore carry large intrinsic errors. The Chen19 loss of $0.2508$ is the smallest among the four, benefiting from its more conservative parallax-error selection, which removes stars with poor distance precision. In general, the clearer the trend and the smaller the scatter between the raw distance of a catalogue and the baseline distance, the better the model performs. The figures in the appendix likewise show that the raw distances carry a systematic bias that varies with the line of sight and the distance, and that after the MLP calibration the residual reaches zero with no systematic trend, which verifies the effectiveness of our distance systematics correction.

\begin{figure*}
	\centering
	\includegraphics[width=0.75\linewidth]{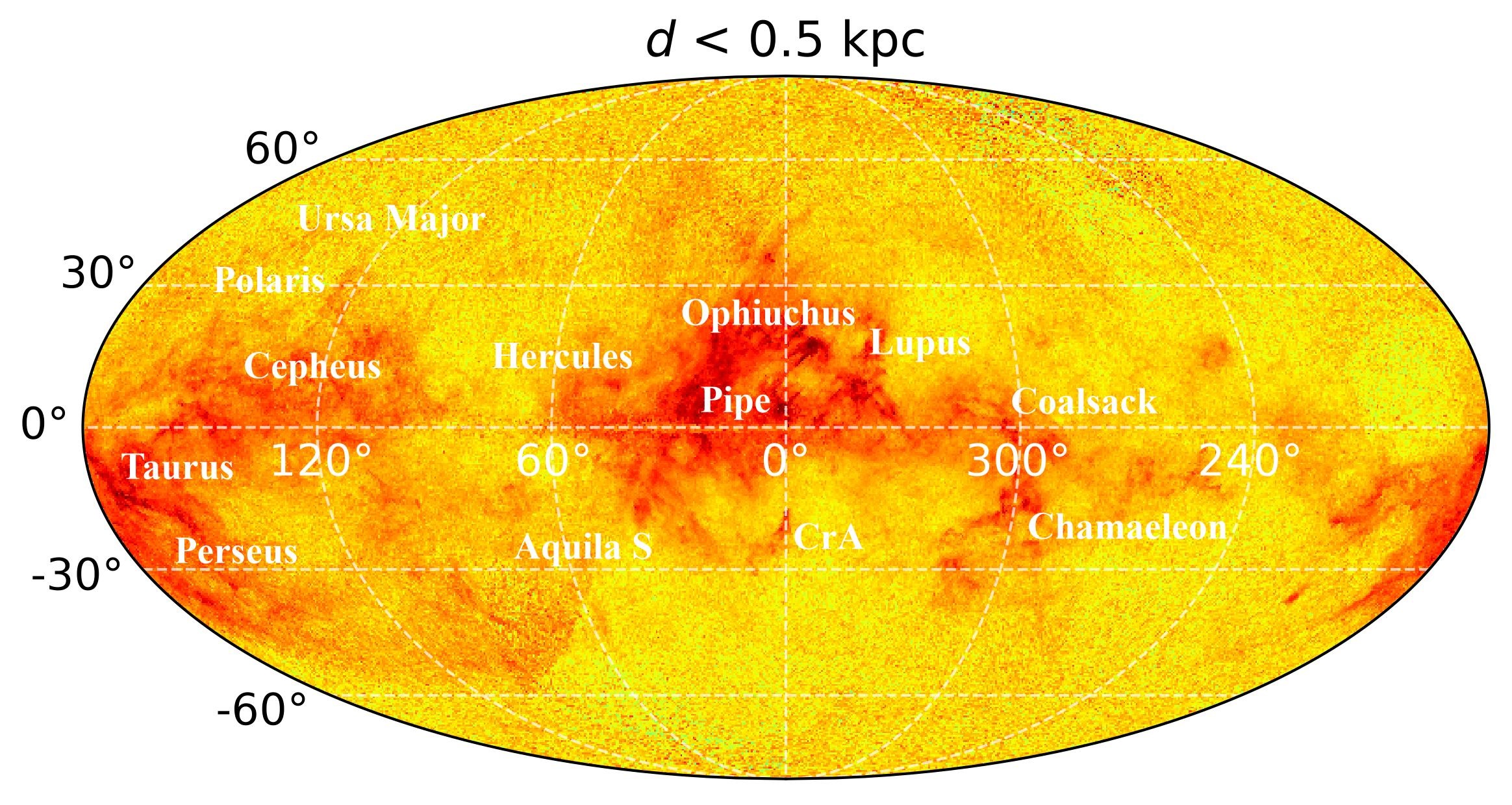}
	\\[0.2em]
	\includegraphics[width=0.75\linewidth]{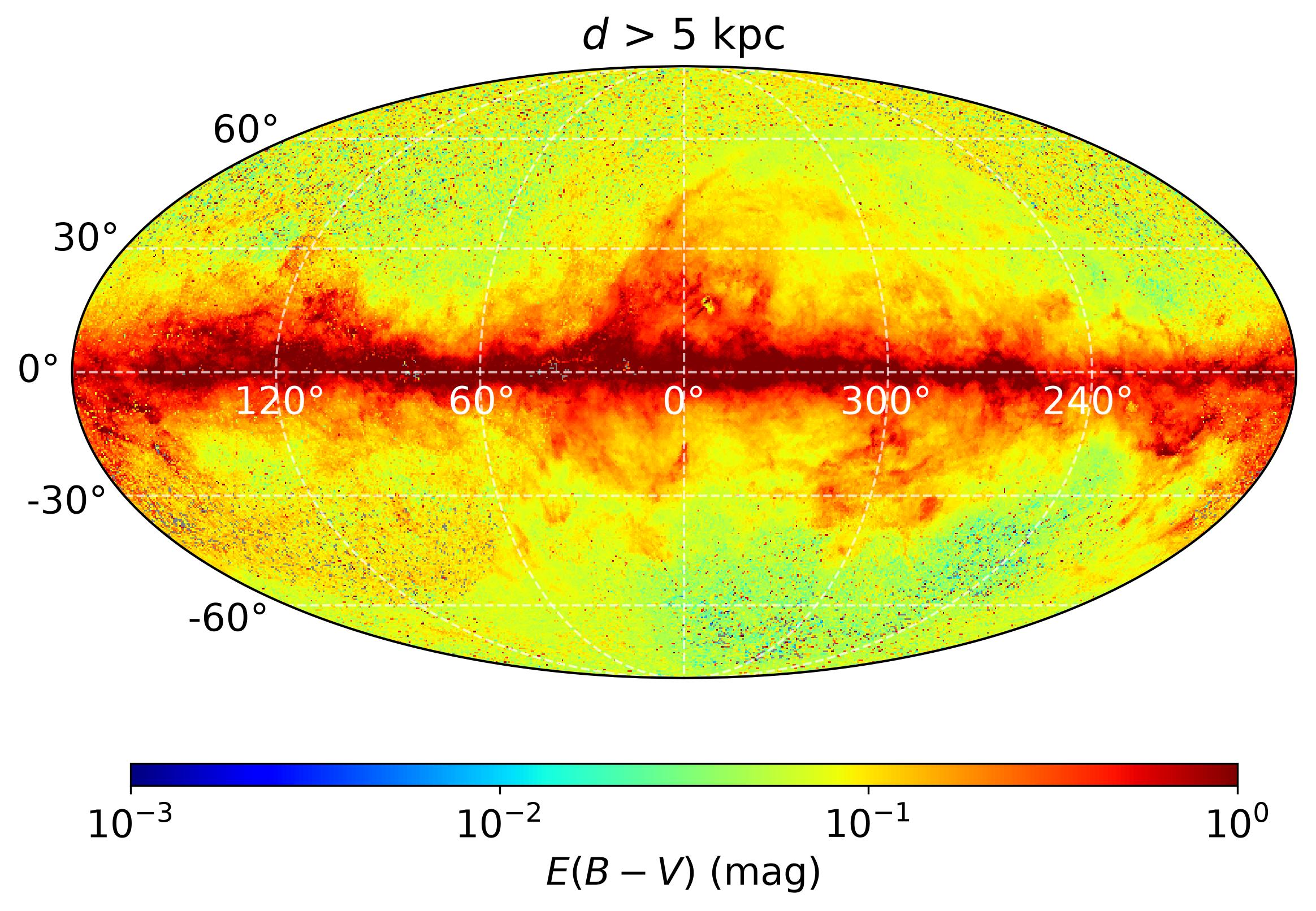}
	\caption{Cumulative $E(B-V)$ extinction sky maps in different distance intervals, shown in Galactic coordinates with a Mollweide projection. The two panels correspond to the ranges $d \textless 0.5$\,kpc (upper) and $d \textgreater 5$\,kpc (bottom). Since extinction is a cumulative quantity, the lower panel (built from stars at \(d > 5\)\,kpc) inherently includes the foreground extinction within 5\,kpc. Each pixel takes the inverse-variance-weighted mean $E(B-V)$ of the valid samples within it, and the colour bar denotes $E(B-V)$ in mag. The data pass a quality selection of $d\_err/d \leq 0.2$, $0 \leq E(B-V)\_err \leq 0.2$, and $E(B-V) \geq 0$.}
	\label{map}
\end{figure*}

\section{The Galactic Stellar Extinction Database}

We apply the trained models to the raw data and obtain a catalogue in which the extinction and distance systematics are corrected. To address potential depth mismatches, the catalogue includes a reliability flag for each entry, indicating whether the data is supervised, weakly supervised, or unsupervised. See Appendix~\ref{appendix_D} for details. The catalogue contains $1{,}935{,}987{,}083$ entries in total, wherein sources in common between catalogues are retained as independent entries and normalized separately, with detailed comparisons and discussions provided in Appendix~\ref{appendix_E}. The extinction maps used in this work already incorporate rigorous astrometric quality control; we thus do not impose an extra generic cut on these already well-vetted datasets.
Table~\ref{gsed_number} summarises the extinction entries from each work, including sample sizes, overlap counts with the baseline map,  and coverage. Notably, sources located in the Magellanic Clouds are retained in the catalogue. Fig.~\ref{map} shows the cumulative extinction sky maps of two distance ranges drawn from our catalogue on a HEALPix grid in Mollweide projection and Galactic coordinates. 
The data pass a quality selection of $d\_err/d \leq 0.2$, $0 \leq E(B-V)\_err \leq 0.2$, and $E(B-V) \geq 0$, and each pixel takes the inverse-variance-weighted mean $E(B-V)$.

\begin{table*}
	\centering
	\begin{threeparttable}
		\caption{Statistics of the extinction data. We present the sample size, overlap counts with baseline map (Zhang25), and coverage.}
		\label{gsed_number}
		\setlength{\tabcolsep}{3.5mm}{
			\begin{tabular}{cccc}
				\hline
				Source & Number&Overlap with Zhang25 &Coverage \\
				\hline
				Chen19 &  50,712,371 & 48,302,417 &	$0\degree<l<360\degree$, $\left| b \right|<10\degree$\\
				Green19 &  798,990,486 & 111,208,036\tnote{*} &	$\delta>-30\degree$	 \\
				Guo21 &  17,356,887 & 16,237,365 &Southern sky \\
				Sun25 &  140,599,779 & 82,582,427 &Southern sky\\
				Zhang25 &  219,197,643 & - &All-sky\\
				Zucker25 &  709,129,917 & 65,307,607	&
				$239\degree<l<6\degree$, $\left| b \right|<10\degree$\\
				\hline
		\end{tabular}}
		\begin{tablenotes}
			\small
			\item[*] The number of extinction data cross-matched with Zhang25. Of these, 109,741,886 entries have data in both the \(g\) and \(r\) bands.
		\end{tablenotes}
	\end{threeparttable}
\end{table*}

At small distances we see many well-known local dust cloud structures, while at large distances the high-extinction regions concentrate near the Galactic plane. The maps are consistent with the known large-scale features of the dust distribution. 
The upper panel shows reddening enhancements toward several well-known nearby cloud regions, including Taurus, Perseus, Ophiuchus, the Pipe Nebula, and Coalsack, whose approximate sky locations are labeled. In particular, the enhancements near the Galactic-centre direction at positive Galactic latitudes are spatially consistent with the locations of Ophiuchus and the Pipe Nebula. These associations provide a qualitative consistency check on the morphology of the reddening distribution. The lower panel shows a prominent band of high cumulative reddening along the Galactic plane for stars at \(d>5\) kpc. However, we note that the high Galactic latitude regions, which appear as the greener, lower-extinction areas in the maps, are subject to sparse stellar sampling, leading to considerably larger statistical uncertainties in the inferred \(E(B-V)\) values. Users should therefore exercise caution when interpreting extinction maps in these high-latitude, data-sparse regions. The reliability of individual pixels depends on the number of contributing stars, their reddening uncertainties, and their distance distribution. Pixels with limited sampling may be particularly sensitive to individual measurements. Isolated pixel-scale fluctuations should therefore be interpreted cautiously.

Storing such a large dataset and supporting real-time queries is a demanding task. For ease of use, we build the GSED with real-time query capability. The database holds ten columns of data, namely the right ascension (RA), declination (Dec), Galactic longitude ($l$), Galactic latitude ($b$), distance ($d$), distance error ($d\_err$), extinction ($E(B-V)$), extinction error ($E(B-V)\_err$), flag, and source. We build the database system through a strategy of reuse and customisation to enable fast queries, with four specific optimisations. First, we reuse HEALPix spatial partitioning and a multi-level tree index to shard the roughly $1.936$ billion extinction entries finely by sky region, so that a query locates only the target shards, which greatly reduces the scan range and accelerates the query. Second, we use a lightweight Python web framework that calls the encapsulated retrieval interface, data-parsing modules, and front-end display modules directly, integrating the data adaptation and the coordinate-system retrieval logic. Third, we reuse a coordinate-adaptation module to convert between the equatorial system, comprising RA and Dec, and the Galactic system, comprising $l$ and $b$, and to retrieve data precisely. Fourth, we reuse an LRU (Least Recently Used) caching technique to keep frequently queried sky regions in memory, further improving the retrieval speed of the integrated multi-source data.

\subsection{Distance--Extinction Curve Fitting}

The GSED database stores the extinction and distance measurements of individual stars, whereas users usually need the extinction at a specific target distance along a given line of sight. We therefore provide the capability to fit the discrete stellar extinction--distance data points along a line of sight into a continuous extinction-growth curve $E(d)$ and then to interpolate or extrapolate the extinction at any distance.

Extinction is essentially the cumulative contribution of the interstellar medium along the line of sight, so the curve $E(d)$ must satisfy the physical constraint of monotonic increase. We accordingly divide the line of sight into $k$ discrete distance intervals and parameterise the extinction increment of each interval as $e^{h_i}$, which is always positive, so that the cumulative extinction is $\sum_{i=0}^{k} e^{h_i}$ and $E(d)$ naturally increases monotonically with distance. For the fitting, we use MCMC sampling to explore the posterior probability of the parameters $\theta$. Considering that the queried extinction data may contain a certain fraction of outliers, such as distance or extinction anomalies caused by stellar misclassification or parallax errors, we adopt a Cauchy-like log-likelihood to reduce the penalty weight of large-residual points,
\begin{equation}
	\ln\mathcal{L} = -\sum_{i=1}^{N}\ln\left(1+\left[\frac{E_i - f(d_i, \theta)}{\sigma E_i}\right]^2\right),
\end{equation}
where $N$ is the number of valid samples and $f(d_i,\theta)$ is the cumulative extinction predicted by the model. Compared with the $\mathcal{L}_2$ norm, the logarithmic term $\ln(1+r^2)$ grows slowly, allowing the model to tolerate outliers without being dominated by them and ensuring that the fitted curve captures the overall trend of the extinction. To improve the convergence efficiency of the MCMC sampling, we adopt a cooperative initialisation strategy, first performing a pre-fit with nonlinear least squares to locate the optimal parameter range quickly and then constructing the initial-state matrix of the Markov chain centred on this result, introducing a moderate random perturbation to increase the dispersion of the initial distribution so that the sampler reaches the global optimum efficiently. 
We note that the “steps” in our fitted distance--extinction curves do not necessarily pinpoint the exact physical extinction jumps. This limitation primarily arises from two factors. First, the MCMC sampling incorporates a strict monotonic prior to ensure robust fitting against outliers in massive datasets, which inherently smooths the curves and suppresses the detection of sharp, localized variations. Second, the use of fixed 0.2\,kpc binning restricts the step positions to predefined boundaries, thereby reducing the positional precision of the inferred jumps. Users interested in the exact locations of extinction jumps can download the search results and perform more detailed independent analyses. 

\begin{figure*}
	\centering
	\includegraphics[width=0.95\linewidth]{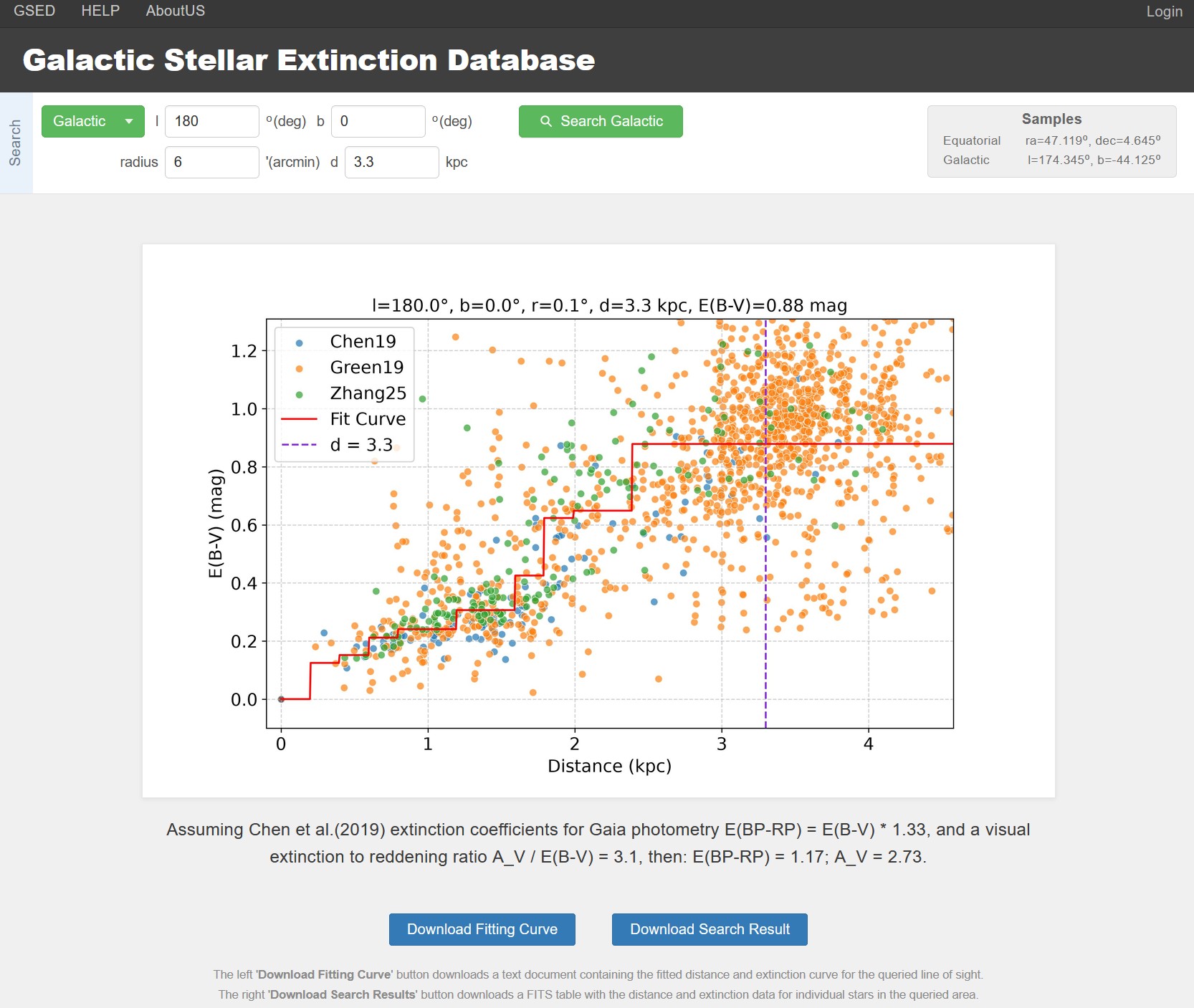}
	\caption{Query result of the GSED website in Galactic coordinates. After a user enters the Galactic longitude $l$=180.0\degree, the Galactic latitude $b$=0.0\degree, the search radius $r$=6$^{\prime}$, and the target distance $d$=3.3\,kpc, the system returns the extinction--distance relation of the stars along this line of sight. The coloured scatter points represent the queried stellar sample, with different colours corresponding to extinction data from different sources, namely Chen19, Green19, and Zhang25; the red step-like curve is the extinction--distance relation obtained by fitting the data; and the purple vertical dashed line marks the target distance $d$=3.3\,kpc entered by the user. The table in the upper-right corner provides an example of the input format; its values are illustrative only and are not intended to match the current search parameters.}
	\label{web}
\end{figure*}

\begin{table}
	\centering
	\caption{MCMC-fitted distance--extinction curve for the same line of sight as in Table~\ref{fits_table}, shown in part. The inputs are $l$=180.0\degree, $b$=0.0\degree, $r$=0.1\degree, and $d$=3.3\,kpc, and the fit returns $E(B-V)$=0.88\,mag.}
	\label{curve_fit}
	\begin{tabular}{cc}
		\hline
		Distance (kpc) & $E(B-V)$ (mag) \\
		\hline
		0.00& 0.00\\
		0.21& 0.12\\
		0.42& 0.15\\
		0.62& 0.21\\
		... & ...\\
		2.91& 0.88\\
		3.12& 0.88\\
		3.33& 0.88\\
		3.54& 0.88\\
		... & ...\\
		\hline
	\end{tabular}
\end{table}

\begin{table*}
	\centering
	\caption{Extinction catalogue obtained for the Galactic coordinate $l$=180.0\degree, $b$=0.0\degree\ with a search radius $r$=0.1\degree, showing the eight randomly selected rows. \footnote{Stars in common between catalogues are retained as independent entries and normalized separately, so a star can have multiple entries, one per source catalogue.}}
	\label{fits_table}
	\setlength{\tabcolsep}{2.5mm}{
		\begin{tabular}{cccccccccc}
			\hline
			RA  & Dec  & $l$ & $b$  & $d$ & $d\_err$ & $E(B-V)$ & $E(B-V)\_err$ & Flag & Source \\
			(deg) & (deg) &  (deg) & (deg) &  (kpc) & (kpc) &(mag) &  (mag) & \\
			\hline
			86.479897 & 28.895094 & 180.069220 & 0.034583 & 1.45 & 0.04 & 0.23 & 0.08 & (2,2) & Chen19 \\
			86.307139 & 28.936930 & 179.954710 & $-0.072681$ & 0.82 & 0.04 & 0.22 & 0.01 & (2,2) & Chen19 \\
			86.326735 & 28.892684 & 180.001406 & $-0.081128$ & 1.73 & 0.31 & 0.59 & 0.07 & (2,2) & Green19 \\
			86.347318 & 28.858186 & 180.040242 & $-0.083741$ & 5.95 & 3.76 & 0.59 & 0.09 & (2,2) & Green19 \\
			86.372967 & 28.857407 & 180.052618 & $-0.064977$ & 1.27 & 0.08 & 0.24 & 0.03 & (2,1) & Green19 \\
			86.474719 & 28.896571 & 180.065599 & 0.031481 & 2.66 & 0.47 & 0.73 & 0.05 & (2,0) & Green19 \\
			86.482277 & 28.994110 & 179.985758 & 0.087897 & 2.87 & 0.22 & 0.82 & 0.02 & (2,2) & Zhang25 \\
			86.452507 & 29.024628 & 179.946147 & 0.081552 & 2.42 & 0.27 & 0.81 & 0.01 & (2,2) & Zhang25 \\
			\hline
	\end{tabular}}
\end{table*}

\subsection{Using the Database}

To make these data easy to obtain, we develop a website accessible at \url{https://nadc.china-vo.org/data/gsed/} that aims to let users obtain extinction information quickly, as shown in Fig.~\ref{web}. From the coordinate entered by the user, which can be switched between the equatorial and Galactic systems, the search radius, and the specified distance, the extinction data within this region are obtained, and the program decides whether to fit the distance--extinction curve according to the amount of data. Specifically, when the queried extinction data number fewer than $30$, the program reports that the data are too few to fit and instead gives a median line computed from the queried extinction data; when the number is between $30$ and $120$, the program divides the data into two to six intervals according to the data amount and invokes MCMC to fit the distance--extinction curve; and when the number exceeds $120$, the program divides the line of sight into intervals of $0.2\,\mathrm{kpc}$ each by distance and invokes MCMC to fit the distance--extinction curve. In addition, when the distance entered by the user exceeds the maximum distance of the queried data, the program reports that the input distance is too large for the model to give a reliable $E(B-V)$ and asks for a smaller distance; in this case the program still invokes MCMC to fit the distance--extinction curve but does not return the $E(B-V)$ at the input distance. The program finally provides $E(BP-RP)$ and $A_V$, which by default are converted using the extinction coefficients of \citet{chen2019mnras}, namely $E(BP-RP)=E(B-V)\times1.33$ and $A_V=E(B-V)\times3.1$, while users may also choose other extinction laws to convert to the extinction in any band according to their scientific needs.

Fig.~\ref{web} shows the query result of the GSED website in Galactic coordinates. The top of the figure lists the input Galactic coordinates $l$=180.0\degree\ and $b$=0.0\degree, the search radius $r$=0.1\degree, the distance $d$=3.3\,kpc, and the fitted $E(B-V)$=0.88\,mag. The red curve is the extinction curve obtained by fitting the data, which increases gradually with distance. In general, the distance--extinction fit agrees well with the variation of the individual stellar extinction with distance and reflects the extinction variation of this region well. The text below the figure describes the conversion based on the extinction coefficients of \citet{chen2019mnras}, which is used to compute the colour excess $E(BP-RP)$ in the Gaia photometric system and the extinction $A_V$. The bottom of the interface provides two buttons, Download Fitting Curve and Download Search Result, which export the fitted distance--extinction curve data as shown in Table~\ref{curve_fit} and the queried extinction catalogue as shown in Table~\ref{fits_table}, respectively. Upon querying a specific sightline, the system displays the initial results and subsequently provides a set of checkboxes. Users can then select or deselect specific catalogues, allowing the combined \(E(B-V)\) estimates to be dynamically recalculated based on their selections.

\begin{figure*}
	\centering
	\includegraphics[width=0.95\linewidth]{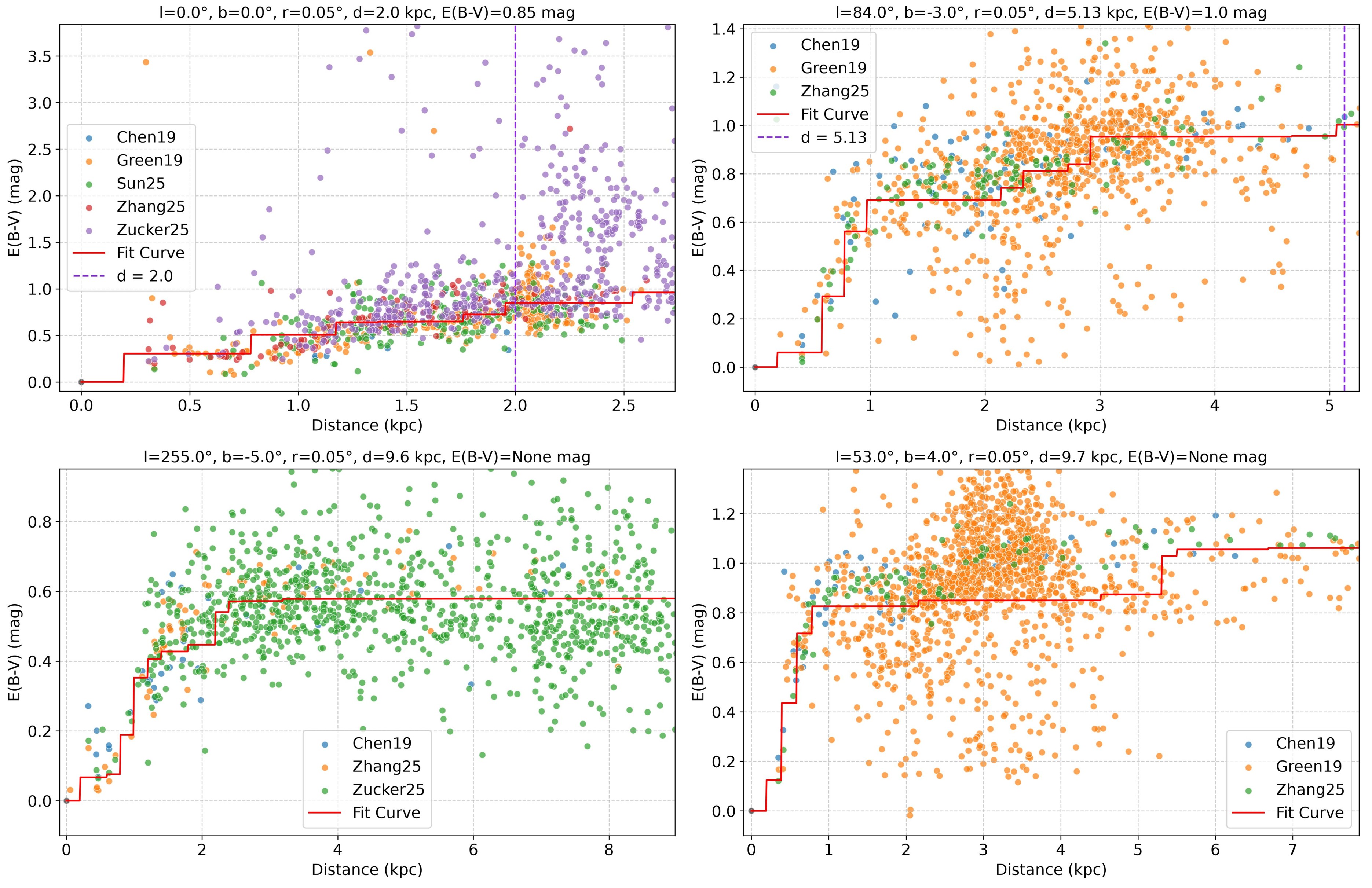}
	\\
	\includegraphics[width=0.95\linewidth]{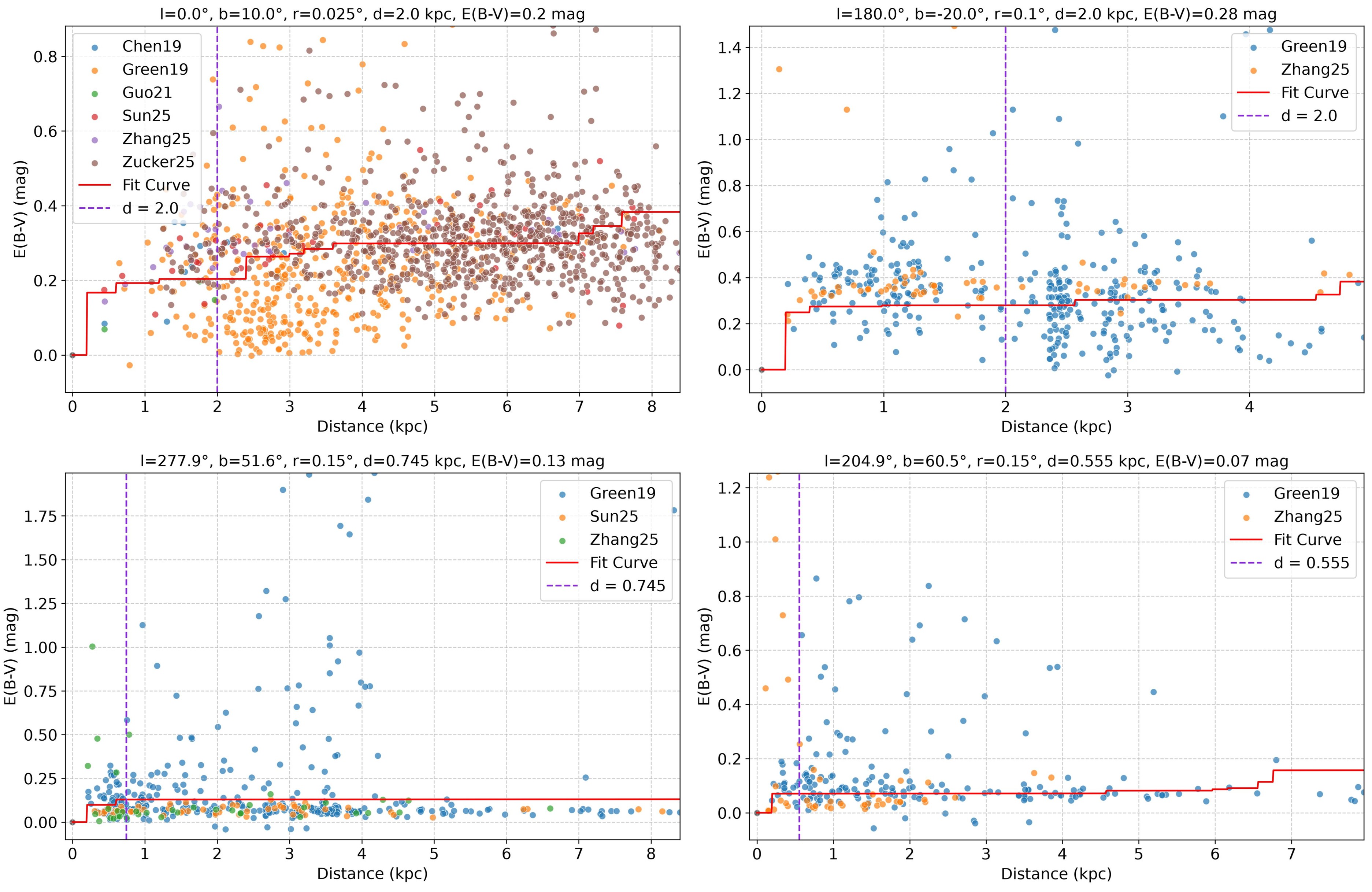}
	\caption{Similar to Fig.~\ref{web}, distance--extinction relations along various sightlines. The top two rows show inner-Galaxy sightlines, namely the Galactic center (default distance 2\,kpc) and GRBs 160623A, 221009A, and 031203, for which we adopt the most prominent dust-cloud distances of 5.13, 9.6, and 9.7\,kpc, respectively, following \citet{Vaia2026AA}. The third row shows two disk directions, and the bottom row shows two high-latitude molecular clouds from \citet{ONeill2026arxiv}, using their cloud distances as target distances. The search radius for each sightline is given in the corresponding panel title.}
	\label{sightlines}
\end{figure*}

Fig.~\ref{sightlines} shows the distance--extinction relations for representative sightlines across an 8-panel layout. The top two rows display inner Galaxy sightlines toward the Galactic center (\(l = 0\degree, b = 0\degree\)) and toward GRBs 160623A, 221009A, and 031203 from \citet{Vaia2026AA}. These sightlines cover inner- and outer-Galaxy disk directions, including regions independently characterized by X-ray dust scattering. 
The search radius for these panels is 0.05\degree. For the Galactic center (\(l = 0\degree, b = 0\degree\)), we adopt a default target distance of 2\,kpc, whereas for the sightlines toward GRBs 160623A, 221009A, and 031203, we adopt their respective dust-cloud distances. As shown in these panels, the prominent extinction jumps are concentrated within \(\sim 1-2\) kpc, and the fitted curves remain essentially flat beyond this range. Because the reliable stellar samples are insufficient to constrain the additional extinction at the target cloud distances for GRBs 221009A and 031203, we do not extrapolate the extinction values beyond our valid data range. 
The bottom two rows present low- and high-latitude sightlines. The third row shows the directions (\(l = 0\degree, b = 10\degree\)) and (\(l = 180\degree, b = -20\degree\)), representing the low-latitude disk and high-latitude outer-Galaxy, respectively, with search radii of 0.025\degree\ and 0.1\degree\, and a default target distance of 2\,kpc. 
The bottom row displays sightlines of high-altitude clouds at \(b > 50\degree\) from \citet{ONeill2026arxiv} at (\(l = 277.9\degree, b = 51.6\degree\)) and (\(l = 204.9\degree, b = 60.5\degree\)), with a search radius of 0.15\degree\ and target distances set to their respective cloud distances. A small step of $\Delta E(B-V)\approx0.03$\,mag is visible toward the former at $\sim0.6$\,kpc. For the latter, although the cloud is located at 0.555\,kpc, a similarly small step appears at a closer distance. This apparent shift may result from the relatively small number of stars at short distances, the intrinsically weak expected increment, and the presence of outliers affecting the fitting. Fitting the data in each panel yields the red step curves, which exhibit a gradual increase in extinction with distance. These distance--extinction fits accurately capture the overall trend of individual stellar extinctions, successfully reflecting the characteristic extinction variations along these diverse sightlines. 

\subsection{Limitations and Future Work}

At present the query function of GSED is oriented mainly toward a single line of sight, where the user enters one coordinate, search radius, and target distance each time and the system returns the extinction curve and extinction value along that direction. This design suits the extinction correction of individual objects or specific lines of sight but cannot efficiently handle the bulk extinction queries needed, for example, correcting extinction for a large stellar sample distributed across diverse sightlines, or generating extinction maps over multiple regions. The main difficulty of bulk queries is that the query for each line of sight requires a real-time database retrieval and an MCMC fit of the distance--extinction curve, so the computational cost grows linearly with the number of queries, while the stellar sample density and extinction morphology differ markedly among lines of sight and are difficult to accelerate with a unified parameterisation. 
We plan to provide a bulk-query entry in future versions, reducing the latency of large-scale queries through parallel computing and caching strategies and exploring a precomputed step-model parameterisation to replace the real-time MCMC fit, thereby balancing accuracy and efficiency. 

Another limitation of the current GSED is that spatial variations in the extinction-curve parameter $R(V)$, as highlighted by \citet{zxy2025Sci} and \citet{Zhang2025ApJL}, are not explicitly considered in the extinction homogenisation process. Although our MLP-based calibration reduces systematic differences among the input catalogues, variations in $R(V)$ may introduce residual biases in the homogenised $E(B-V)$ measurements and the resulting distance–extinction relations, particularly in regions with unusual dust properties. In addition, the default conversion from $E(B-V)$ to other extinction quantities relies on fixed extinction coefficients, which may introduce further uncertainties. Future versions of GSED will explore incorporating spatially varying $R(V)$ measurements to provide wavelength-dependent extinction information and thereby improve the accuracy and reliability of extinction estimates.

In addition, we plan to continuously incorporate newly released extinction catalogues to extend the coverage and distance depth of GSED, and to develop a Python package that allows users to call the query and fitting functions of GSED directly from their local programs. To ensure transparency and reproducibility, future updates of GSED will follow a version control scheme: each release is assigned a unique version number (e.g., GSED v1.0, v2.0), and its documentation records the baseline dataset, model configuration, release date, and a changelog.

\section{Summary}

We construct the GSED and normalise six representative 3D extinction datasets onto a common baseline of the colour excess $E(B-V)$ and parallax distance. To place the six datasets onto a common system, we design a six-layer MLP framework with two identical branches supervised respectively by the $E(B-V)$ and parallax-based distances from Zhang25. Specifically, the extinction branch takes the raw extinction and available stellar parameters or intrinsic colours as input, while the distance branch takes the Galactic coordinates and raw distance as input. 
Both branches effectively learn and remove the systematic differences inherent in the heterogeneous input catalogues. The colour excesses and distances produced by the model show no apparent systematic difference relative to the supervision values, indicating that the framework learns the complex mappings among heterogeneous datasets effectively. Applying the trained models to the raw data, we obtain a homogenised stellar extinction catalogue of $1{,}935{,}987{,}083$ entries in total and use it to draw an all-sky 3D extinction map of the Galaxy. Unlike traditional extinction maps that output only voxelised extinction values, GSED uses raw stellar data as its storage unit, supports a user-defined query radius, automatically invokes an MCMC fit of the distance--extinction curve, and provides $E(B-V)$, $E(G_{\rm BP}-G_{\rm RP})$, and $A_V$ directly.

The core advantage of GSED lies in its extensibility, as the systematic-correction and distance-correction framework established here can be reused for future extinction datasets without rebuilding from scratch. At the functional level, the current query is oriented mainly toward a single line of sight, and in the future we will provide a bulk-query interface that reduces the computational cost of large-scale extinction correction through parallel computing and pre-fitting strategies, together with a Python package that supports local use. At the data level, as new-generation surveys such as Gaia DR4, CSST, and LSST advance, higher-precision parallaxes and deeper multi-band photometry will further improve the quality of extinction measurements and distance estimates, and the data volume, coverage depth, and parameter richness of GSED will grow accordingly. More accurate extinction data not only help to improve the resolution and reliability of the 3D dust structure of the Galaxy, but also provide a data basis for finer studies of the interstellar medium, such as the spatial variation of the dust extinction curve and the internal extinction law of molecular clouds. We expect GSED to become a common infrastructure for extinction correction in the Galaxy and to evolve continuously as observational data and functionality grow richer.

\begin{acknowledgments}
	We thank the anonymous referee for their detailed and constructive comments, which have significantly improved the manuscript. This work is supported by the National Key Research and Development Program of China 2024YFA1611601, the National Natural Science Foundation of China 12322304 and 12273077, National Natural Science Foundation of Yunnan Province 202301AV070002, and the Xingdian talent support programme of Yunnan Province. We acknowledge the science research grants from the China Manned Space Project with no. CMS-CSST-2025-A11. 
	
	Data resources are supported by China National Astronomical Data Center (NADC) and Chinese Virtual Observatory (China-VO). 
\end{acknowledgments}

\bibliography{sample702}{}

\begin{thebibliography}{}
\expandafter\ifx\csname natexlab\endcsname\relax\def\natexlab#1{#1}\fi
\providecommand{\url}[1]{\href{#1}{#1}}
\providecommand{\dodoi}[1]{doi:~\href{http://doi.org/#1}{\nolinkurl{#1}}}
\providecommand{\doeprint}[1]{\href{http://ascl.net/#1}{\nolinkurl{http://ascl.net/#1}}}
\providecommand{\doarXiv}[1]{\href{https://arxiv.org/abs/#1}{\nolinkurl{https://arxiv.org/abs/#1}}}

\bibitem[{E.~B. {Am{\^o}res} {et~al.}(2021){Am{\^o}res}, {Jesus}, {Moitinho},
  {Arsenijevic}, {Levenhagen}, {Marshall}, {Kerber}, {K{\"u}nzel}, \&
  {Moura}}]{2021GALExtin}
{Am{\^o}res}, E.~B., {Jesus}, R.~M., {Moitinho}, A., {et~al.} 2021,
  \bibinfo{title}{{GALExtin: an alternative online tool to determine the
  interstellar extinction in the Milky Way},} \mnras, 508, 1788,
  \dodoi{10.1093/mnras/stab2248}

\bibitem[{M. {Berry} {et~al.}(2012){Berry}, {Ivezi{\'c}}, {Sesar}, {Juri{\'c}},
  {Schlafly}, {Bellovary}, {Finkbeiner}, {Vrbanec}, {Beers}, {Brooks},
  {Schneider}, {Gibson}, {Kimball}, {Jones}, {Yoachim}, {Krughoff}, {Connolly},
  {Loebman}, {Bond}, {Schlegel}, {Dalcanton}, {Yanny}, {Majewski}, {Knapp},
  {Gunn}, {Allyn Smith}, {Fukugita}, {Kent}, {Barentine}, {Krzesinski}, \&
  {Long}}]{Berry2012ApJ}
{Berry}, M., {Ivezi{\'c}}, {\v{Z}}., {Sesar}, B., {et~al.} 2012,
  \bibinfo{title}{{The Milky Way Tomography with Sloan Digital Sky Survey. IV.
  Dissecting Dust},} \apj, 757, 166, \dodoi{10.1088/0004-637X/757/2/166}

\bibitem[{K.~C. {Chambers} {et~al.}(2016){Chambers}, {Magnier}, {Metcalfe},
  {Flewelling}, {Huber}, {Waters}, {Denneau}, {Draper}, {Farrow}, {Finkbeiner},
  {Holmberg}, {Koppenhoefer}, {Price}, {Rest}, {Saglia}, {Schlafly}, {Smartt},
  {Sweeney}, {Wainscoat}, {Burgett}, {Chastel}, {Grav}, {Heasley}, {Hodapp},
  {Jedicke}, {Kaiser}, {Kudritzki}, {Luppino}, {Lupton}, {Monet}, {Morgan},
  {Onaka}, {Shiao}, {Stubbs}, {Tonry}, {White}, {Ba{\~n}ados}, {Bell},
  {Bender}, {Bernard}, {Boegner}, {Boffi}, {Botticella}, {Calamida},
  {Casertano}, {Chen}, {Chen}, {Cole}, {Deacon}, {Frenk}, {Fitzsimmons},
  {Gezari}, {Gibbs}, {Goessl}, {Goggia}, {Gourgue}, {Goldman}, {Grant},
  {Grebel}, {Hambly}, {Hasinger}, {Heavens}, {Heckman}, {Henderson}, {Henning},
  {Holman}, {Hopp}, {Ip}, {Isani}, {Jackson}, {Keyes}, {Koekemoer}, {Kotak},
  {Le}, {Liska}, {Long}, {Lucey}, {Liu}, {Martin}, {Masci}, {McLean}, {Mindel},
  {Misra}, {Morganson}, {Murphy}, {Obaika}, {Narayan}, {Nieto-Santisteban},
  {Norberg}, {Peacock}, {Pier}, {Postman}, {Primak}, {Rae}, {Rai}, {Riess},
  {Riffeser}, {Rix}, {R{\"o}ser}, {Russel}, {Rutz}, {Schilbach}, {Schultz},
  {Scolnic}, {Strolger}, {Szalay}, {Seitz}, {Small}, {Smith}, {Soderblom},
  {Taylor}, {Thomson}, {Taylor}, {Thakar}, {Thiel}, {Thilker}, {Unger},
  {Urata}, {Valenti}, {Wagner}, {Walder}, {Walter}, {Watters}, {Werner},
  {Wood-Vasey}, \& {Wyse}}]{2016arXiv161205560C}
{Chambers}, K.~C., {Magnier}, E.~A., {Metcalfe}, N., {et~al.} 2016,
  \bibinfo{title}{{The Pan-STARRS1 Surveys},} arXiv e-prints, arXiv:1612.05560,
  \dodoi{10.48550/arXiv.1612.05560}

\bibitem[{B.-Q. {Chen} {et~al.}(2015){Chen}, {Liu}, {Yuan}, {Huang}, \&
  {Xiang}}]{Chen2015MNRAS}
{Chen}, B.-Q., {Liu}, X.-W., {Yuan}, H.-B., {Huang}, Y., \& {Xiang}, M.-S.
  2015, \bibinfo{title}{{Dust-to-gas ratio, X$_{CO}$ factor and CO-dark gas in
  the Galactic anticentre: an observational study},} \mnras, 448, 2187,
  \dodoi{10.1093/mnras/stv103}

\bibitem[{B.~Q. {Chen} {et~al.}(2013){Chen}, {Schultheis}, {Jiang}, {Gonzalez},
  {Robin}, {Rejkuba}, \& {Minniti}}]{chen2013AA3dmap}
{Chen}, B.~Q., {Schultheis}, M., {Jiang}, B.~W., {et~al.} 2013,
  \bibinfo{title}{{Three-dimensional interstellar extinction map toward the
  Galactic bulge},} \aap, 550, A42, \dodoi{10.1051/0004-6361/201219682}

\bibitem[{B.-Q. {Chen} {et~al.}(2014){Chen}, {Liu}, {Yuan}, {Zhang},
  {Schultheis}, {Jiang}, {Huang}, {Xiang}, {Zhao}, {Yao}, \&
  {Lu}}]{chen2014mnrasanti}
{Chen}, B.-Q., {Liu}, X.-W., {Yuan}, H.-B., {et~al.} 2014, \bibinfo{title}{{A
  three-dimensional extinction map of the Galactic anticentre from multiband
  photometry},} \mnras, 443, 1192, \dodoi{10.1093/mnras/stu1192}

\bibitem[{B.-Q. {Chen} {et~al.}(2019){Chen}, {Huang}, {Yuan}, {Wang}, {Fan},
  {Xiang}, {Zhang}, {Tian}, \& {Liu}}]{chen2019mnras}
{Chen}, B.-Q., {Huang}, Y., {Yuan}, H.-B., {et~al.} 2019,
  \bibinfo{title}{{Three-dimensional interstellar dust reddening maps of the
  Galactic plane},} \mnras, 483, 4277, \dodoi{10.1093/mnras/sty3341}

\bibitem[{B.~T. {Draine}(2003){Draine}}]{Draine2003ARAA}
{Draine}, B.~T. 2003, \bibinfo{title}{{Interstellar Dust Grains},} \araa, 41,
  241, \dodoi{10.1146/annurev.astro.41.011802.094840}

\bibitem[{G. {Edenhofer} {et~al.}(2024){Edenhofer}, {Zucker}, {Frank},
  {Saydjari}, {Speagle}, {Finkbeiner}, \& {En{\ss}lin}}]{Edenhofer2024AA}
{Edenhofer}, G., {Zucker}, C., {Frank}, P., {et~al.} 2024, \bibinfo{title}{{A
  parsec-scale Galactic 3D dust map out to 1.25 kpc from the Sun},} \aap, 685,
  A82, \dodoi{10.1051/0004-6361/202347628}

\bibitem[{{}{Gaia Collaboration} {et~al.}(2018){Gaia Collaboration}, {Brown},
  {Vallenari}, {Prusti}, {de Bruijne}, {Babusiaux}, \&
  {Bailer-Jones}}]{gaia2018AA}
{Gaia Collaboration}, {Brown}, A.~G.~A., {Vallenari}, A., {et~al.} 2018,
  \bibinfo{title}{{Gaia Data Release 2. Summary of the contents and survey
  properties},} \aap, 616, A1, \dodoi{10.1051/0004-6361/201833051}

\bibitem[{{}{Gaia Collaboration} {et~al.}(2023){Gaia Collaboration},
  {Vallenari}, {Brown}, {Prusti}, {de Bruijne}, {Arenou}, {Babusiaux},
  {Biermann}, {Creevey}, {Ducourant}, {Evans}, {Eyer}, {Guerra}, {Hutton},
  {Jordi}, {Klioner}, {Lammers}, {Lindegren}, {Luri}, {Mignard}, {Panem},
  {Pourbaix}, {Randich}, {Sartoretti}, {Soubiran}, {Tanga}, {Walton},
  {Bailer-Jones}, {Bastian}, {Drimmel}, {Jansen}, {Katz}, {Lattanzi}, {van
  Leeuwen}, {Bakker}, {Cacciari}, {Casta{\~n}eda}, {De Angeli}, {Fabricius},
  {Fouesneau}, {Fr{\'e}mat}, {Galluccio}, {Guerrier}, {Heiter}, {Masana},
  {Messineo}, {Mowlavi}, {Nicolas}, {Nienartowicz}, {Pailler}, {Panuzzo},
  {Riclet}, {Roux}, {Seabroke}, {Sordo}, {Th{\'e}venin}, {Gracia-Abril},
  {Portell}, {Teyssier}, {Altmann}, {Andrae}, {Audard}, {Bellas-Velidis},
  {Benson}, {Berthier}, {Blomme}, {Burgess}, {Busonero}, {Busso},
  {C{\'a}novas}, {Carry}, {Cellino}, {Cheek}, {Clementini}, {Damerdji},
  {Davidson}, {de Teodoro}, {Nu{\~n}ez Campos}, {Delchambre}, {Dell'Oro},
  {Esquej}, {Fern{\'a}ndez-Hern{\'a}ndez}, {Fraile}, {Garabato},
  {Garc{\'\i}a-Lario}, {Gosset}, {Haigron}, {Halbwachs}, {Hambly}, {Harrison},
  {Hern{\'a}ndez}, {Hestroffer}, {Hodgkin}, {Holl}, {Jan{\ss}en}, {Jevardat de
  Fombelle}, {Jordan}, {Krone-Martins}, {Lanzafame}, {L{\"o}ffler}, {Marchal},
  {Marrese}, {Moitinho}, {Muinonen}, {Osborne}, {Pancino}, {Pauwels},
  {Recio-Blanco}, {Reyl{\'e}}, {Riello}, {Rimoldini}, {Roegiers}, {Rybizki},
  {Sarro}, {Siopis}, {Smith}, {Sozzetti}, {Utrilla}, {van Leeuwen}, {Abbas},
  {{\'A}brah{\'a}m}, {Abreu Aramburu}, {Aerts}, {Aguado}, {Ajaj},
  {Aldea-Montero}, {Altavilla}, {{\'A}lvarez}, {Alves}, {Anders}, {Anderson},
  {Anglada Varela}, {Antoja}, {Baines}, {Baker}, {Balaguer-N{\'u}{\~n}ez},
  {Balbinot}, {Balog}, {Barache}, {Barbato}, {Barros}, {Barstow},
  {Bartolom{\'e}}, {Bassilana}, {Bauchet}, {Becciani}, {Bellazzini},
  {Berihuete}, {Bernet}, {Bertone}, {Bianchi}, {Binnenfeld}, {Blanco-Cuaresma},
  {Blazere}, {Boch}, {Bombrun}, {Bossini}, {Bouquillon}, {Bragaglia},
  {Bramante}, {Breedt}, {Bressan}, {Brouillet}, {Brugaletta}, {Bucciarelli},
  {Burlacu}, {Butkevich}, {Buzzi}, {Caffau}, {Cancelliere}, {Cantat-Gaudin},
  {Carballo}, {Carlucci}, {Carnerero}, {Carrasco}, {Casamiquela}, {Castellani},
  {Castro-Ginard}, {Chaoul}, {Charlot}, {Chemin}, {Chiaramida}, {Chiavassa},
  {Chornay}, {Comoretto}, {Contursi}, {Cooper}, {Cornez}, {Cowell}, {Crifo},
  {Cropper}, {Crosta}, {Crowley}, {Dafonte}, {Dapergolas}, {David}, {David},
  {de Laverny}, {De Luise}, \& {De March}}]{gaia2023A&A}
{Gaia Collaboration}, {Vallenari}, A., {Brown}, A.~G.~A., {et~al.} 2023,
  \bibinfo{title}{{Gaia Data Release 3. Summary of the content and survey
  properties},} \aap, 674, A1, \dodoi{10.1051/0004-6361/202243940}

\bibitem[{G.~A. {Gontcharov} {et~al.}(2025){Gontcharov}, {Marchuk},
  {Savchenko}, {Mosenkov}, {Il'in}, {Poliakov}, {Smirnov}, \&
  {Krayani}}]{Gontcharov2025RAA}
{Gontcharov}, G.~A., {Marchuk}, A.~A., {Savchenko}, S.~S., {et~al.} 2025,
  \bibinfo{title}{{Foreground Extinction to Extended Celestial Objects. I. New
  Extinction Maps},} Research in Astronomy and Astrophysics, 25, 125016,
  \dodoi{10.1088/1674-4527/ae12a6}

\bibitem[{O.~A. {Gonzalez} {et~al.}(2012){Gonzalez}, {Rejkuba}, {Zoccali},
  {Valenti}, {Minniti}, {Schultheis}, {Tobar}, \& {Chen}}]{Gonzalez2012AA}
{Gonzalez}, O.~A., {Rejkuba}, M., {Zoccali}, M., {et~al.} 2012,
  \bibinfo{title}{{Reddening and metallicity maps of the Milky Way bulge from
  VVV and 2MASS. II. The complete high resolution extinction map and
  implications for Galactic bulge studies},} \aap, 543, A13,
  \dodoi{10.1051/0004-6361/201219222}

\bibitem[{G. {Green}(2018){Green}}]{2018dustmaps}
{Green}, G. 2018, \bibinfo{title}{{dustmaps: A Python interface for maps of
  interstellar dust},} The Journal of Open Source Software, 3, 695,
  \dodoi{10.21105/joss.00695}

\bibitem[{G.~M. {Green} {et~al.}(2019){Green}, {Schlafly}, {Zucker}, {Speagle},
  \& {Finkbeiner}}]{green2019ApJ}
{Green}, G.~M., {Schlafly}, E., {Zucker}, C., {Speagle}, J.~S., \&
  {Finkbeiner}, D. 2019, \bibinfo{title}{{A 3D Dust Map Based on Gaia,
  Pan-STARRS 1, and 2MASS},} \apj, 887, 93, \dodoi{10.3847/1538-4357/ab5362}

\bibitem[{G.~M. {Green} {et~al.}(2015){Green}, {Schlafly}, {Finkbeiner}, {Rix},
  {Martin}, {Burgett}, {Draper}, {Flewelling}, {Hodapp}, {Kaiser}, {Kudritzki},
  {Magnier}, {Metcalfe}, {Price}, {Tonry}, \& {Wainscoat}}]{green2015a3dmap}
{Green}, G.~M., {Schlafly}, E.~F., {Finkbeiner}, D.~P., {et~al.} 2015,
  \bibinfo{title}{{A Three-dimensional Map of Milky Way Dust},} \apj, 810, 25,
  \dodoi{10.1088/0004-637X/810/1/25}

\bibitem[{G.~M. {Green} {et~al.}(2018){Green}, {Schlafly}, {Finkbeiner}, {Rix},
  {Martin}, {Burgett}, {Draper}, {Flewelling}, {Hodapp}, {Kaiser}, {Kudritzki},
  {Magnier}, {Metcalfe}, {Tonry}, {Wainscoat}, \& {Waters}}]{green2018MNRAS}
{Green}, G.~M., {Schlafly}, E.~F., {Finkbeiner}, D., {et~al.} 2018,
  \bibinfo{title}{{Galactic reddening in 3D from stellar photometry - an
  improved map},} \mnras, 478, 651, \dodoi{10.1093/mnras/sty1008}

\bibitem[{H.-L. {Guo} {et~al.}(2021){Guo}, {Chen}, {Yuan}, {Huang}, {Liu},
  {Yang}, {Li}, {Sun}, \& {Liu}}]{guo2021ApJ}
{Guo}, H.-L., {Chen}, B.-Q., {Yuan}, H.-B., {et~al.} 2021,
  \bibinfo{title}{{Three-dimensional Distribution of the Interstellar Dust in
  the Milky Way},} \apj, 906, 47, \dodoi{10.3847/1538-4357/abc68a}

\bibitem[{R.~J. {Hanson} {et~al.}(2016){Hanson}, {Bailer-Jones}, {Burgett},
  {Chambers}, {Hodapp}, {Kaiser}, {Tonry}, {Wainscoat}, \&
  {Waters}}]{Hanson2016MNRAS}
{Hanson}, R.~J., {Bailer-Jones}, C.~A.~L., {Burgett}, W.~S., {et~al.} 2016,
  \bibinfo{title}{{Dust in three dimensions in the Galactic plane},} \mnras,
  463, 3604, \dodoi{10.1093/mnras/stw2240}

\bibitem[{L. {Lindegren} {et~al.}(2021){Lindegren}, {Bastian}, {Biermann},
  {Bombrun}, {de Torres}, {Gerlach}, {Geyer}, {Hern{\'a}ndez}, {Hilger},
  {Hobbs}, {Klioner}, {Lammers}, {McMillan}, {Ramos-Lerate},
  {Steidelm{\"u}ller}, {Stephenson}, \& {van Leeuwen}}]{Lindegren2020AA}
{Lindegren}, L., {Bastian}, U., {Biermann}, M., {et~al.} 2021,
  \bibinfo{title}{{Gaia Early Data Release 3. Parallax bias versus magnitude,
  colour, and position},} \aap, 649, A4, \dodoi{10.1051/0004-6361/202039653}

\bibitem[{D.~J. {Marshall} {et~al.}(2006){Marshall}, {Robin}, {Reyl{\'e}},
  {Schultheis}, \& {Picaud}}]{marshall2006AA}
{Marshall}, D.~J., {Robin}, A.~C., {Reyl{\'e}}, C., {Schultheis}, M., \&
  {Picaud}, S. 2006, \bibinfo{title}{{Modelling the Galactic interstellar
  extinction distribution in three dimensions},} \aap, 453, 635,
  \dodoi{10.1051/0004-6361:20053842}

\bibitem[{T.~J. {O'Neill} {et~al.}(2026){O'Neill}, {Saydjari}, {Zucker},
  {Koch}, {Benjamin}, {Frank}, \& {Yoshida}}]{ONeill2026arxiv}
{O'Neill}, T.~J., {Saydjari}, A.~K., {Zucker}, C., {et~al.} 2026,
  \bibinfo{title}{{The 3D Structure and Kinematics of the Local Disk-Halo
  Interface: Intermediate-velocity Clouds are the Minority of High-altitude
  Clouds in the Solar Neighborhood},} arXiv e-prints, arXiv:2605.24342,
  \dodoi{10.48550/arXiv.2605.24342}

\bibitem[{{}{Planck Collaboration} {et~al.}(2014){Planck Collaboration},
  {Abergel}, {Ade}, {Aghanim}, {Alves}, {Aniano}, {Armitage-Caplan}, {Arnaud},
  {Ashdown}, {Atrio-Barandela}, {Aumont}, {Baccigalupi}, {Banday}, {Barreiro},
  {Bartlett}, {Battaner}, {Benabed}, {Beno{\^\i}t}, {Benoit-L{\'e}vy},
  {Bernard}, {Bersanelli}, {Bielewicz}, {Bobin}, {Bock}, {Bonaldi}, {Bond},
  {Borrill}, {Bouchet}, {Boulanger}, {Bridges}, {Bucher}, {Burigana}, {Butler},
  {Cardoso}, {Catalano}, {Chamballu}, {Chary}, {Chiang}, {Chiang},
  {Christensen}, {Church}, {Clemens}, {Clements}, {Colombi}, {Colombo},
  {Combet}, {Couchot}, {Coulais}, {Crill}, {Curto}, {Cuttaia}, {Danese},
  {Davies}, {Davis}, {de Bernardis}, {de Rosa}, {de Zotti}, {Delabrouille},
  {Delouis}, {D{\'e}sert}, {Dickinson}, {Diego}, {Dole}, {Donzelli},
  {Dor{\'e}}, {Douspis}, {Draine}, {Dupac}, {Efstathiou}, {En{\ss}lin},
  {Eriksen}, {Falgarone}, {Finelli}, {Forni}, {Frailis}, {Fraisse},
  {Franceschi}, {Galeotta}, {Ganga}, {Ghosh}, {Giard}, {Giardino},
  {Giraud-H{\'e}raud}, {Gonz{\'a}lez-Nuevo}, {G{\'o}rski}, {Gratton},
  {Gregorio}, {Grenier}, {Gruppuso}, {Guillet}, {Hansen}, {Hanson}, {Harrison},
  {Helou}, {Henrot-Versill{\'e}}, {Hern{\'a}ndez-Monteagudo}, {Herranz},
  {Hildebrandt}, {Hivon}, {Hobson}, {Holmes}, {Hornstrup}, {Hovest},
  {Huffenberger}, {Jaffe}, {Jaffe}, {Jewell}, {Joncas}, {Jones}, {Juvela},
  {Keih{\"a}nen}, {Keskitalo}, {Kisner}, {Knoche}, {Knox}, {Kunz},
  {Kurki-Suonio}, {Lagache}, {L{\"a}hteenm{\"a}ki}, {Lamarre}, {Lasenby},
  {Laureijs}, {Lawrence}, {Leonardi}, {Le{\'o}n-Tavares}, {Lesgourgues},
  {Levrier}, {Liguori}, {Lilje}, {Linden-V{\o}rnle}, {L{\'o}pez-Caniego},
  {Lubin}, {Mac{\'\i}as-P{\'e}rez}, {Maffei}, {Maino}, {Mandolesi}, {Maris},
  {Marshall}, {Martin}, {Mart{\'\i}nez-Gonz{\'a}lez}, {Masi}, {Massardi},
  {Matarrese}, {Matthai}, {Mazzotta}, {McGehee}, {Melchiorri}, {Mendes},
  {Mennella}, {Migliaccio}, {Mitra}, {Miville-Desch{\^e}nes}, {Moneti},
  {Montier}, {Morgante}, {Mortlock}, {Munshi}, {Murphy}, {Naselsky}, {Nati},
  {Natoli}, {Netterfield}, {N{\o}rgaard-Nielsen}, {Noviello}, {Novikov},
  {Novikov}, {Osborne}, {Oxborrow}, {Paci}, {Pagano}, {Pajot}, {Paladini},
  {Paoletti}, {Pasian}, {Patanchon}, {Perdereau}, {Perotto}, {Perrotta},
  {Piacentini}, {Piat}, {Pierpaoli}, {Pietrobon}, {Plaszczynski},
  {Pointecouteau}, {Polenta}, {Ponthieu}, {Popa}, {Poutanen}, {Pratt},
  {Pr{\'e}zeau}, {Prunet}, {Puget}, {Rachen}, {Reach}, {Rebolo}, {Reinecke},
  {Remazeilles}, {Renault}, {Ricciardi}, \& {Riller}}]{Planck2014AA}
{Planck Collaboration}, {Abergel}, A., {Ade}, P.~A.~R., {et~al.} 2014,
  \bibinfo{title}{{Planck 2013 results. XI. All-sky model of thermal dust
  emission},} \aap, 571, A11, \dodoi{10.1051/0004-6361/201323195}

\bibitem[{A.~K. {Saydjari} {et~al.}(2023){Saydjari}, {Schlafly}, {Lang},
  {Meisner}, {Green}, {Zucker}, {Zelko}, {Speagle}, {Daylan}, {Lee}, {Valdes},
  {Schlegel}, \& {Finkbeiner}}]{Saydjari2023ApJS}
{Saydjari}, A.~K., {Schlafly}, E.~F., {Lang}, D., {et~al.} 2023,
  \bibinfo{title}{{The Dark Energy Camera Plane Survey 2 (DECaPS2): More Sky,
  Less Bias, and Better Uncertainties},} \apjs, 264, 28,
  \dodoi{10.3847/1538-4365/aca594}

\bibitem[{D.~J. {Schlegel} {et~al.}(1998){Schlegel}, {Finkbeiner}, \&
  {Davis}}]{sfd1998ApJ}
{Schlegel}, D.~J., {Finkbeiner}, D.~P., \& {Davis}, M. 1998,
  \bibinfo{title}{{Maps of Dust Infrared Emission for Use in Estimation of
  Reddening and Cosmic Microwave Background Radiation Foregrounds},} \apj, 500,
  525, \dodoi{10.1086/305772}

\bibitem[{M. {Schultheis} {et~al.}(1999){Schultheis}, {Ganesh}, {Simon},
  {Omont}, {Alard}, {Borsenberger}, {Copet}, {Epchtein}, {Fouqu{\'e}}, \&
  {Habing}}]{Schultheis1999AA}
{Schultheis}, M., {Ganesh}, S., {Simon}, G., {et~al.} 1999,
  \bibinfo{title}{{Interstellar extinction towards the inner Galactic Bulge},}
  \aap, 349, L69, \dodoi{10.48550/arXiv.astro-ph/9908349}

\bibitem[{M. {Schultheis} {et~al.}(2014){Schultheis}, {Chen}, {Jiang},
  {Gonzalez}, {Enokiya}, {Fukui}, {Torii}, {Rejkuba}, \&
  {Minniti}}]{Schultheis2014AA}
{Schultheis}, M., {Chen}, B.~Q., {Jiang}, B.~W., {et~al.} 2014,
  \bibinfo{title}{{Mapping the Milky Way bulge at high resolution: the 3D dust
  extinction, CO, and X factor maps},} \aap, 566, A120,
  \dodoi{10.1051/0004-6361/201322788}

\bibitem[{H. {Shen} {et~al.}(2022){Shen}, {Chen}, {Guo}, {Yuan}, {Sun}, \&
  {Li}}]{Shen2022MNRAS}
{Shen}, H., {Chen}, B.-Q., {Guo}, H.-L., {et~al.} 2022,
  \bibinfo{title}{{Estimating accurate reddening values of LAMOST M dwarfs},}
  \mnras, 514, 4398, \dodoi{10.1093/mnras/stac1615}

\bibitem[{M.~F. {Skrutskie} {et~al.}(2006){Skrutskie}, {Cutri}, {Stiening},
  {Weinberg}, {Schneider}, {Carpenter}, {Beichman}, {Capps}, {Chester},
  {Elias}, {Huchra}, {Liebert}, {Lonsdale}, {Monet}, {Price}, {Seitzer},
  {Jarrett}, {Kirkpatrick}, {Gizis}, {Howard}, {Evans}, {Fowler}, {Fullmer},
  {Hurt}, {Light}, {Kopan}, {Marsh}, {McCallon}, {Tam}, {Van Dyk}, \&
  {Wheelock}}]{Skrutskie2006AJ}
{Skrutskie}, M.~F., {Cutri}, R.~M., {Stiening}, R., {et~al.} 2006,
  \bibinfo{title}{{The Two Micron All Sky Survey (2MASS)},} \aj, 131, 1163,
  \dodoi{10.1086/498708}

\bibitem[{M. {Sun} {et~al.}(2023){Sun}, {Chen}, {Guo}, {Zhao}, {Yang}, \&
  {Cui}}]{Sun2023AJ}
{Sun}, M., {Chen}, B., {Guo}, H., {et~al.} 2023, \bibinfo{title}{{SPar:
  Estimating Stellar Parameters from Multiband Photometries with Empirical
  Stellar Libraries},} \aj, 166, 126, \dodoi{10.3847/1538-3881/ace5ab}

\bibitem[{M. {Sun} {et~al.}(2025){Sun}, {Chen}, {Sun}, {Wang}, {Yu}, {Zhang},
  {Zhang}, {Bao}, {Zeng}, {Yang}, \& {Cui}}]{sun2025RAA}
{Sun}, M., {Chen}, B., {Sun}, B., {et~al.} 2025, \bibinfo{title}{{Stellar
  Parameters, Extinction, and Distances for Stars in SMSS DR2 by the SPar
  Method},} Research in Astronomy and Astrophysics, 25, 057002,
  \dodoi{10.1088/1674-4527/adc5e1}

\bibitem[{R.~J. {Trumpler}(1930){Trumpler}}]{Trumpler1930PASP}
{Trumpler}, R.~J. 1930, \bibinfo{title}{{Absorption of Light in the Galactic
  System},} \pasp, 42, 214, \dodoi{10.1086/124039}

\bibitem[{B. {Vaia} {et~al.}(2026){Vaia}, {Fornasiero}, {Tiengo}, {Bracco},
  {Jeli{\'c}}, {Bo{\v{s}}njak}, {Pintore}, \& {Esposito}}]{Vaia2026AA}
{Vaia}, B., {Fornasiero}, I., {Tiengo}, A., {et~al.} 2026,
  \bibinfo{title}{{Accurate distances of the Galactic spiral arms from
  dust-scattered X-ray emission of gamma-ray bursts},} \aap, 710, A196,
  \dodoi{10.1051/0004-6361/202557431}

\bibitem[{T. {Wang} {et~al.}(2025){Wang}, {Yuan}, {Chen}, {Xiang}, {Zhang},
  {Huang}, {Gu}, {Wang}, \& {Li}}]{wang2025ApJS}
{Wang}, T., {Yuan}, H., {Chen}, B., {et~al.} 2025, \bibinfo{title}{{An All-sky
  3D Dust Map Based on Gaia and LAMOST},} \apjs, 280, 15,
  \dodoi{10.3847/1538-4365/adea39}

\bibitem[{C. {Wolf} {et~al.}(2018){Wolf}, {Onken}, {Luvaul}, {Schmidt},
  {Bessell}, {Chang}, {Da Costa}, {Mackey}, {Martin-Jones}, {Murphy},
  {Preston}, {Scalzo}, {Shao}, {Smillie}, {Tisserand}, {White}, \&
  {Yuan}}]{Wolf2018PASA}
{Wolf}, C., {Onken}, C.~A., {Luvaul}, L.~C., {et~al.} 2018,
  \bibinfo{title}{{SkyMapper Southern Survey: First Data Release (DR1)},}
  \pasa, 35, e010, \dodoi{10.1017/pasa.2018.5}

\bibitem[{J. {Yu} {et~al.}(2026){Yu}, {Casagrande}, {Taylor}, {Ciuc{\u{a}}},
  {Cordoni}, {Drimmel}, {Khanna}, {Nguyen}, {R{\'o}{\.z}a{\'n}ski}, {Stello},
  {Yuan}, \& {Yuan}}]{Yu2026MNRAS}
{Yu}, J., {Casagrande}, L., {Taylor}, J.~A., {et~al.} 2026,
  \bibinfo{title}{{High-fidelity stellar extinction with Gaia and APOGEE ─ I.
  The method and a new extinction curve},} \mnras, 549, stag848,
  \dodoi{10.1093/mnras/stag848}

\bibitem[{R. {Zhang} {\&} H. {Yuan}(2023){Zhang} \& {Yuan}}]{ZhangYuan2023ApJS}
{Zhang}, R., \& {Yuan}, H. 2023, \bibinfo{title}{{Empirical Temperature- and
  Extinction-dependent Extinction Coefficients for the GALEX, Pan-STARRS 1,
  Gaia, SDSS, 2MASS, and WISE Passbands},} \apjs, 264, 14,
  \dodoi{10.3847/1538-4365/ac9dfa}

\bibitem[{X. {Zhang} {\&} G.~M. {Green}(2025){Zhang} \& {Green}}]{zxy2025Sci}
{Zhang}, X., \& {Green}, G.~M. 2025, \bibinfo{title}{{Three-dimensional maps of
  the interstellar dust extinction curve within the Milky Way galaxy},}
  Science, 387, 1209, \dodoi{10.1126/science.ado9787}

\bibitem[{X. {Zhang} {et~al.}(2023){Zhang}, {Green}, \& {Rix}}]{zxy2023MNRAS}
{Zhang}, X., {Green}, G.~M., \& {Rix}, H.-W. 2023, \bibinfo{title}{{Parameters
  of 220 million stars from Gaia BP/RP spectra},} \mnras, 524, 1855,
  \dodoi{10.1093/mnras/stad1941}

\bibitem[{X. {Zhang} {et~al.}(2025{\natexlab{a}}){Zhang}, {Hensley}, \&
  {Green}}]{Zhang2025ApJL}
{Zhang}, X., {Hensley}, B.~S., \& {Green}, G.~M. 2025{\natexlab{a}},
  \bibinfo{title}{{Dust-extinction-curve Variation in the Translucent
  Interstellar Medium Is Driven by Polycyclic Aromatic Hydrocarbon Growth},}
  \apjl, 979, L17, \dodoi{10.3847/2041-8213/ada28f}

\bibitem[{G. {Zhao} {et~al.}(2012){Zhao}, {Zhao}, {Chu}, {Jing}, \&
  {Deng}}]{Zhao2012RAA}
{Zhao}, G., {Zhao}, Y.-H., {Chu}, Y.-Q., {Jing}, Y.-P., \& {Deng}, L.-C. 2012,
  \bibinfo{title}{{LAMOST spectral survey {\textemdash} An overview},} Research
  in Astronomy and Astrophysics, 12, 723, \dodoi{10.1088/1674-4527/12/7/002}

\bibitem[{C. {Zucker} {et~al.}(2025){Zucker}, {Saydjari}, {Speagle},
  {Schlafly}, {Green}, {Benjamin}, {Peek}, {Edenhofer}, {Goodman}, {Kuhn}, \&
  {Finkbeiner}}]{Zucker2025ApJ}
{Zucker}, C., {Saydjari}, A.~K., {Speagle}, J.~S., {et~al.} 2025,
  \bibinfo{title}{{A Deep, High-angular-resolution 3D Dust Map of the Southern
  Galactic Plane},} \apj, 992, 39, \dodoi{10.3847/1538-4357/adfbe6}

\end{thebibliography}
\bibliographystyle{aasjournalv7.1}

\appendix

\section{Per-Catalogue Results of the Extinction-Systematics Correction}\label{appendix_A}

\subsection{Green19}\label{apxA_green19}

\begin{figure*}
	\centering
	\includegraphics[width=0.53\linewidth]{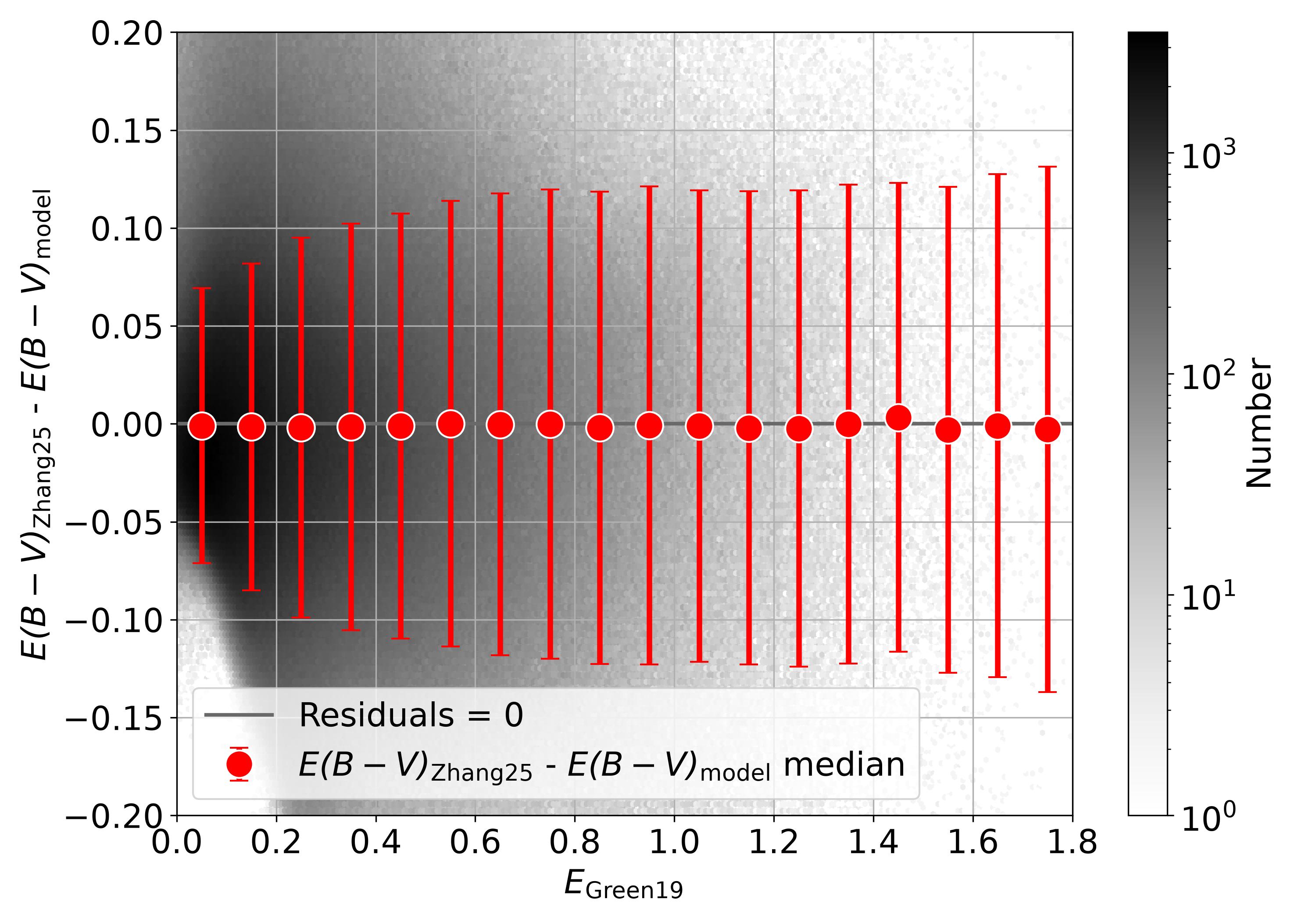}%
	\includegraphics[width=0.53\linewidth]{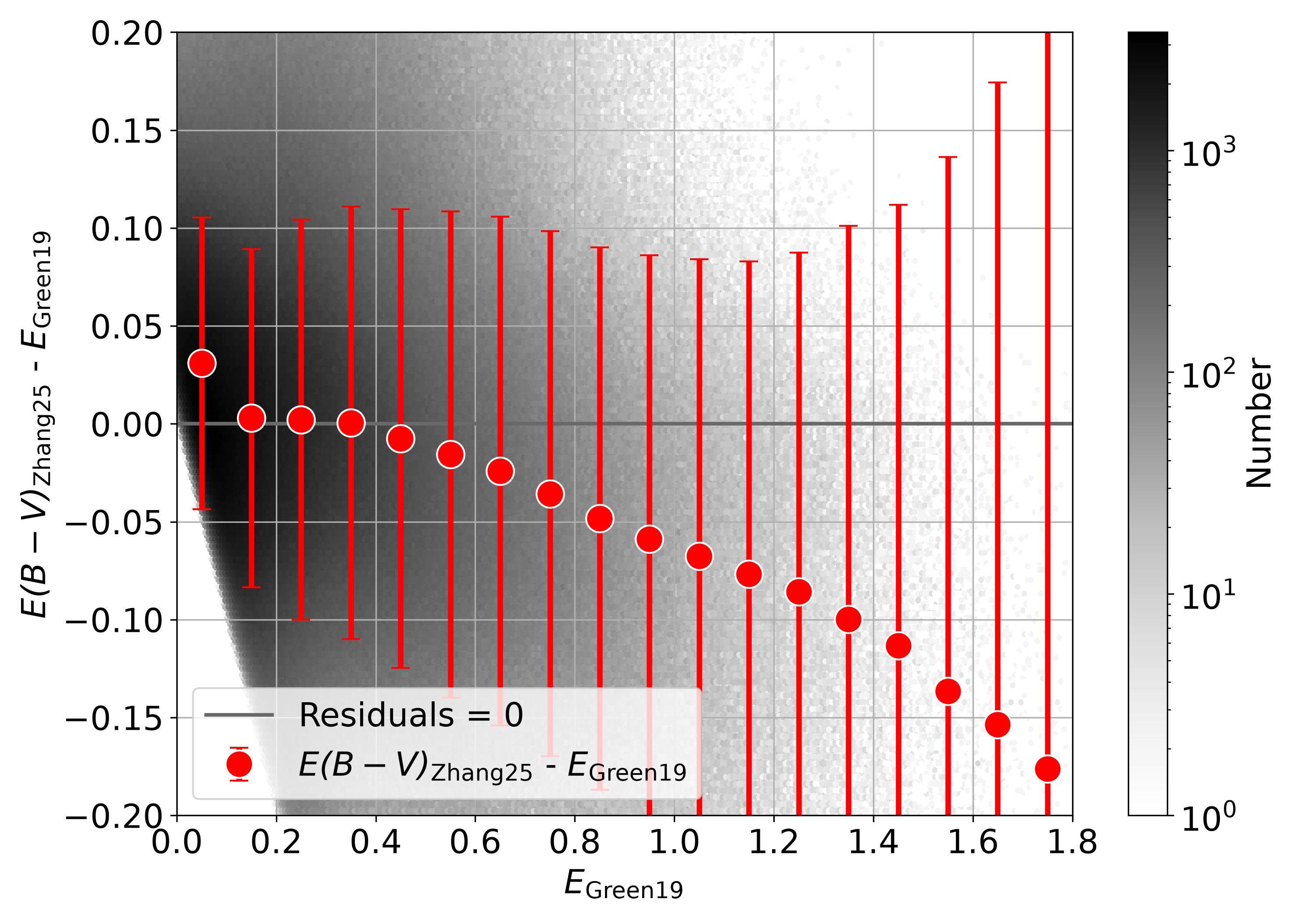}
	\\[0.2em]
	\includegraphics[width=0.53\linewidth]{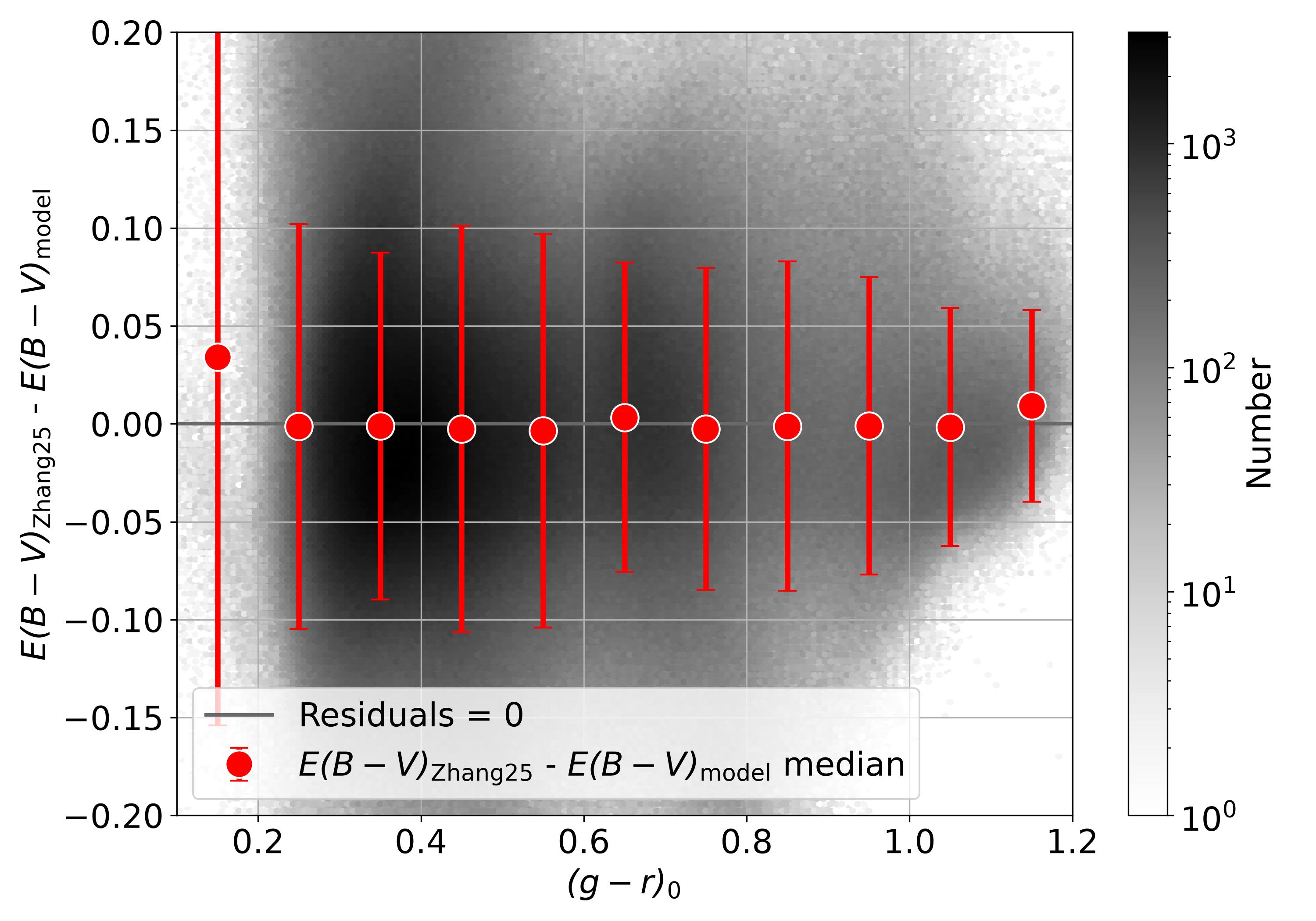}%
	\includegraphics[width=0.53\linewidth]{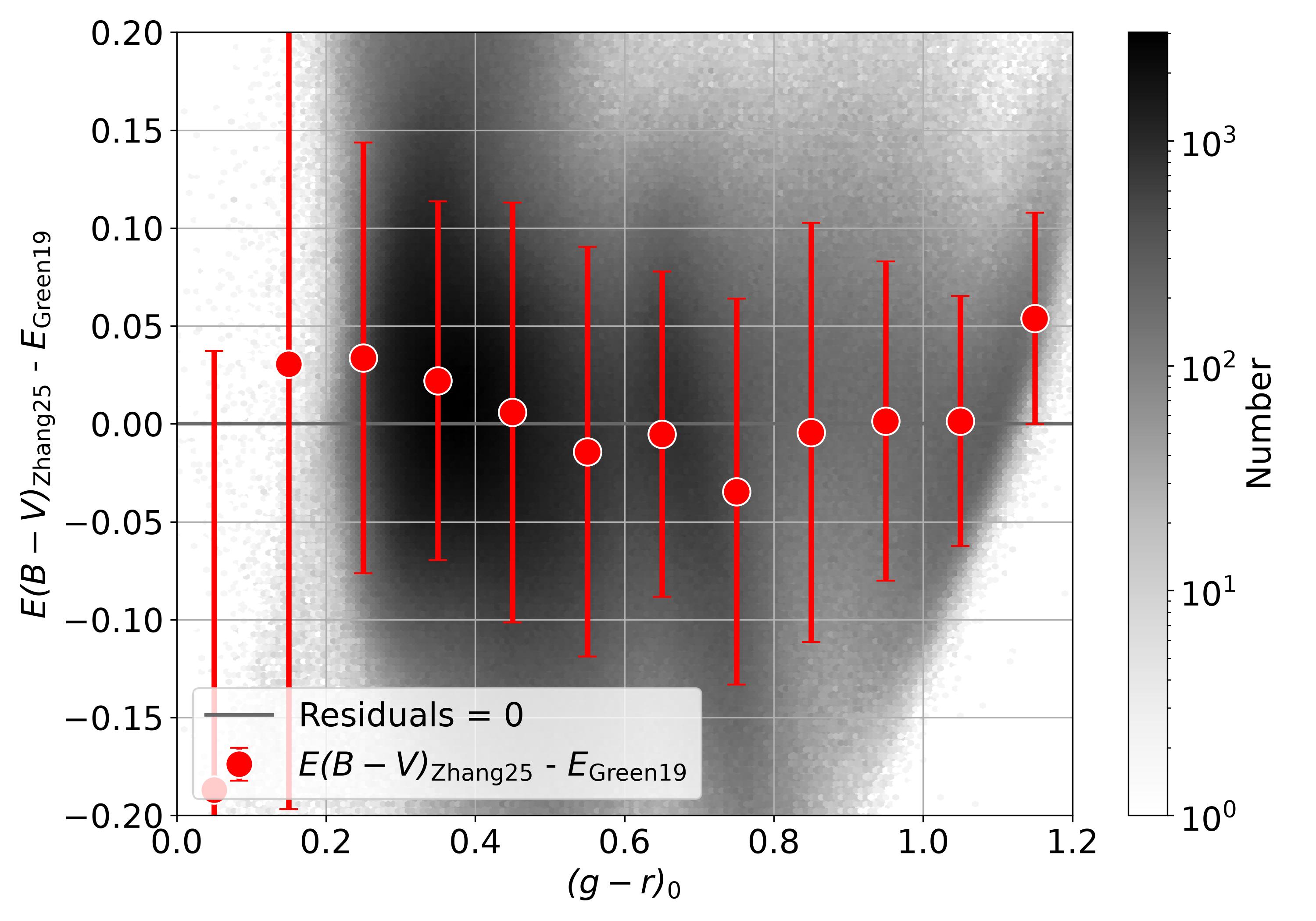}
	\\[0.2em]
	\includegraphics[width=0.53\linewidth]{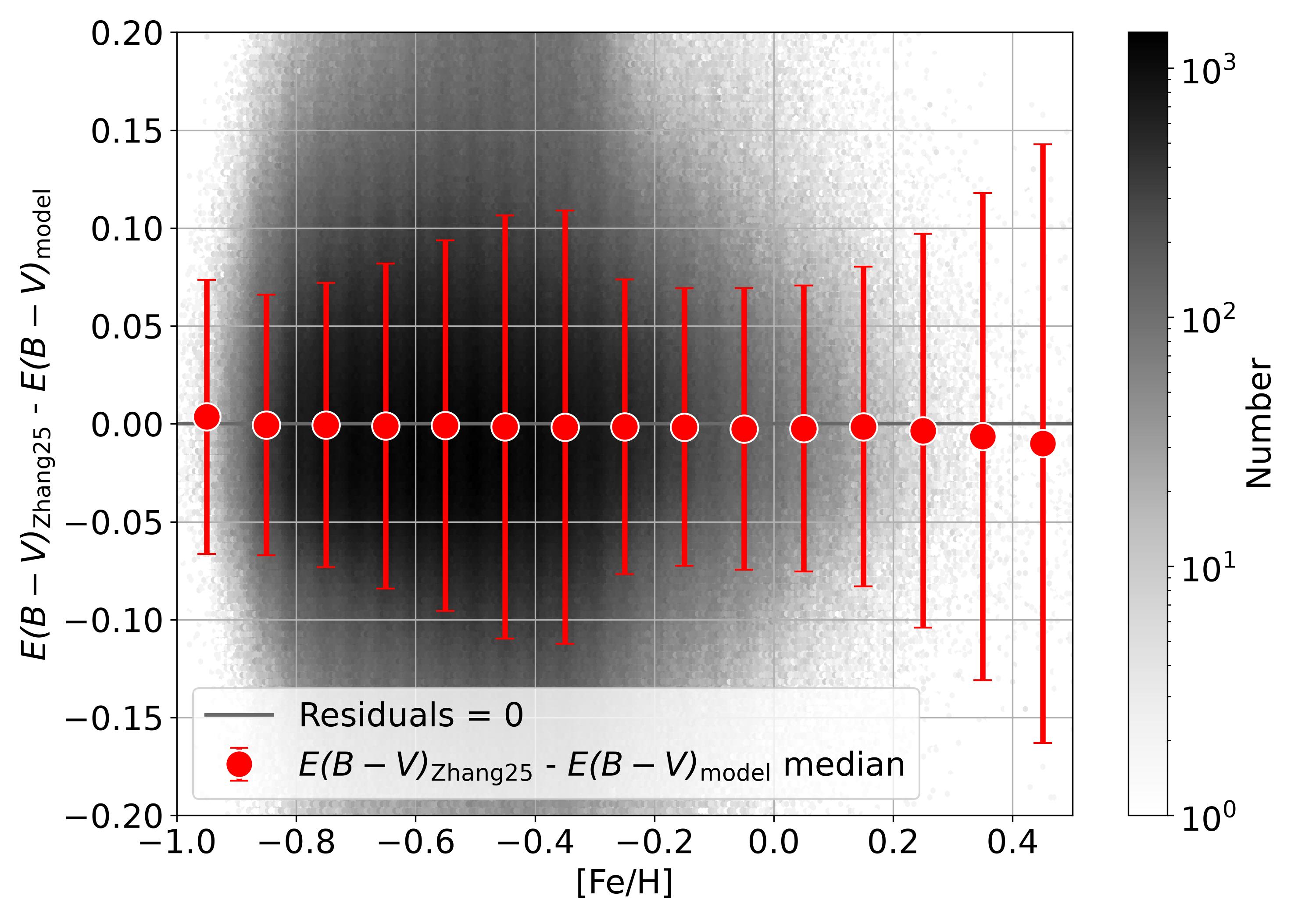}%
	\includegraphics[width=0.53\linewidth]{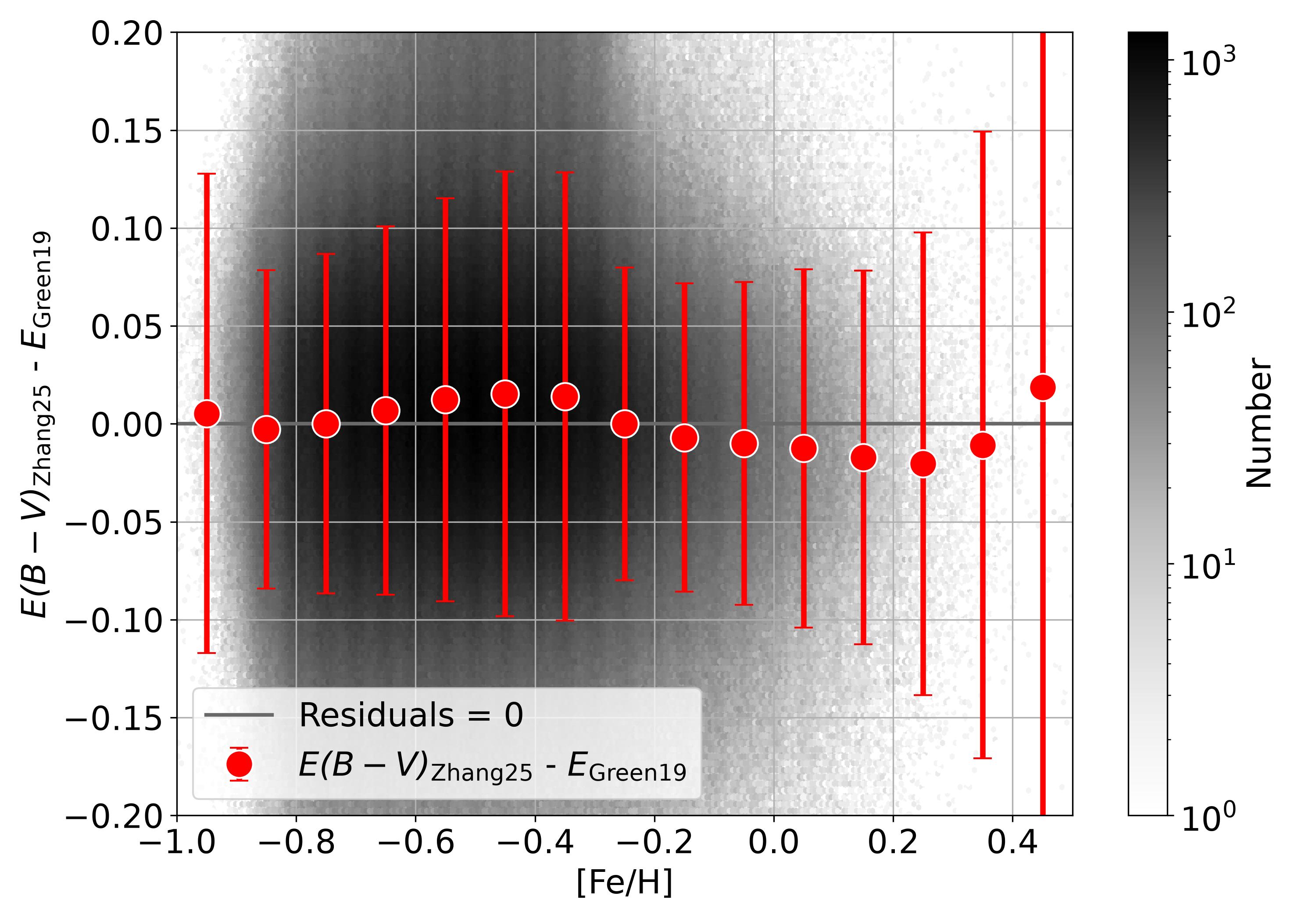}
	\caption{Median residuals (dots) and residual density distributions of the target \(E(B-V)_{\rm Zhang25}\) relative to the model prediction \(E(B-V)_{\rm model}\) (left column) and to \(E_{\rm Green19}\) (right column). From top to bottom, the residuals are plotted as functions of \(E_{\rm Green19}\), intrinsic colour \((g-r)_0\), and metallicity \([\rm Fe/H]\), respectively. In all panels, the data are binned in intervals of 0.1\,mag (or the relevant unit), the error bars denote the standard deviation, and the grey horizontal line marks the zero residual.}
	\label{green19e-gr0-feh}
\end{figure*}

We first applied stringent quality cuts to the Zhang25 catalogue to ensure reliable stellar parameters, requiring \texttt{quality\_flags < 8}, confidence levels for \(T_{\rm eff}\), \(\log g\), and \(\rm [Fe/H]\) greater than 0.5, parameter uncertainties of \(\sigma_{T_{\rm eff}} < 0.5\), \(\sigma_{\log g} < 0.3\), and \(\sigma_{\rm [Fe/H]} < 0.3\), along with a non-negative extinction (\(E(B-V) \ge 0\)). We then cross-matched the Green19 extinction data, which include the intrinsic colour \((g-r)_{\rm 0}\) and satisfy an [Fe/H] error limit of \(\sigma_{\rm [Fe/H]} < 0.3\), with this filtered Zhang25 catalogue. This procedure yields \(59{,}828{,}391\) common entries, which we split randomly into training and test sets in a ratio of $8\!:\!2$. The upper two panels of Fig.~\ref{green19e-gr0-feh} compares the extinction residuals before and after the correction as a function of $E_{\rm Green19}$. After the MLP correction, the median residual stays near zero and shows no systematic drift as $E_{\rm Green19}$ increases, indicating that the model prediction $E(B-V)_{\rm model}$ is stable. In contrast, the raw $E_{\rm Green19}$ carries a marked extinction-dependent systematic difference relative to the baseline $E(B-V)_{\rm Zhang25}$, with an acceptable residual when $E_{\rm Green19}<0.8$\,mag but a residual exceeding $0.1$\,mag in the high-extinction regime where $E_{\rm Green19} \textgreater 1.4$\,mag, which reflects a severe underestimate. 
The two middle panels of Fig.~\ref{green19e-gr0-feh} compares the residuals as a function of $(g-r)_{\rm 0}$. The MLP-corrected value agrees closely with the baseline across the full range of $(g-r)_{\rm 0}$, whereas the raw Green19 extinction is systematically low at the blue end where $(g-r)_{\rm 0} \textless 0.1$. The two bottom panels of Fig.~\ref{green19e-gr0-feh} compares the residuals as a function of [Fe/H]. After the correction the residual shows no clear trend within $[Fe/H]\in[-1.0,0.5]$, while the raw data exhibit a slight bias, and the model corrects this bias effectively. For a few stars without $(g-r)_{\rm 0}$ information, we use the samples that have colour information to establish a linear relation between $E(B-V)_{\rm model}$ and $E_{\rm Green19}$, applying iterative $3\sigma$ clipping to remove outliers until convergence. The best-fit line is $Y=0.91X+0.0338$, with which the systematic correction is completed.

\subsection{Chen19}

\begin{figure*}
	\centering
	\includegraphics[width=0.53\linewidth]{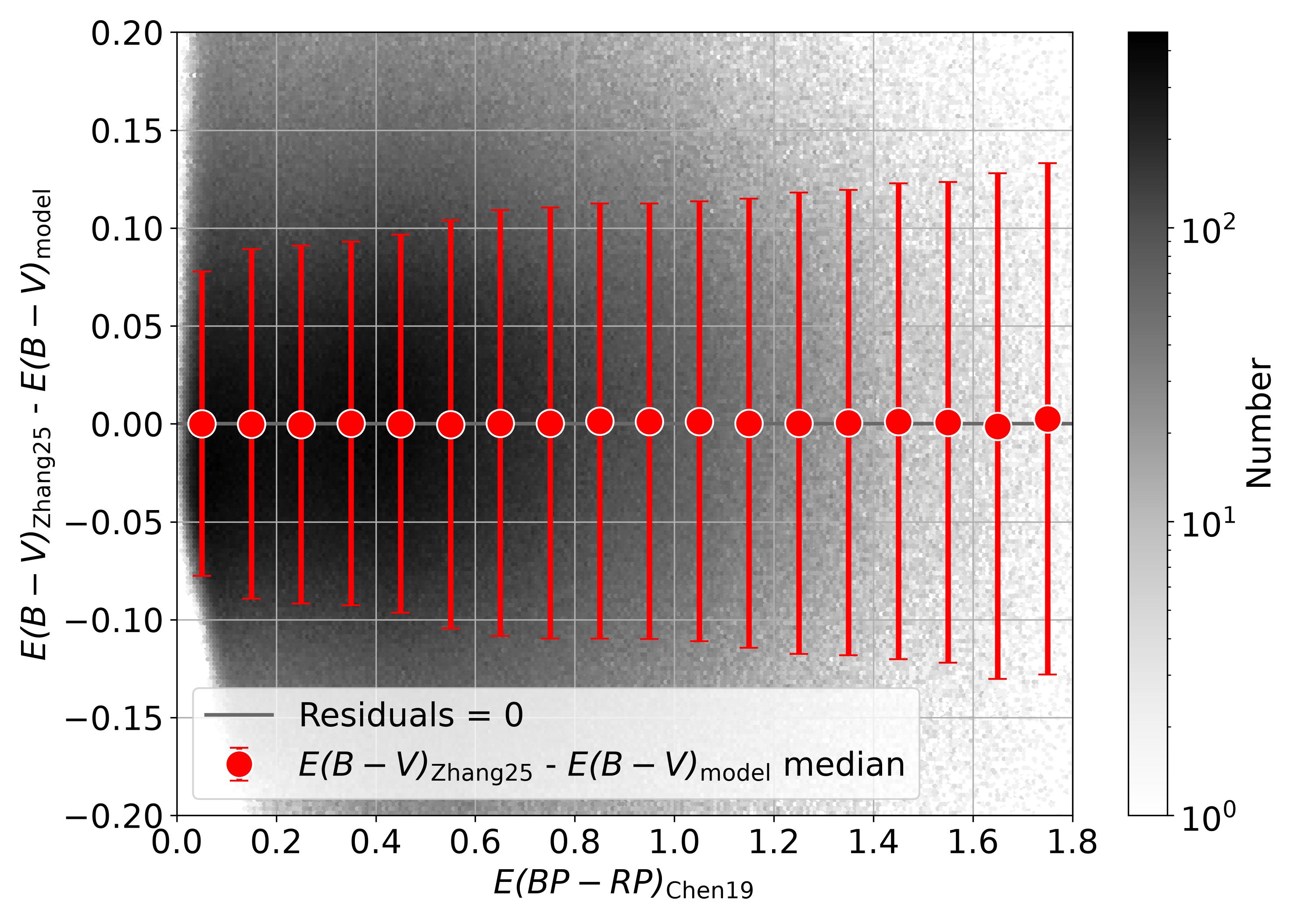}%
	\includegraphics[width=0.53\linewidth]{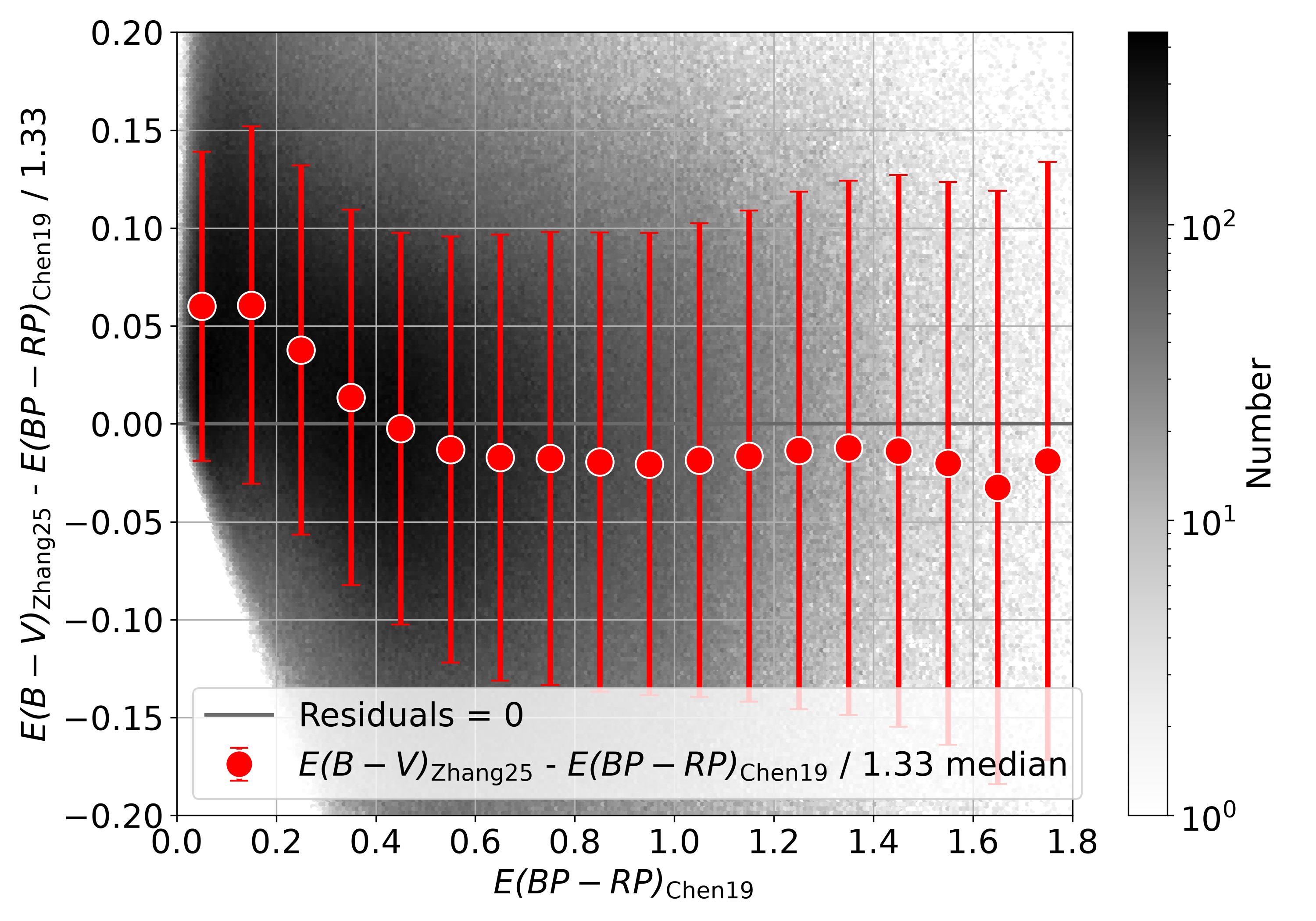}
	\\[0.2em]
	\includegraphics[width=0.53\linewidth]{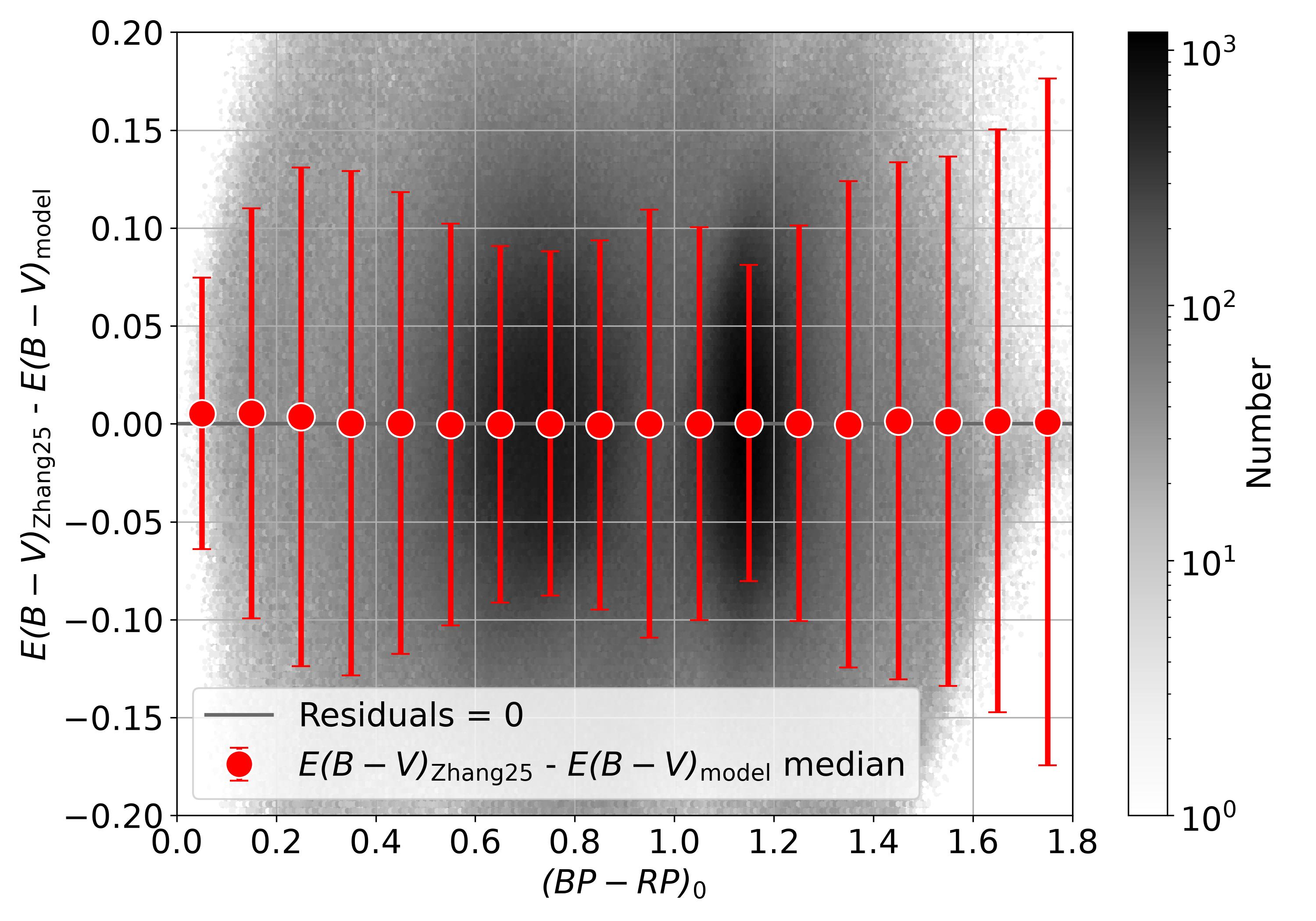}%
	\includegraphics[width=0.53\linewidth]{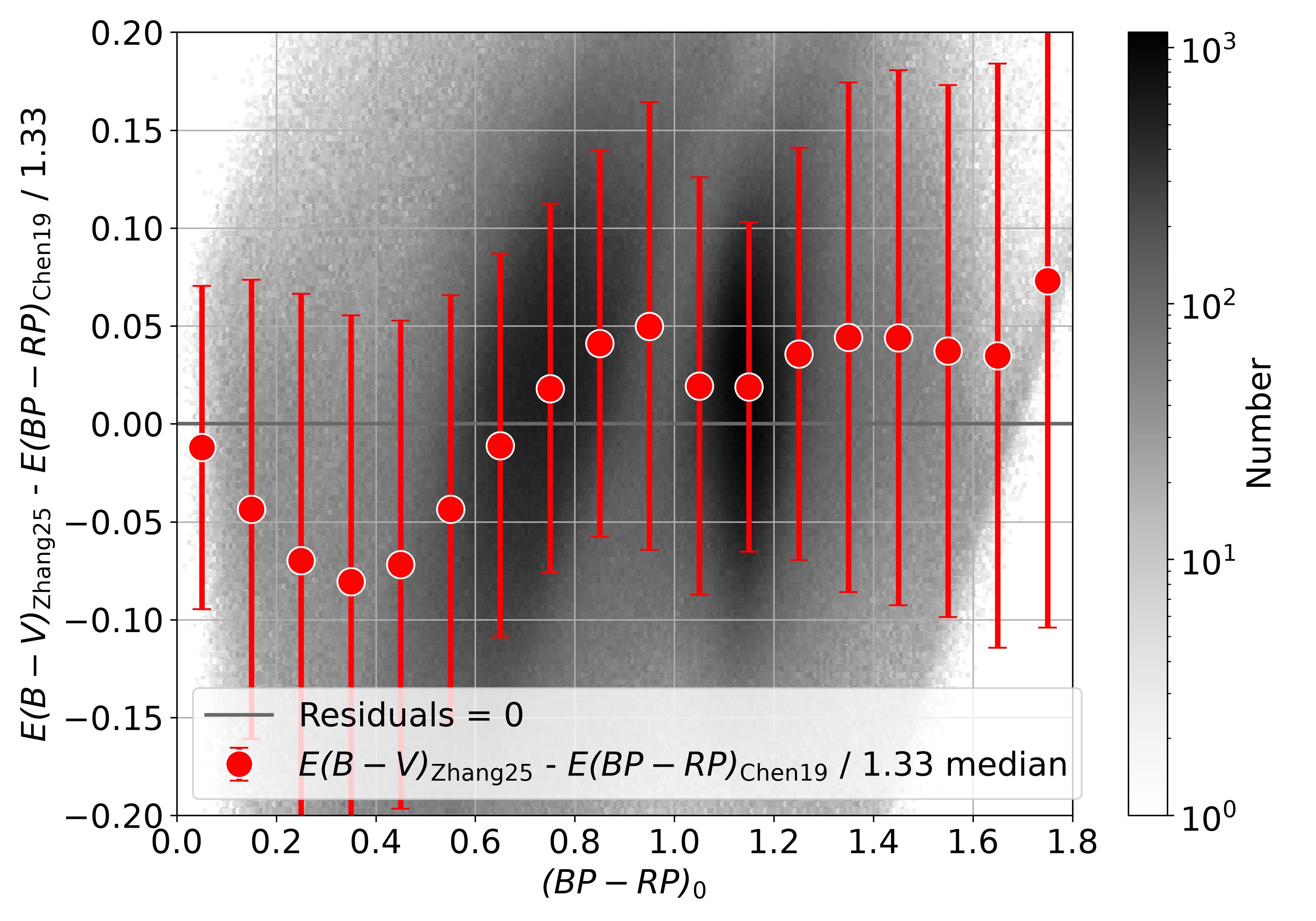}
	\caption{Median residuals (dots) and residual density distributions of the target \(E(B-V)_{\rm Zhang25}\) relative to the model prediction \(E(B-V)_{\rm model}\) (left column) and to \(E(BP-RP)_{\rm Chen19}/1.33\) (right column). The top row shows the residuals as a function of \(E(BP-RP)_{\rm Chen19}\), while the bottom row shows them as a function of the intrinsic colour \((BP-RP)_0\).}
	\label{chen19ebr-bprp0}
\end{figure*}

Following the same stringent quality cuts applied to the Zhang25 catalogue in Section~\ref{apxA_green19}, we cross-matched the Chen19 extinction data with this filtered catalogue. We further applied the constraint \(E(BP-RP)_{\rm Chen19} \ge 0\) to the matched data. After this processing, \(35{,}121{,}339\) common sources remained, which we randomly split into training and test sets with an \(8\!:\!2\) ratio.
The upper two panels of Fig.~\ref{chen19ebr-bprp0} compares the residuals as a function of $E(BP-RP)_{\rm Chen19}$. After the MLP correction the median residual lies close to the zero line, whereas the raw $E(BP-RP)_{\rm Chen19}$ still deviates slightly from the baseline after conversion with the extinction coefficient, and the model corrects this bias effectively. The two bottom panels of Fig.~\ref{chen19ebr-bprp0} compares the residuals as a function of \((BP-RP)_\mathrm{0}\). While the MLP-corrected residuals remain consistent across different spectral types, the raw data exhibit a pronounced nonlinear color-dependent deviation. Specifically, the raw residuals show a significant negative bias in the blue and intermediate color regimes and gradually shift to a slight positive offset at the red end, a trend that is effectively removed by the model.

\subsection{Guo21}

\begin{figure*}
	\centering
	\includegraphics[width=0.53\linewidth]{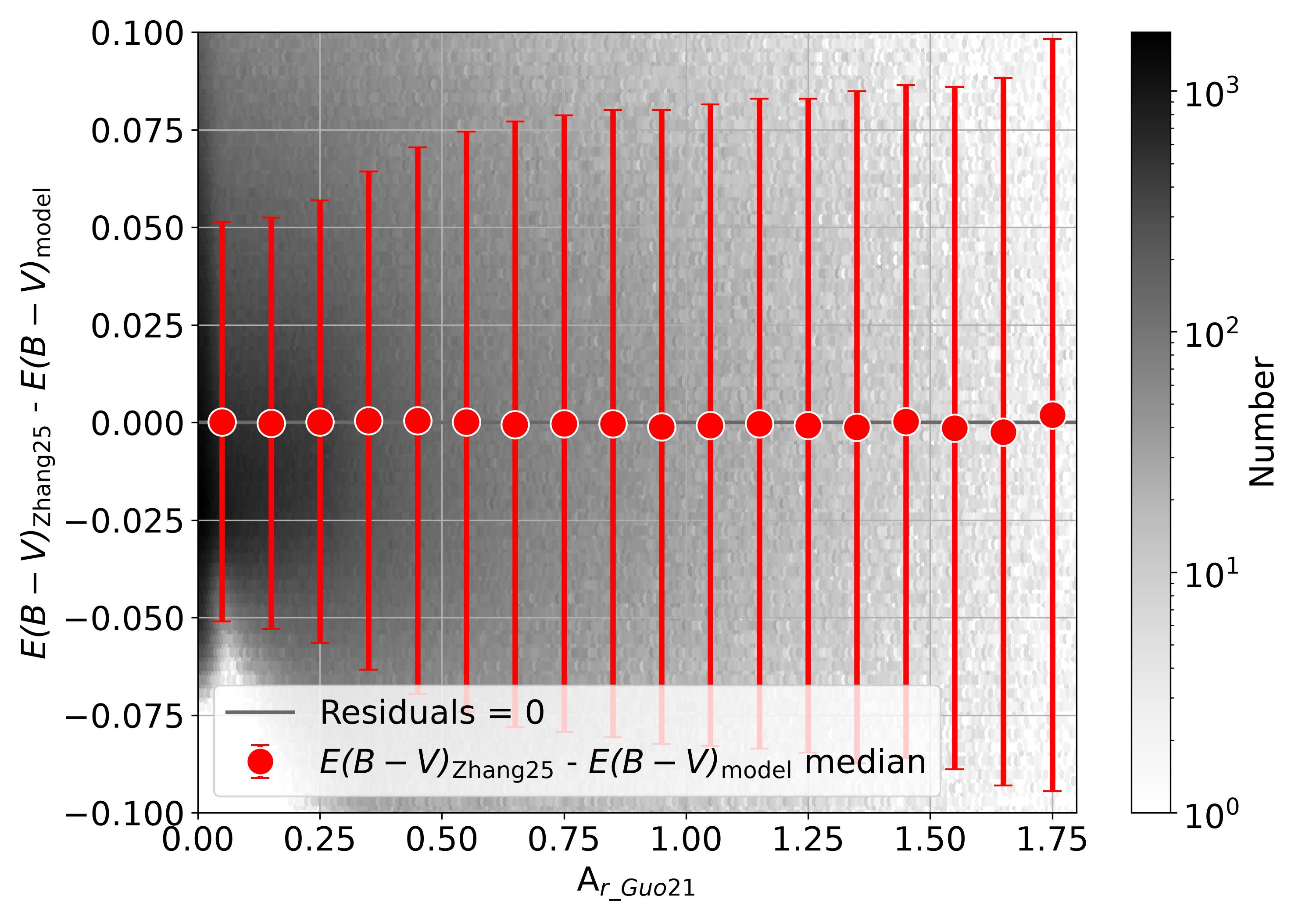}%
	\includegraphics[width=0.53\linewidth]{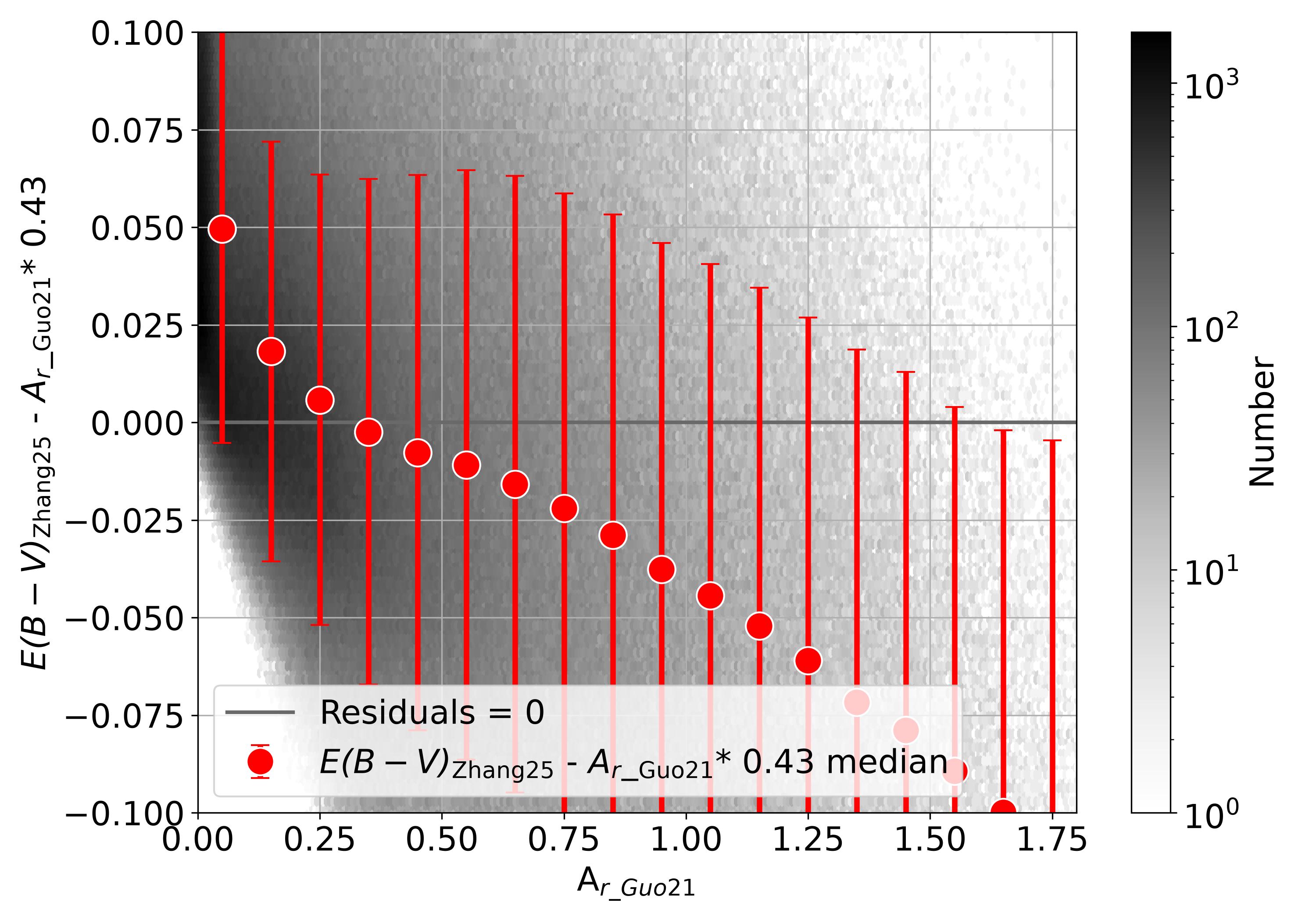}
	\\[0.2em]
	\includegraphics[width=0.53\linewidth]{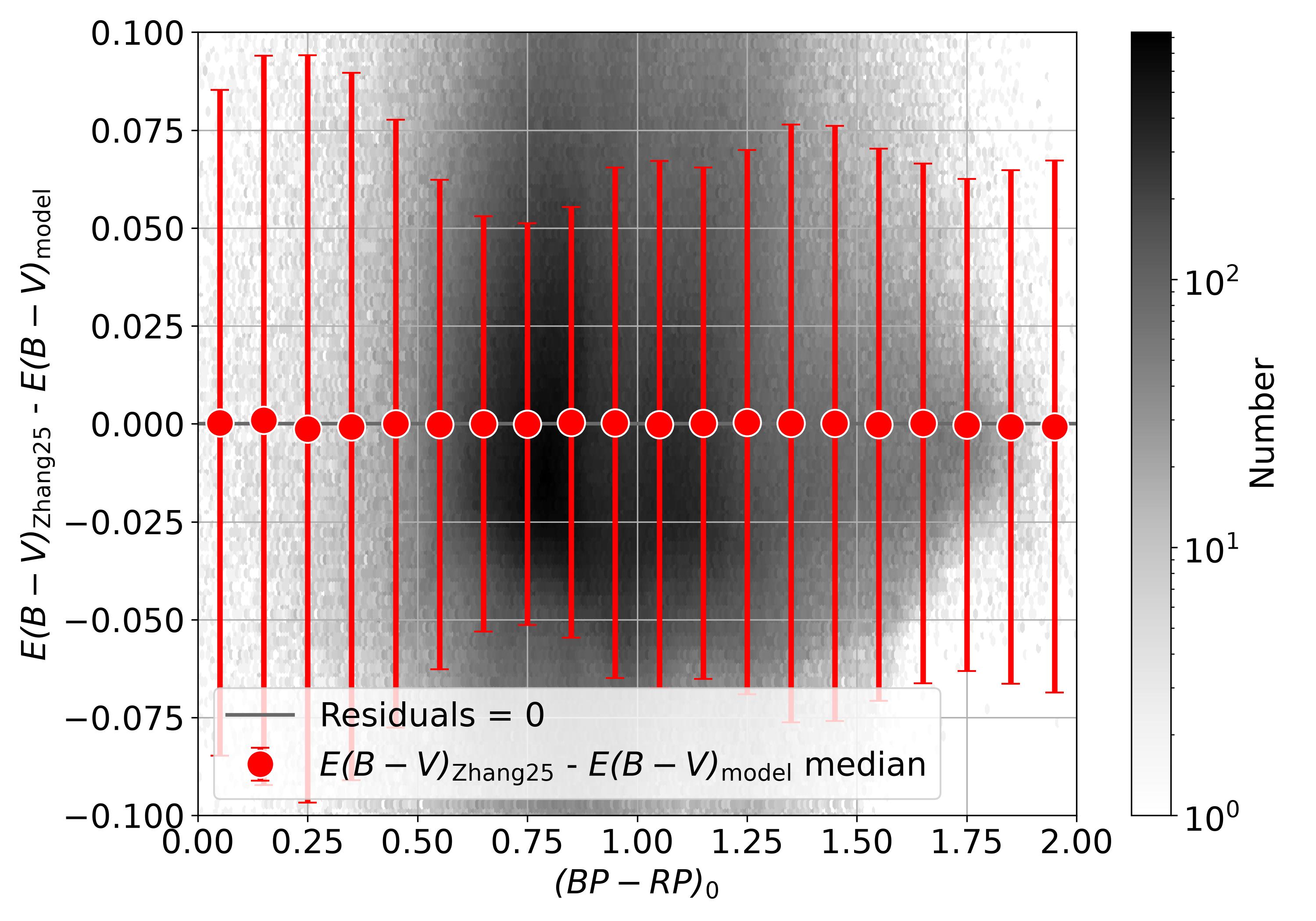}%
	\includegraphics[width=0.53\linewidth]{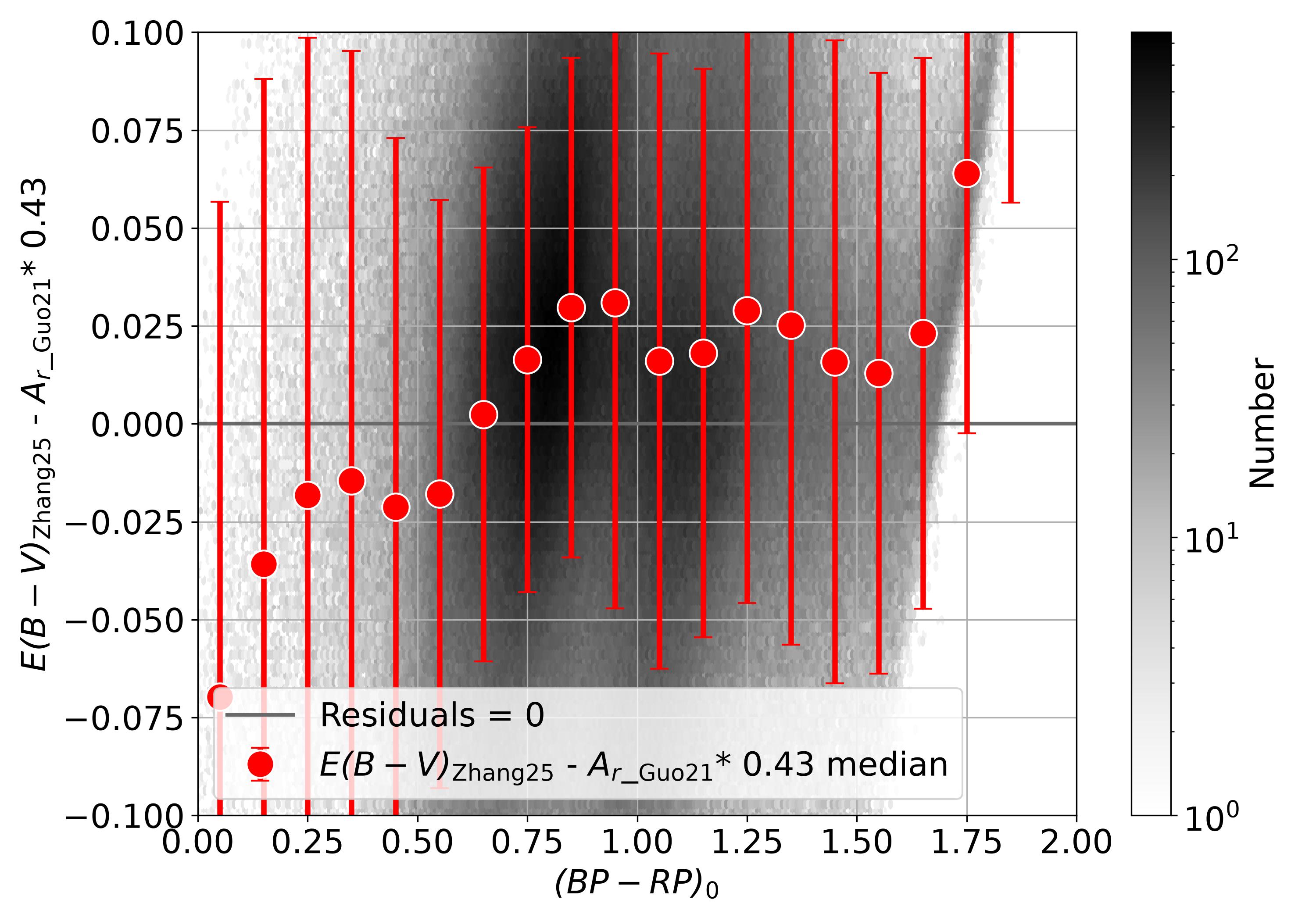}
	\caption{Median residuals (dots) and residual density distributions of the target \(E(B-V)_{\rm Zhang25}\) relative to the model prediction \(E(B-V)_{\rm model}\) (left column) and to \(A_{r\_\rm Guo21}\times0.43\) (right column). The top row shows the residuals as a function of \(A_{r\_\rm Guo21}\), while the bottom row shows them as a function of the intrinsic colour \((BP-RP)_0\).}
	\label{guo21Ar-bprp0}
\end{figure*}

Following the same stringent quality cuts applied to the Zhang25 catalogue in Section~\ref{apxA_green19}, we cross-matched the Guo21 extinction data with this filtered catalogue. We further applied the constraint \(A_{r\_\mathrm{Guo21}}\ge 0\) to the matched data. After this processing, \(14{,}276{,}717\) common sources remained.
The upper two panels of Fig.~\ref{guo21Ar-bprp0} compares the residuals as a function of $A_{r\_\mathrm{Guo21}}$. The MLP-corrected residuals lie close to the zero line across the full range of extinction, whereas the raw $A_{r\_\mathrm{Guo21}}$ shows a systematic negative offset as the extinction increases, which the model corrects effectively. The two bottom panels of Fig.~\ref{guo21Ar-bprp0} compares the residuals as a function of $(BP-RP)_\mathrm{0}$. After the correction the residuals of stars of different spectral types are well consistent, and both the scatter and the systematic offset of the raw data are improved.

\subsection{Sun25}

\begin{figure*}
	\centering
	\includegraphics[width=0.53\linewidth]{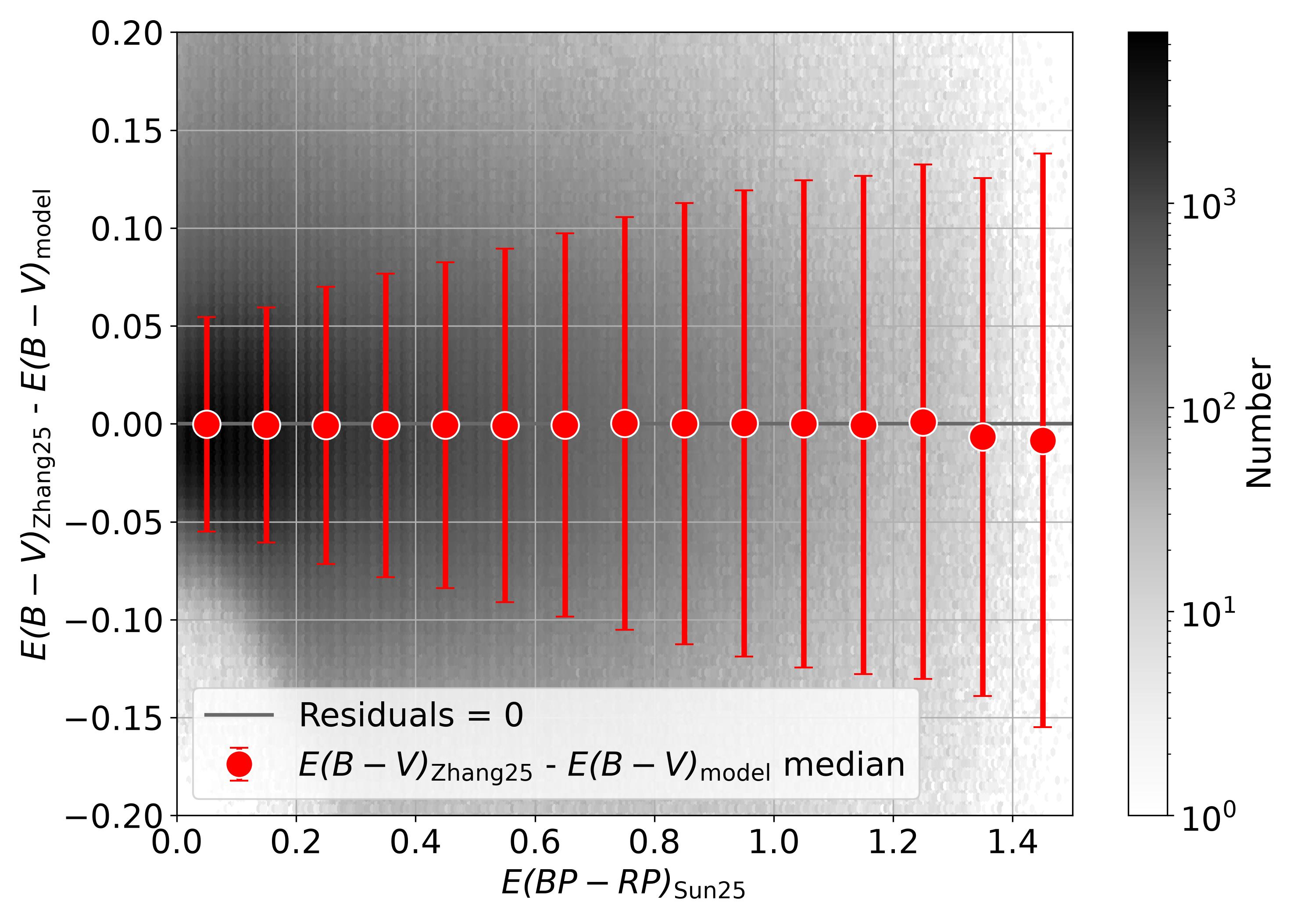}%
	\includegraphics[width=0.53\linewidth]{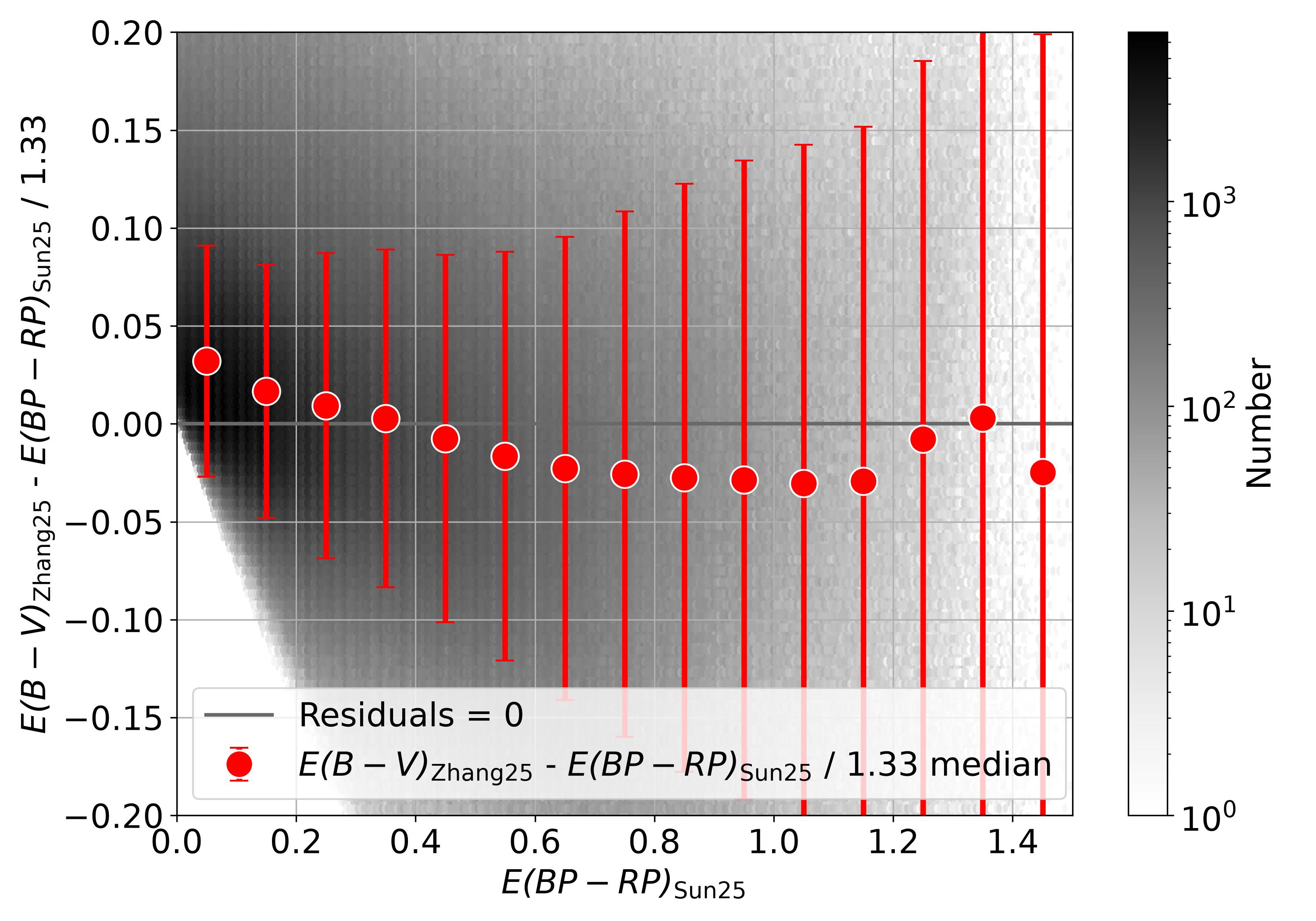}
	\\[0.2em]
	\includegraphics[width=0.53\linewidth]{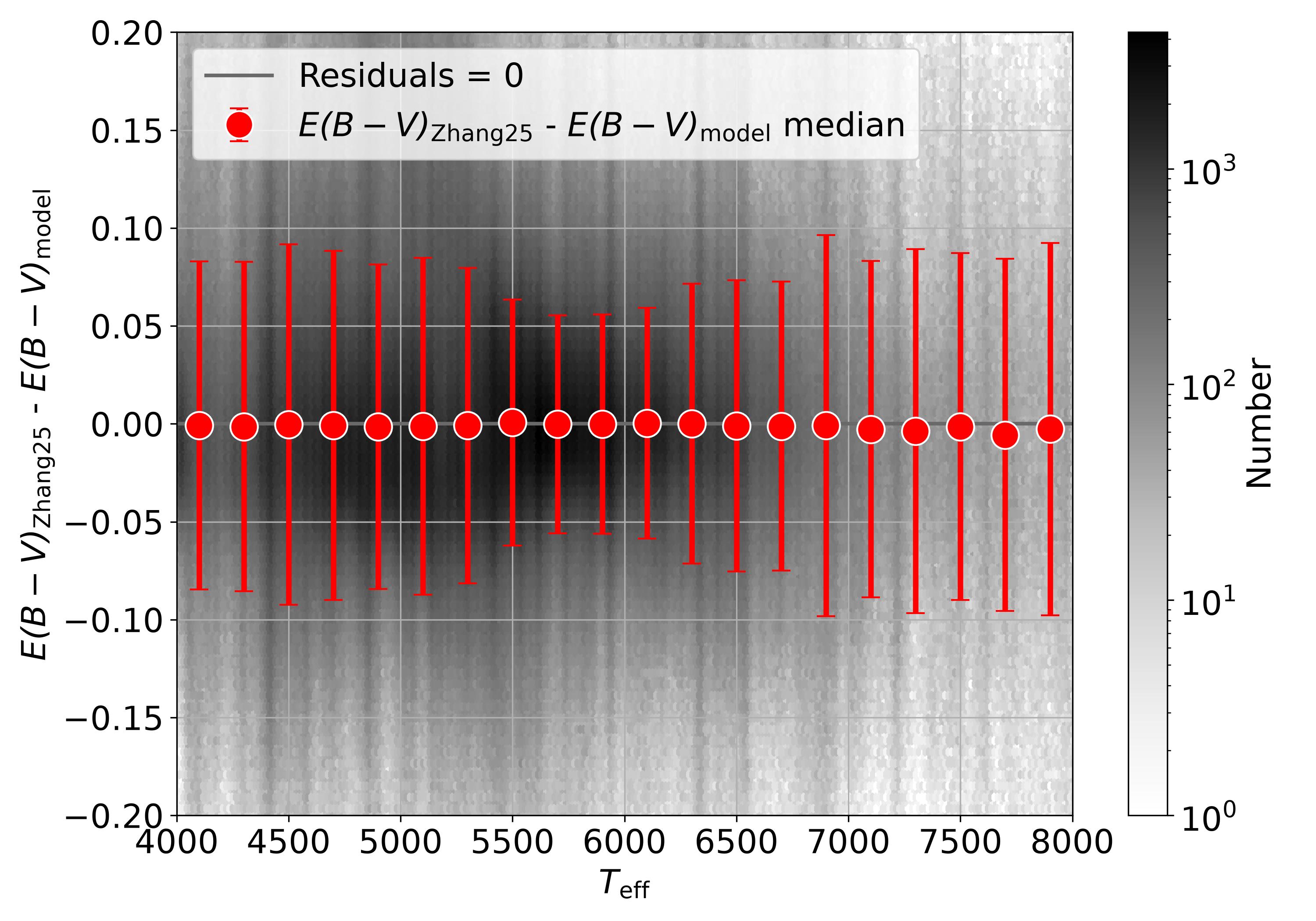}%
	\includegraphics[width=0.53\linewidth]{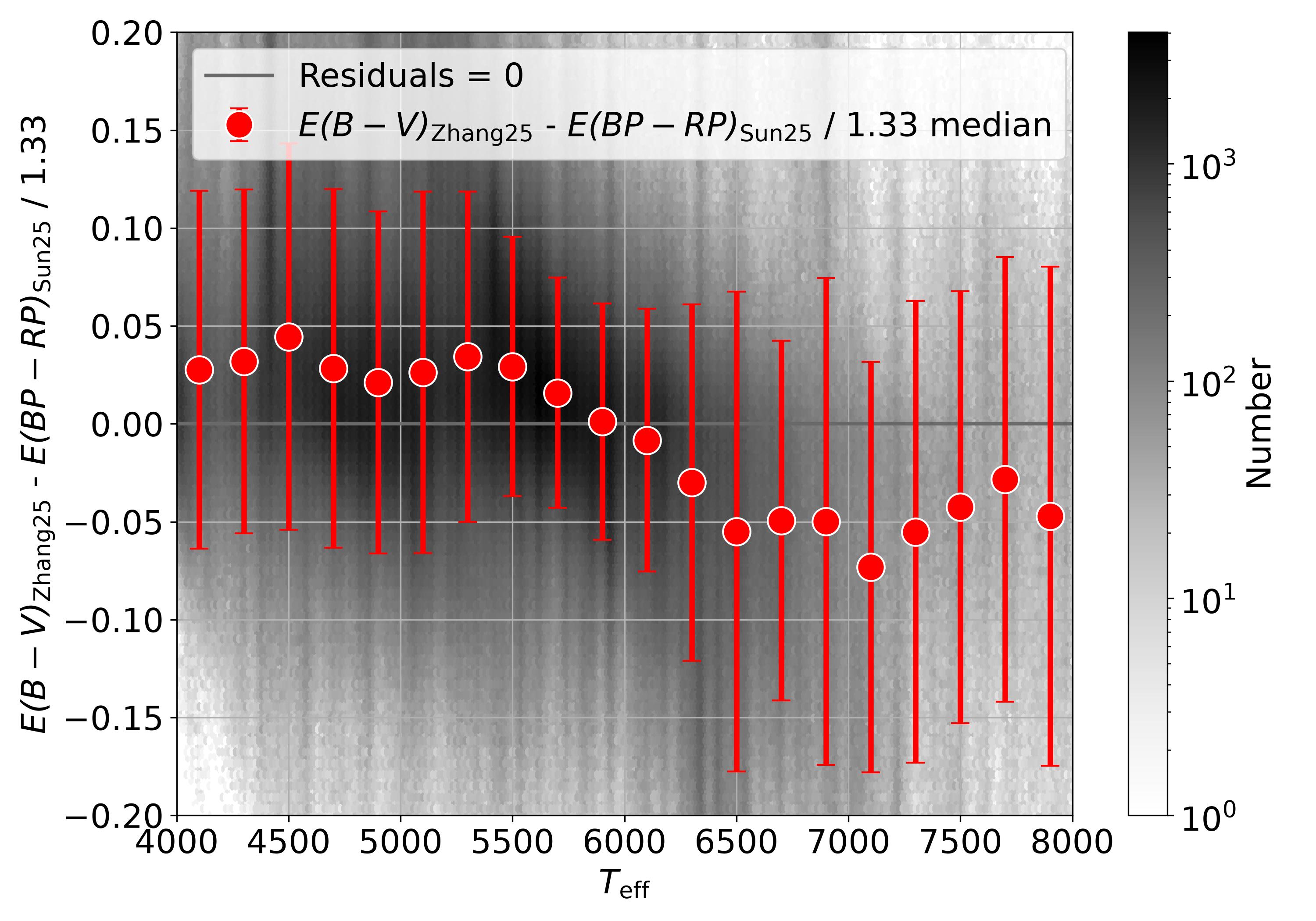}
	\\[0.2em]
	\includegraphics[width=0.53\linewidth]{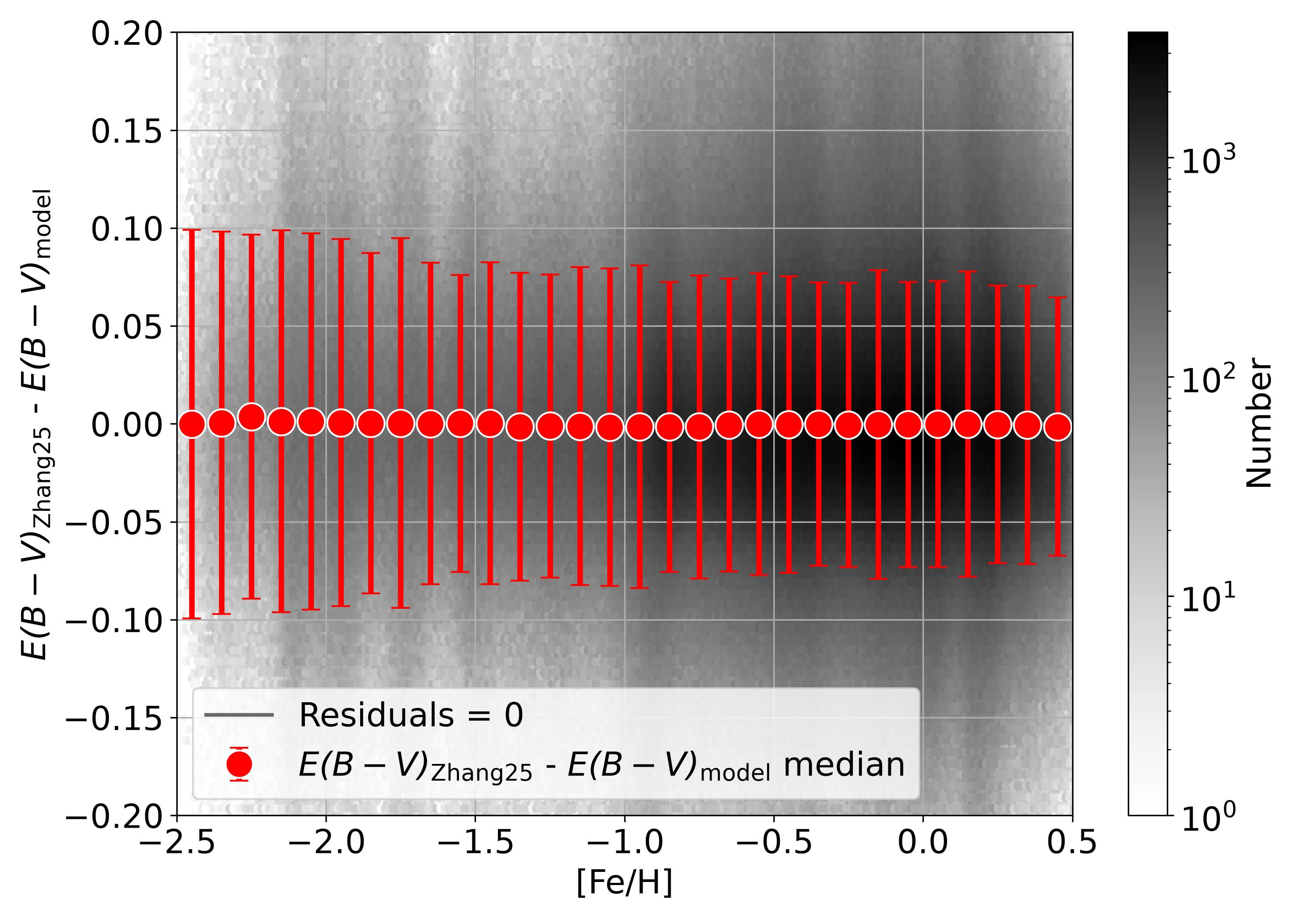}%
	\includegraphics[width=0.53\linewidth]{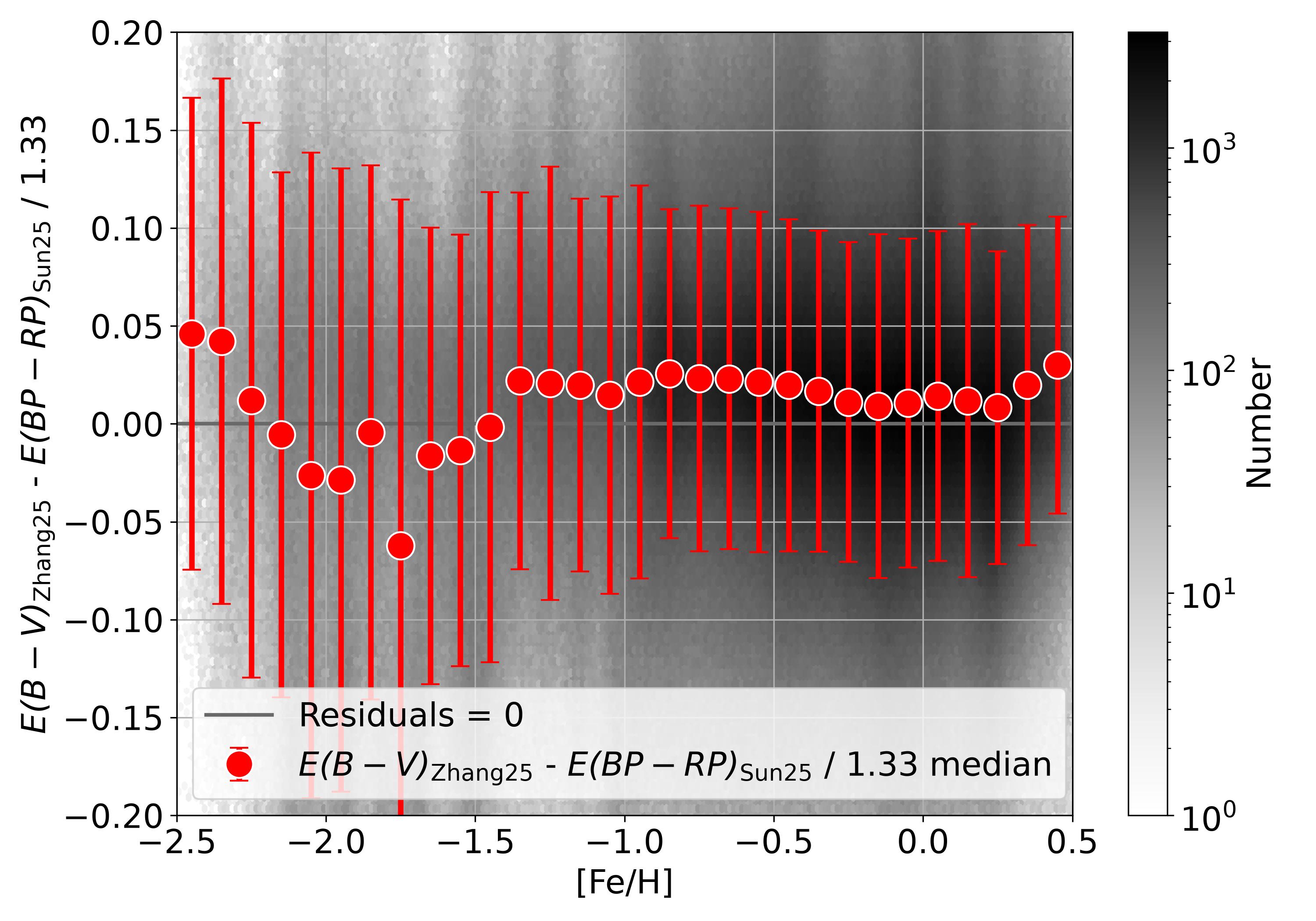}
	\caption{Median residuals (dots) and residual density distributions of the target \(E(B-V)_{\rm Zhang25}\) relative to the model prediction \(E(B-V)_{\rm model}\) (left column) and to \(E(BP-RP)_{\rm Sun25}/1.33\) (right column). From top to bottom, the residuals are plotted as functions of \(E(BP-RP)_{\rm Sun25}\), effective temperature $T_{\rm eff}$, and metallicity \([\rm Fe/H]\), respectively.}
	\label{sun25ebr-teff-feh}
\end{figure*}

Following the same stringent quality cuts applied to the Zhang25 catalogue in Section~\ref{apxA_green19}, we applied corresponding cuts to the Sun25 extinction data (i.e., \(E(BP-RP)_{\rm Sun25} \ge 0\), \(\sigma_{T_{\rm eff}} < 500\,{\rm K}\), and \(\sigma_{\rm [Fe/H]} < 0.3\)). We then cross-matched this filtered Sun25  extinction data with the Zhang25 catalogue, yielding \(57{,}015{,}806\) common entries. The upper two panels of Fig.~\ref{sun25ebr-teff-feh} compares the residuals as a function of \(E(BP-RP)_{\rm Sun25}\). The MLP-corrected median residual lies essentially on the zero line, whereas the raw Sun25 extinction exhibits a systematic positive offset in the low-extinction regime, transitioning to a slight negative offset at moderate extinctions. The two middle panels of Fig.~\ref{sun25ebr-teff-feh} compares the residuals as a function of \(T_{\rm eff}\). After the correction the residuals are evenly distributed within the range of \(4000\) to \(8000\,\mathrm{K}\), whereas at the high-temperature end (\(T_{\rm eff} > 6000\,\mathrm{K}\)), the raw data exhibit a pronounced systematic negative offset and a significantly increasing scatter, which the model effectively suppresses. The two bottom panels of Fig.~\ref{sun25ebr-teff-feh} compares the residuals as a function of [Fe/H]. After the MLP correction, the residuals are well-centered around zero. In contrast, the raw data show complex systematic biases dependent on [Fe/H]—featuring large fluctuations at the metal-poor end and a slight positive offset at the metal-rich end—all of which are effectively eliminated by the model.

\subsection{Zucker25}

\begin{figure*}
	\centering
	\includegraphics[width=0.53\linewidth]{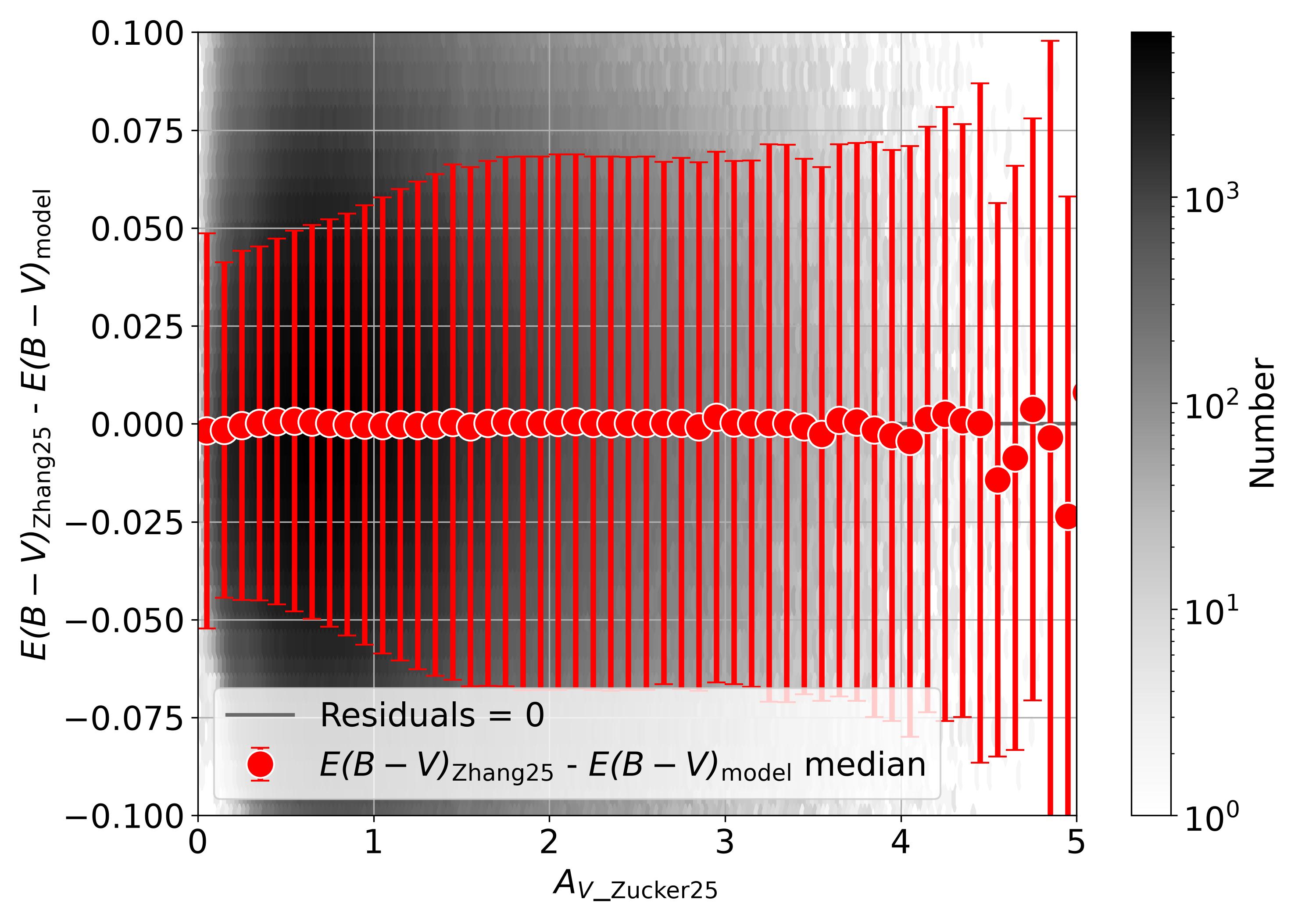}%
	\includegraphics[width=0.53\linewidth]{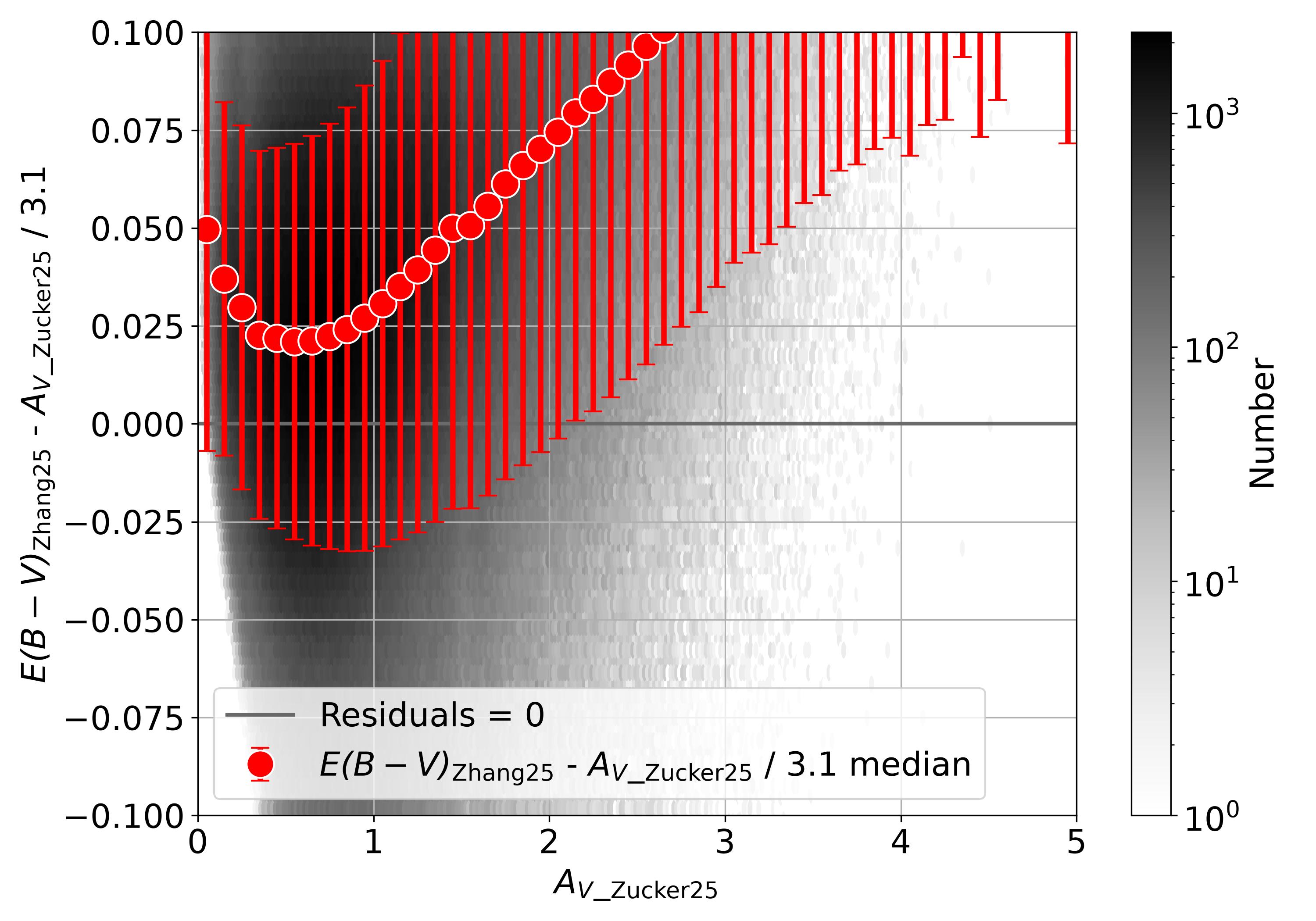}
	\\[0.2em]
	\includegraphics[width=0.53\linewidth]{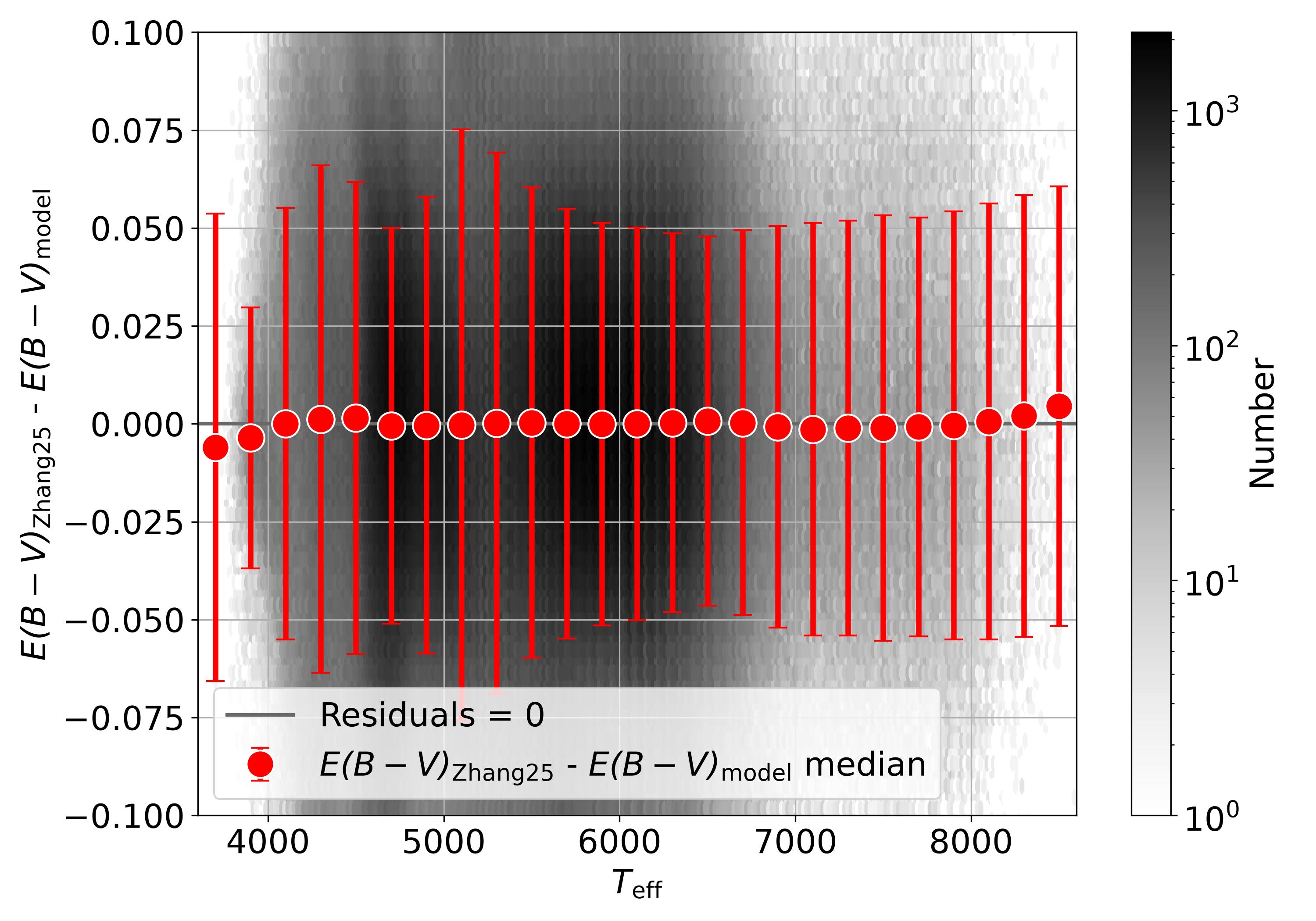}%
	\includegraphics[width=0.53\linewidth]{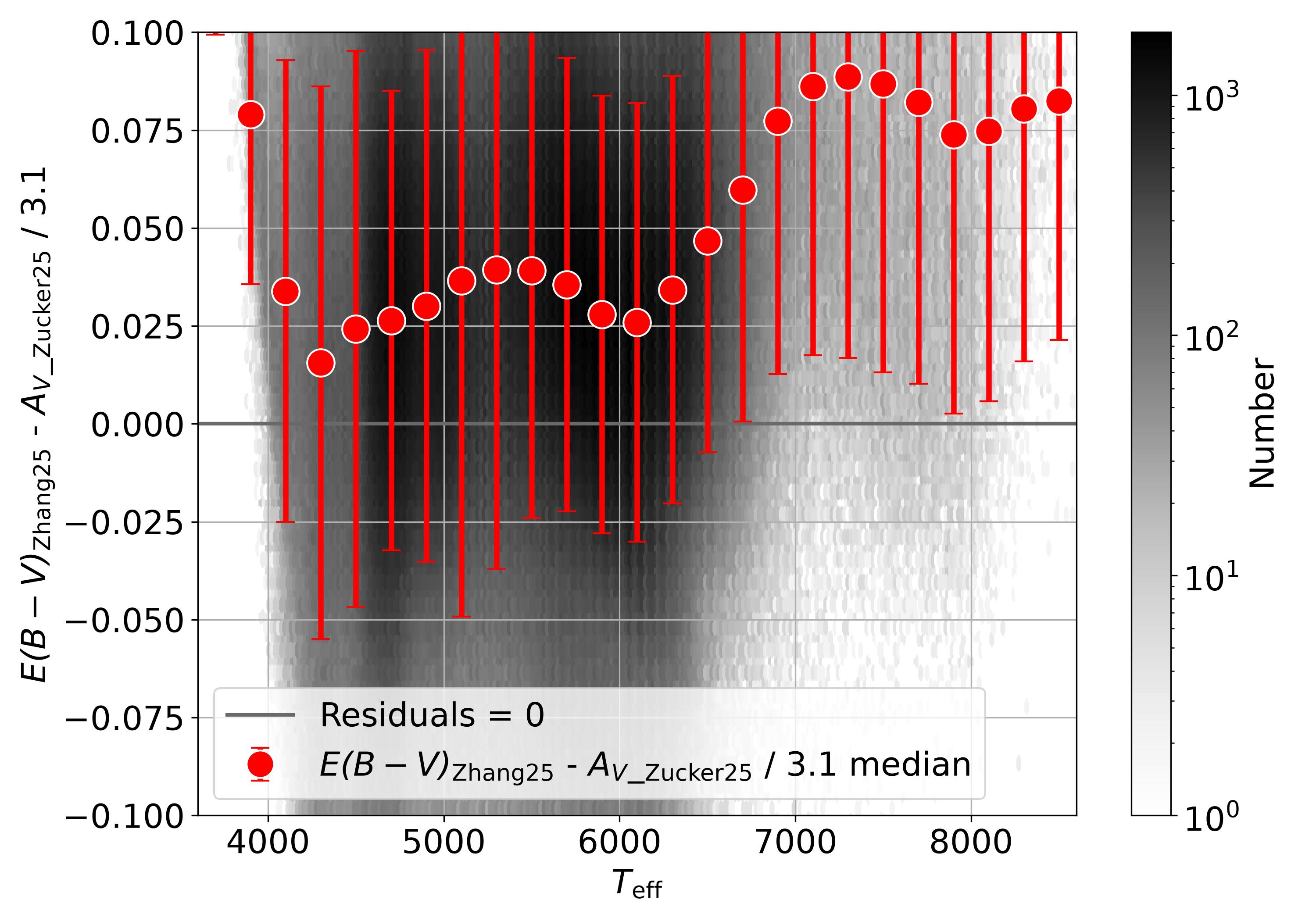}
	\\[0.2em]
	\includegraphics[width=0.53\linewidth]{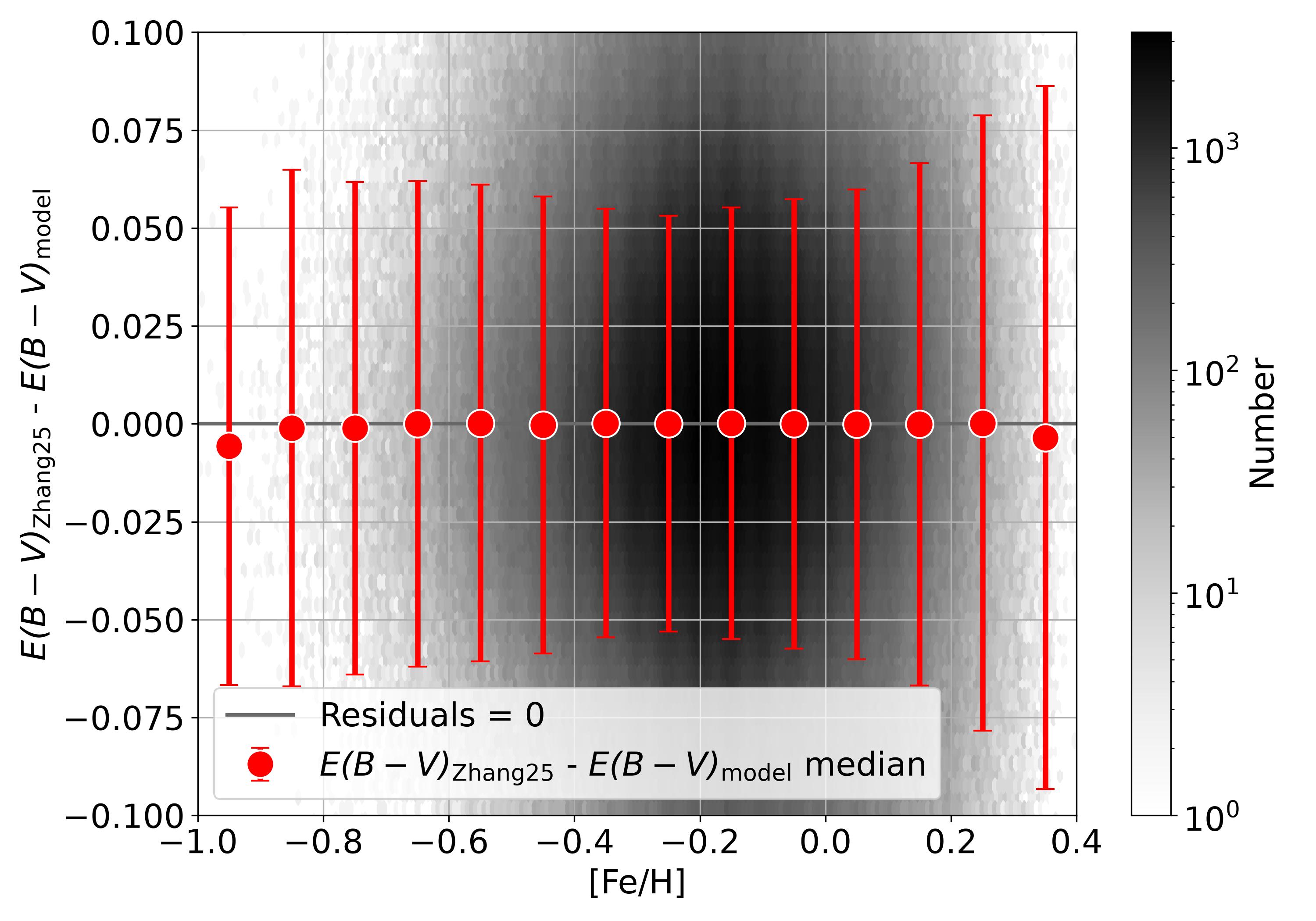}%
	\includegraphics[width=0.53\linewidth]{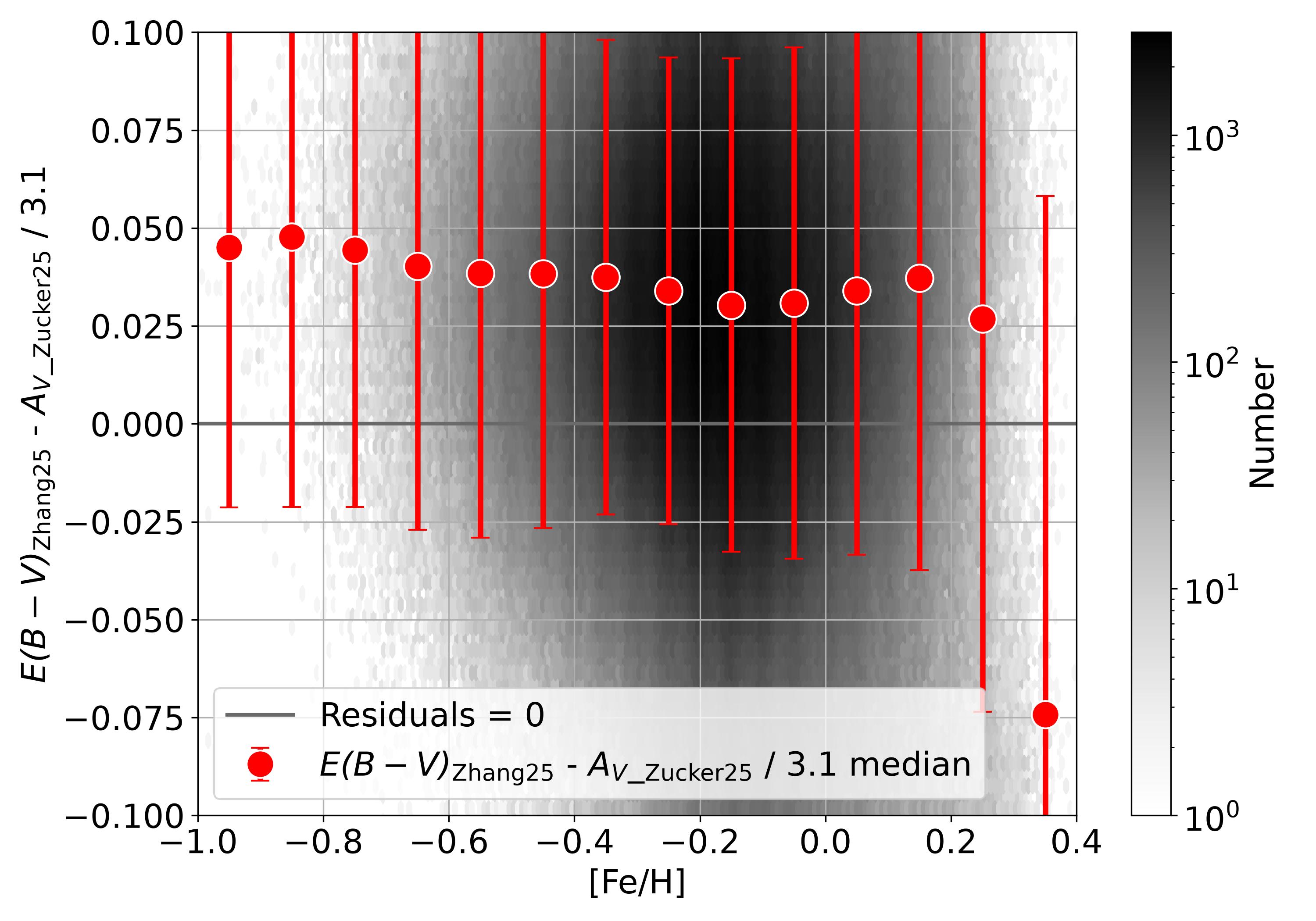}
	\caption{Median residuals (dots) and residual density distributions of the target \(E(B-V)_{\rm Zhang25}\) relative to the model prediction \(E(B-V)_{\rm model}\) (left column) and to \(A_{V\_\mathrm{Zucker25}}/3.1\) (right column). From top to bottom, the residuals are plotted as functions of \(A_{V\_\mathrm{Zucker25}}\), effective temperature $T_{\rm eff}$, and metallicity \([\rm Fe/H]\), respectively.}
	\label{zucker25Av-teff-feh}
\end{figure*}

To ensure consistency with the stringent quality cuts applied to the Zhang25 catalogue in Section~\ref{apxA_green19}, we filtered the Zucker25 extinction data using the following criteria: \(A_{V\_\mathrm{Zucker25}} \ge 0\), \(\sigma_{T_{\rm eff}} < 500\,{\rm K}\), \(\sigma_{\rm [Fe/H]} < 0.3\), and \(\sigma_{\log g} < 0.3\). Cross-matching this filtered sample with the Zhang25 catalogue yields \(27{,}701{,}931\) common entries.
Fig.~\ref{zucker25Av-teff-feh} compares the residuals as a function of $A_{V\_\mathrm{Zucker25}}$. The MLP-corrected median residual lies close to the zero line, whereas the raw $A_{V\_\mathrm{Zucker25}}$ shows a marked systematic positive offset relative to the baseline after conversion with a fixed extinction coefficient, which is especially prominent in the high-extinction regime. Fig.~\ref{zucker25Av-teff-feh} compares the residuals as a function of $T_{\rm eff}$. After the correction the residuals show no clear trend across the full range of $T_{\rm eff}$, and both the systematic bias and the scatter of the raw data are suppressed effectively. Fig.~\ref{zucker25Av-teff-feh} compares the residuals as a function of [Fe/H]. After the MLP correction the residuals concentrate around zero, whereas the raw residuals are markedly positive and fluctuate considerably, and the model removes this bias effectively.

\section{Per-Catalogue Results of the Distance-Systematics Correction}\label{appendix_B}

The distance baseline is taken from the Zhang25 catalogue, so for every catalogue to be calibrated we first cross-match it with Zhang25 and then select a clean baseline sample by imposing the same set of quality constraints on the Zhang25 columns, namely \texttt{quality\_flags < 8}, confidence levels for \(T_{\rm eff}\), \(\log g\), and \(\rm [Fe/H]\) greater than 0.5. These constraints retain only the Zhang25 stars with well-determined parameters and reliable distances, so that the baseline distance $D_{Zhang25}$ is accurate enough to serve as the supervision value. In the following subsections we therefore report only the additional constraints imposed on each calibrated catalogue itself, without repeating the Zhang25 baseline selection. In every case, the figures compare the residuals before and after the calibration as a function of the raw distance; unless noted otherwise, the MLP-calibrated median residual lies close to the zero line, with good consistency at the near end ($d<5$\,kpc) and growing scatter at the far end while the overall trend still follows the baseline, indicating that the model effectively removes the systematic bias of the raw distance.

\begin{figure*}
	\centering
	\includegraphics[width=0.53\linewidth]{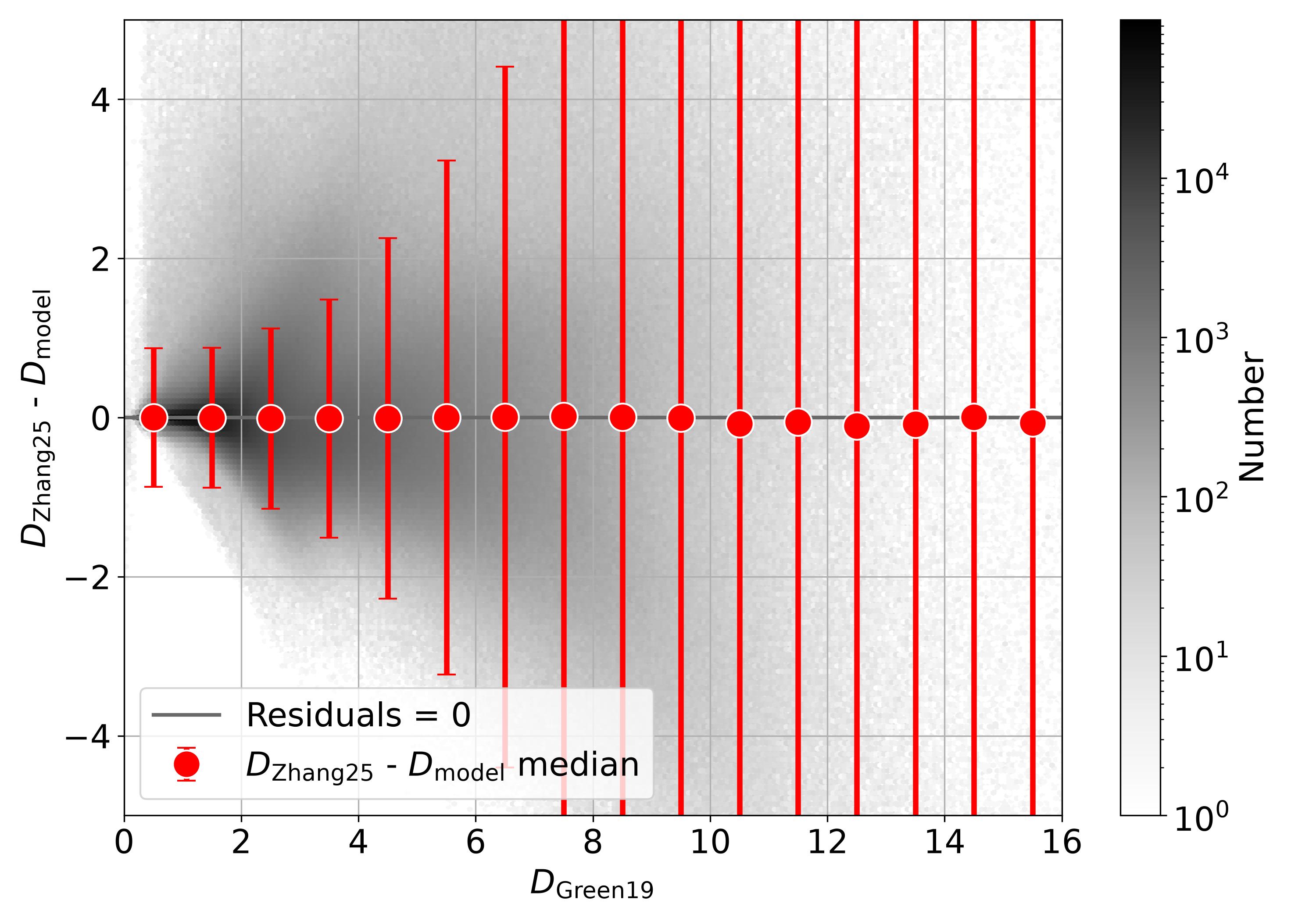}%
	\includegraphics[width=0.53\linewidth]{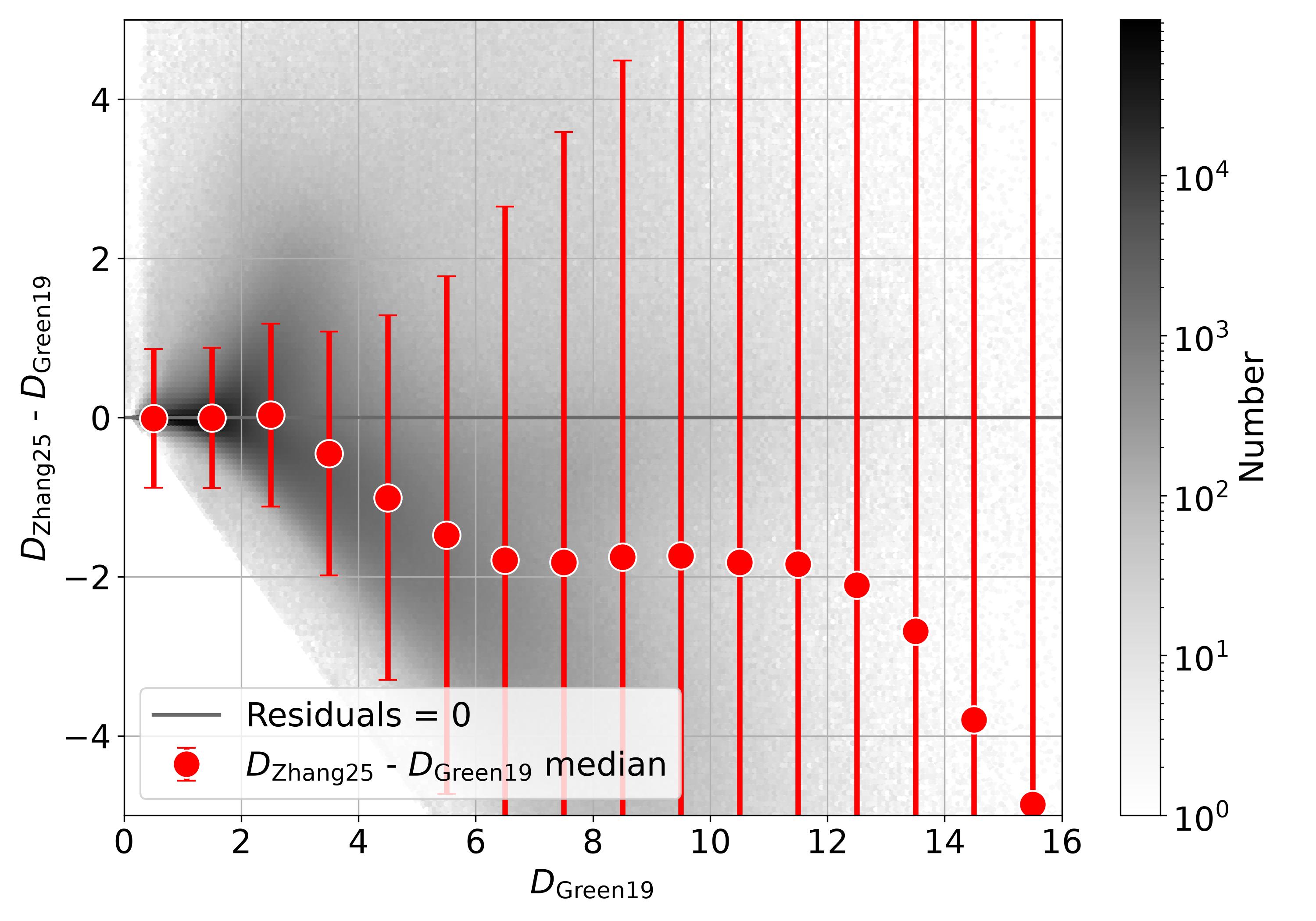}
	\caption{Median residuals and density distributions of the target $D_{\rm Zhang25}$ relative to the model prediction $D_{\rm model}$ in the upper panel and to $D_{\rm Green19}$ in the bottom panel, as a function of $D_{\rm Green19}$. The data are grouped by $D_{\rm Green19}$ in intervals of $1$\,kpc.}
	\label{Green19d}
\end{figure*}

\begin{figure*}
	\centering
	\includegraphics[width=0.53\linewidth]{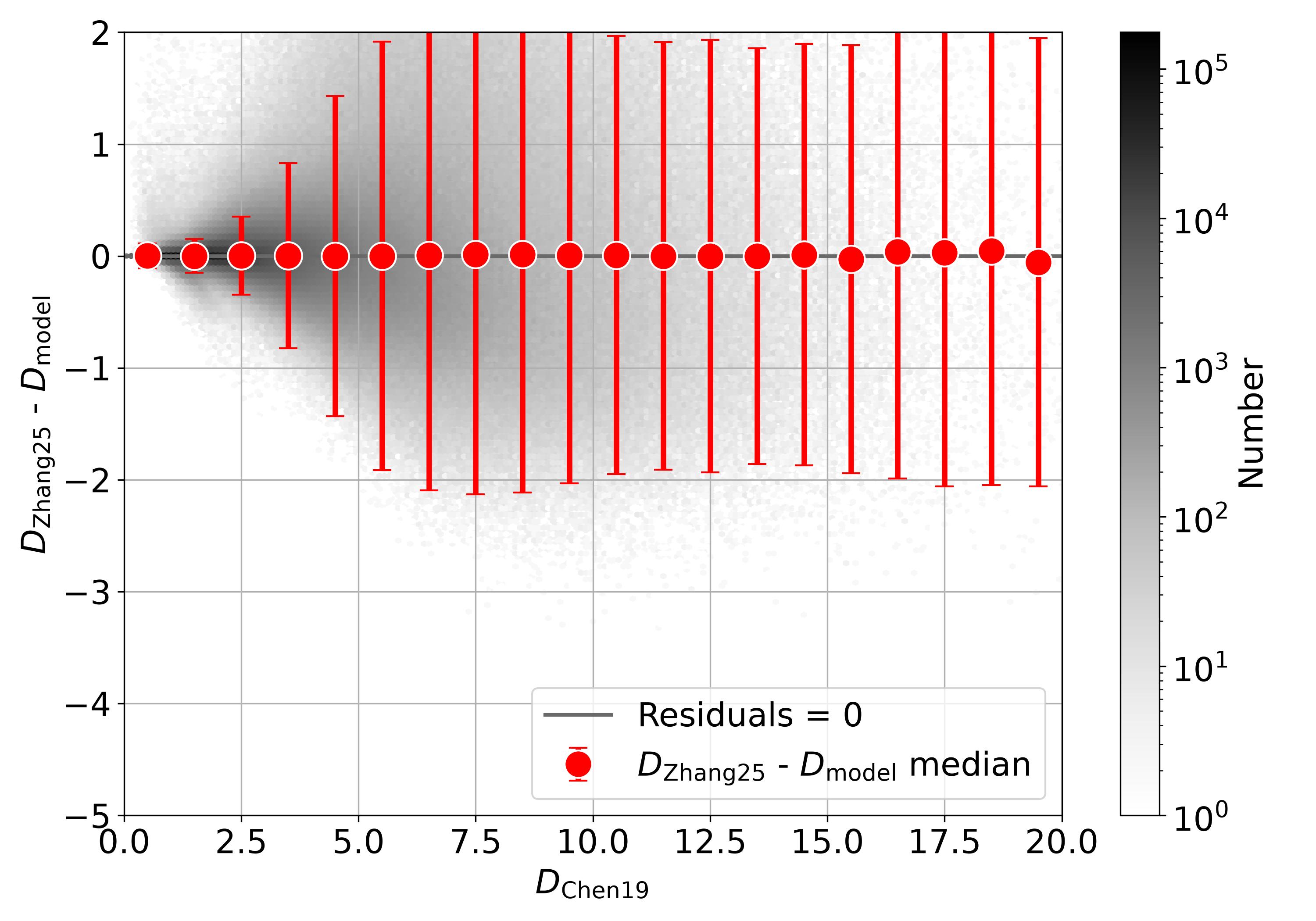}%
	\includegraphics[width=0.53\linewidth]{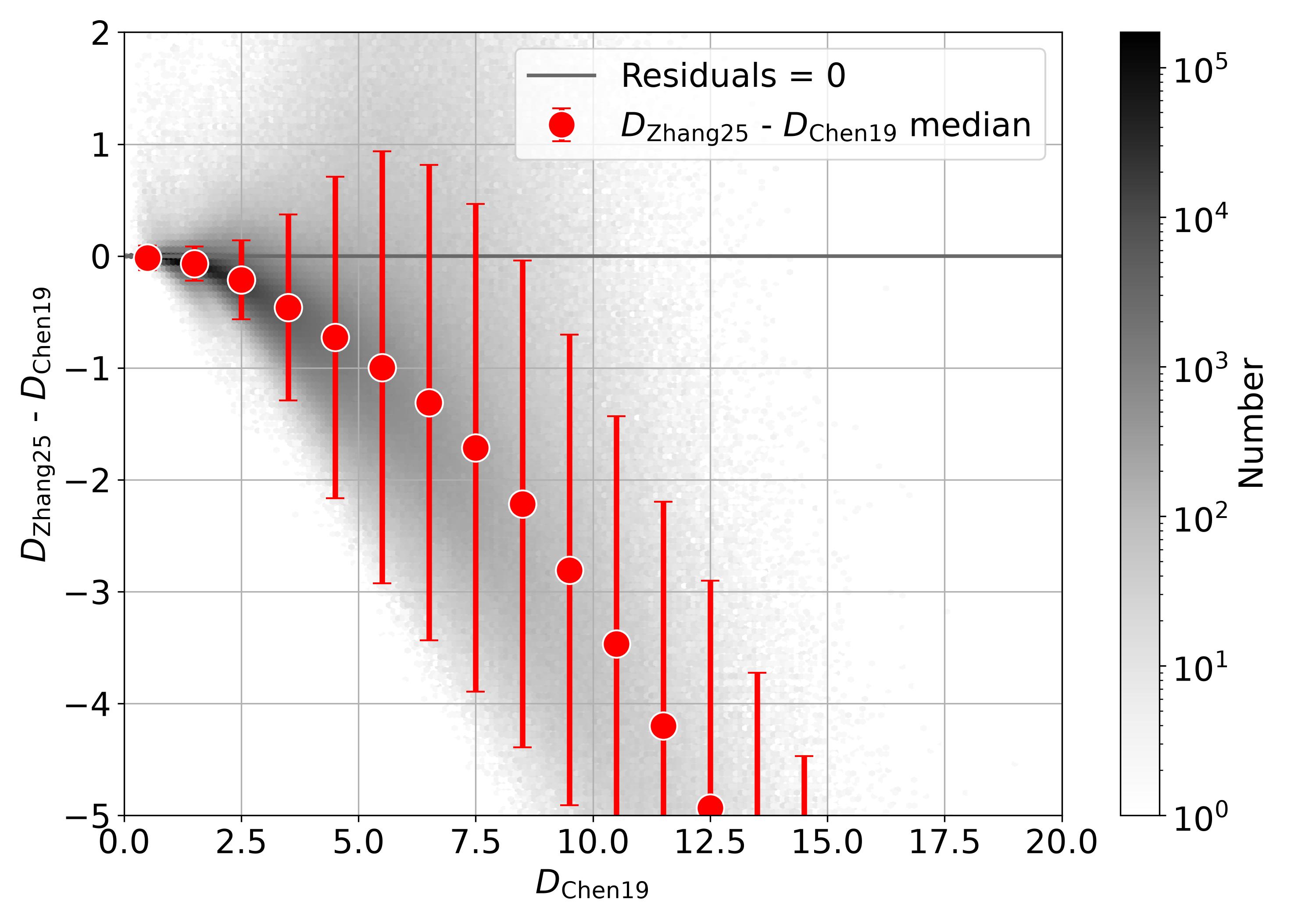}
	\caption{Same as Fig.~\ref{Green19d}, but for the distance systematics correction of Chen19.}
	\label{chen19d}
\end{figure*}

\begin{figure*}
	\centering
	\includegraphics[width=0.53\linewidth]{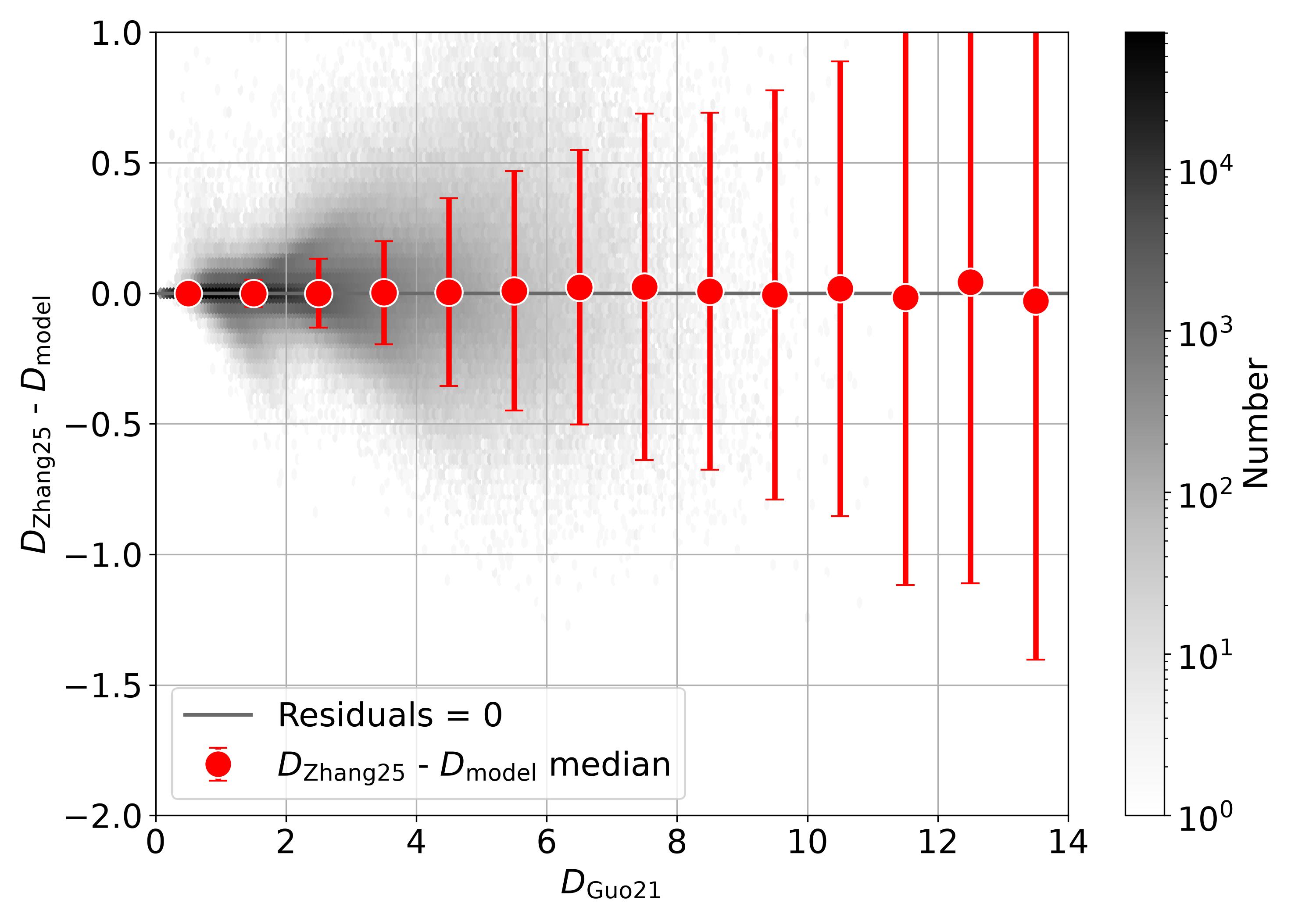}%
	\includegraphics[width=0.53\linewidth]{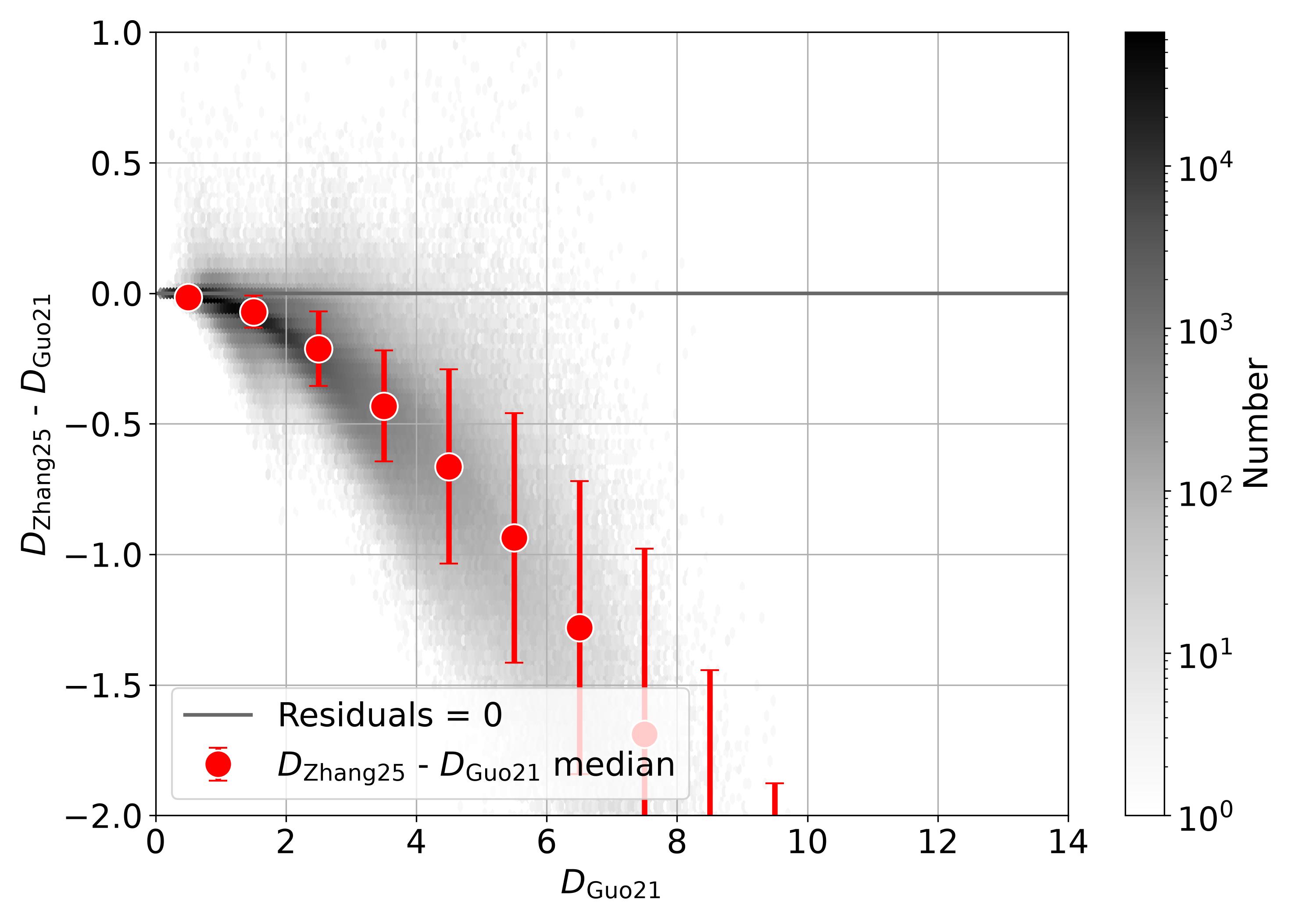}
	\caption{Same as Fig.~\ref{Green19d}, but for the distance systematics correction of Guo21.}
	\label{guo21d}
\end{figure*}

\begin{figure*}
	\centering
	\includegraphics[width=0.53\linewidth]{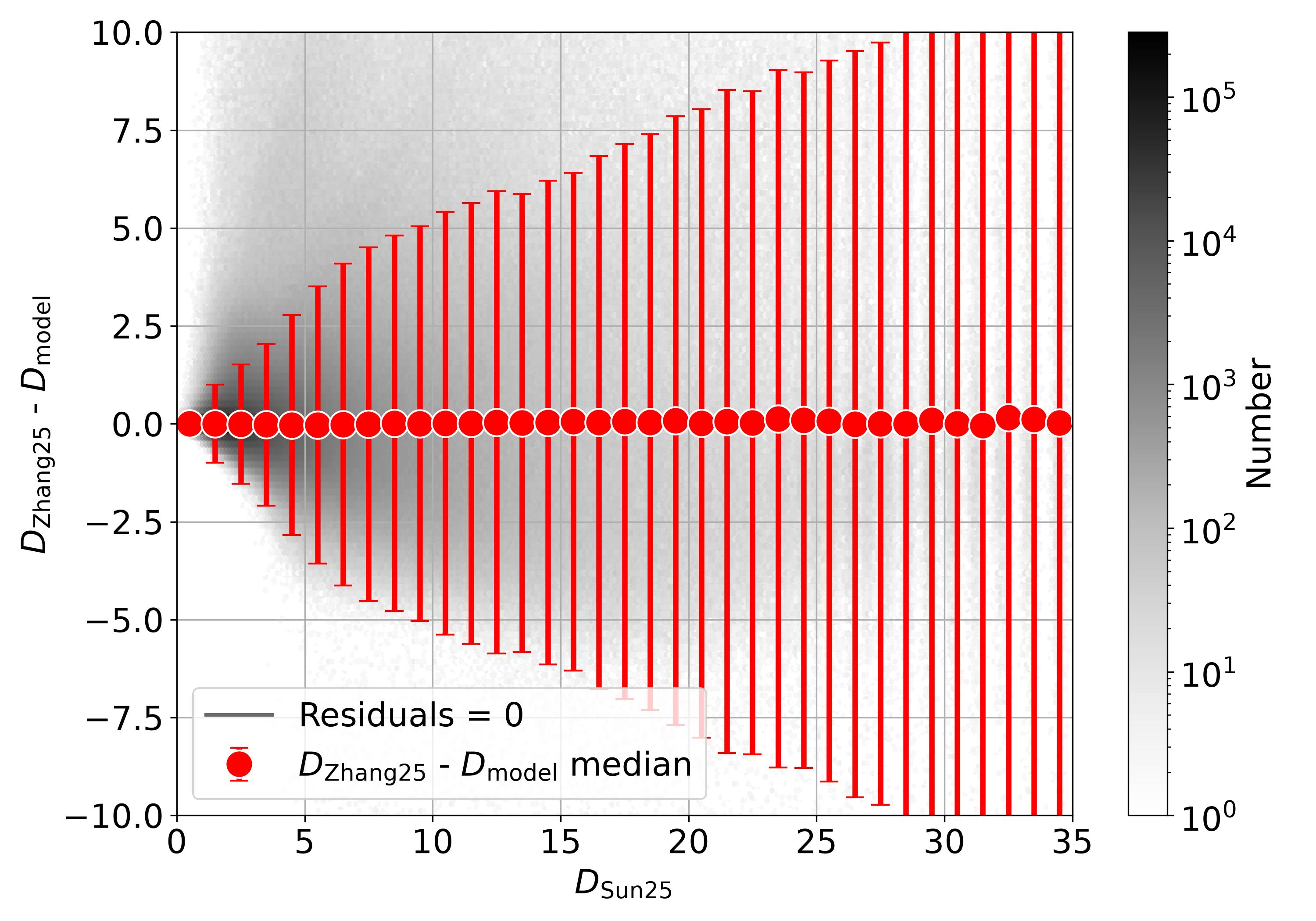}%
	\includegraphics[width=0.53\linewidth]{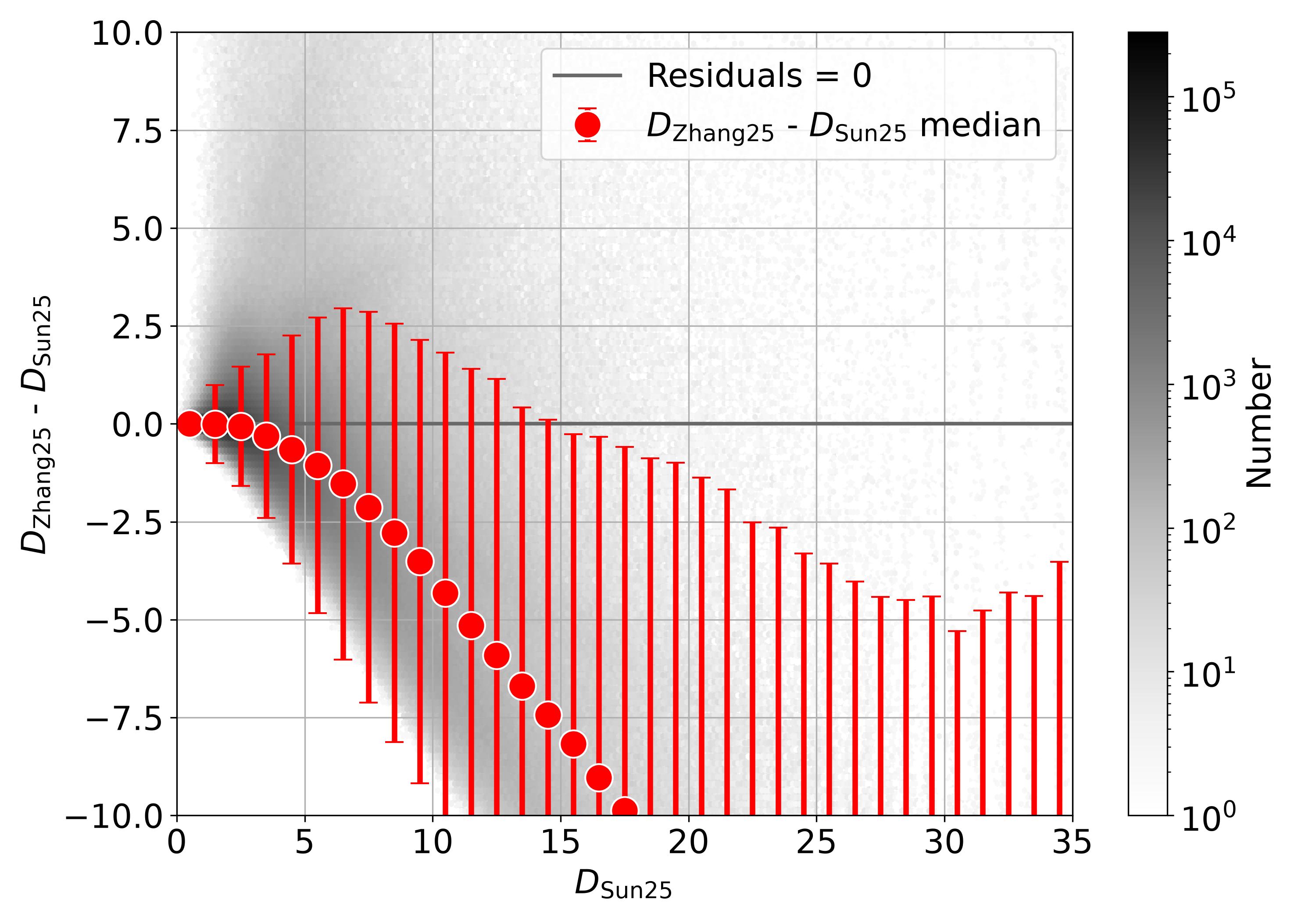}
	\caption{Same as Fig.~\ref{Green19d}, but for the distance systematics correction of Sun25.}
	\label{sun25d}
\end{figure*}

\begin{figure*}
	\centering
	\includegraphics[width=0.53\linewidth]{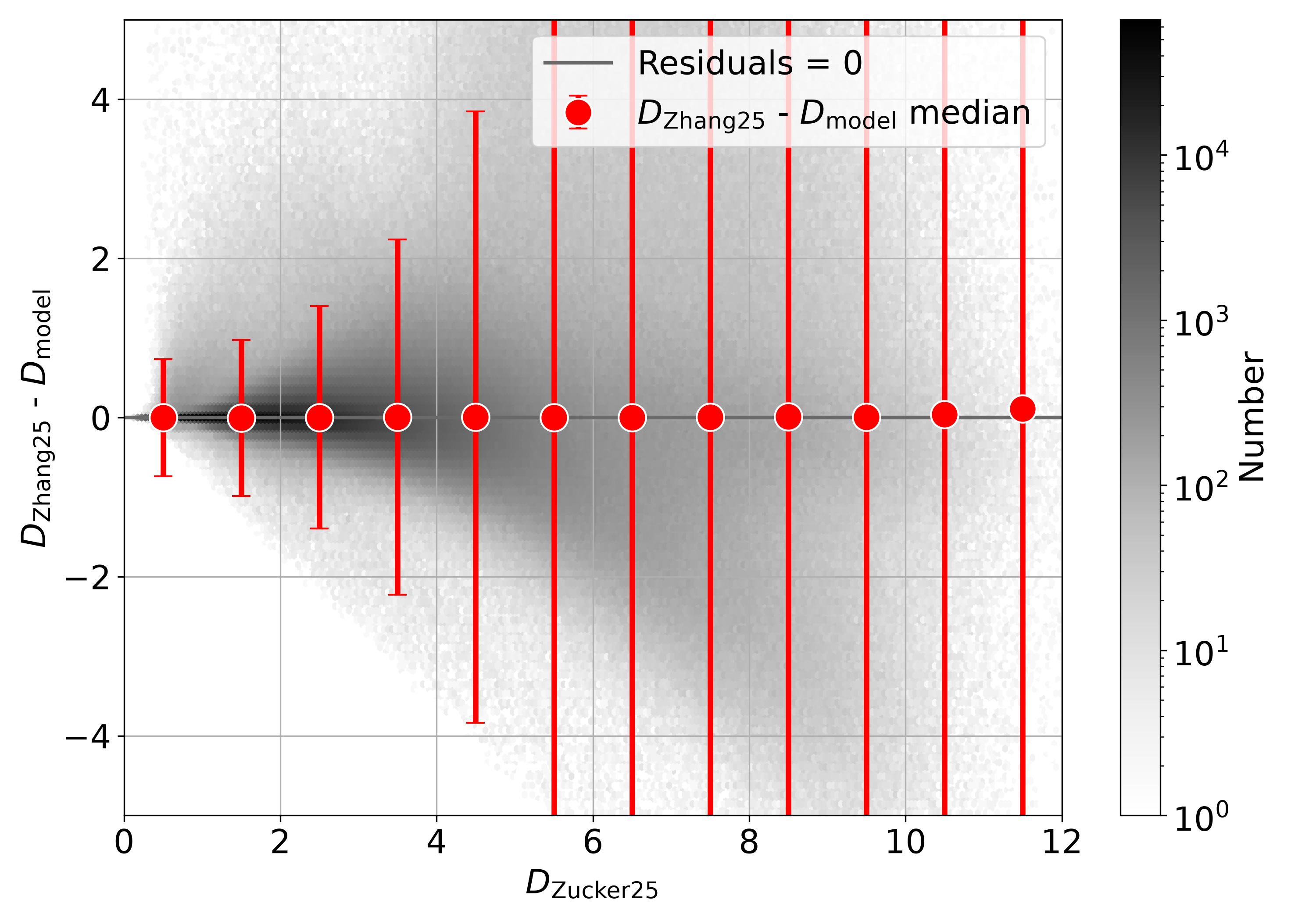}%
	\includegraphics[width=0.53\linewidth]{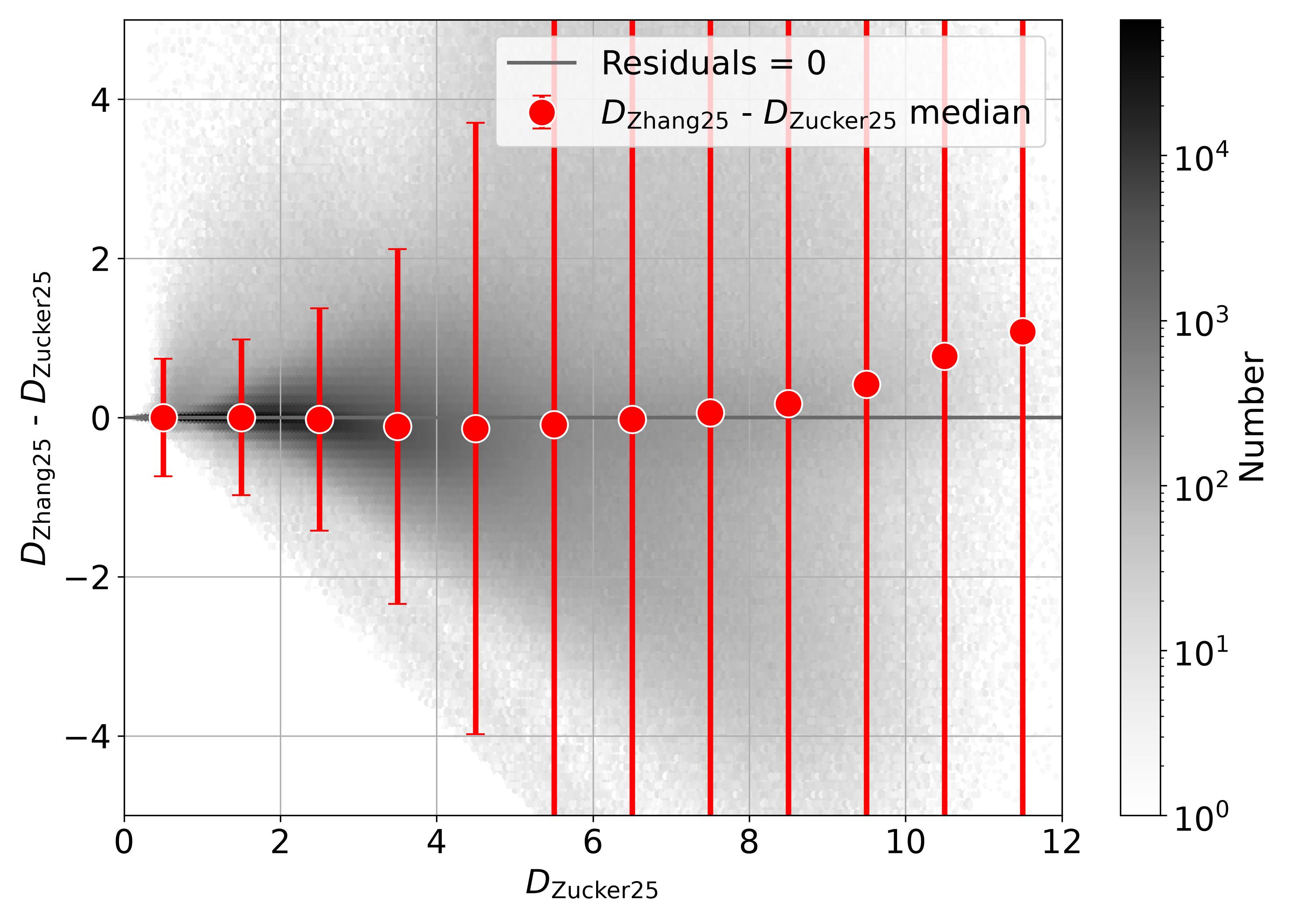}
	\caption{Same as Fig.~\ref{Green19d}, but for the distance systematics correction of Zucker25.}
	\label{zucker25d}
\end{figure*}

Cross-matching the Green19 catalogue with the Zhang25 baseline sample yields $88{,}740{,}824$ records. As shown in Fig.~\ref{Green19d}, the raw $D_{\rm Green19}$ is markedly too large at the far end, and the MLP calibration removes this bias effectively.
Cross-matching the Chen19 catalogue with the Zhang25 baseline sample yields $36{,}707{,}318$ records. As shown in Fig.~\ref{chen19d}, the raw $D_{\rm Chen19}$ is markedly too large at the far end, and the MLP calibration removes this bias effectively.
Cross-matching the Guo21 catalogue with the Zhang25 baseline sample yields $14{,}370{,}710$ records, and no additional constraint is imposed on Guo21 itself. As shown in Fig.~\ref{guo21d}, the raw $D_{\rm Guo21}$ is markedly too large at the far end, and the MLP calibration removes this bias effectively.
Cross-matching the Sun25 catalogue with the Zhang25 baseline sample yields $74{,}134{,}112$ records. As shown in Fig.~\ref{sun25d}, the raw $D_{\rm Sun25}$ is markedly too large at the far end, and the MLP calibration removes this bias effectively.
Cross-matching the Zucker25 catalogue with the Zhang25 baseline sample yields $42{,}310{,}568$ records. As shown in Fig.~\ref{zucker25d}, the raw Zucker25 distance shows the smallest systematic bias among the five catalogues, with its median residual already lying near the zero line, and after the MLP calibration the residual remains centred on zero with a slightly reduced scatter.

\section{Effect of Metallicity as an Input Feature}\label{appendix_C}

\begin{figure*}
	\centering
	\includegraphics[width=0.53\linewidth]{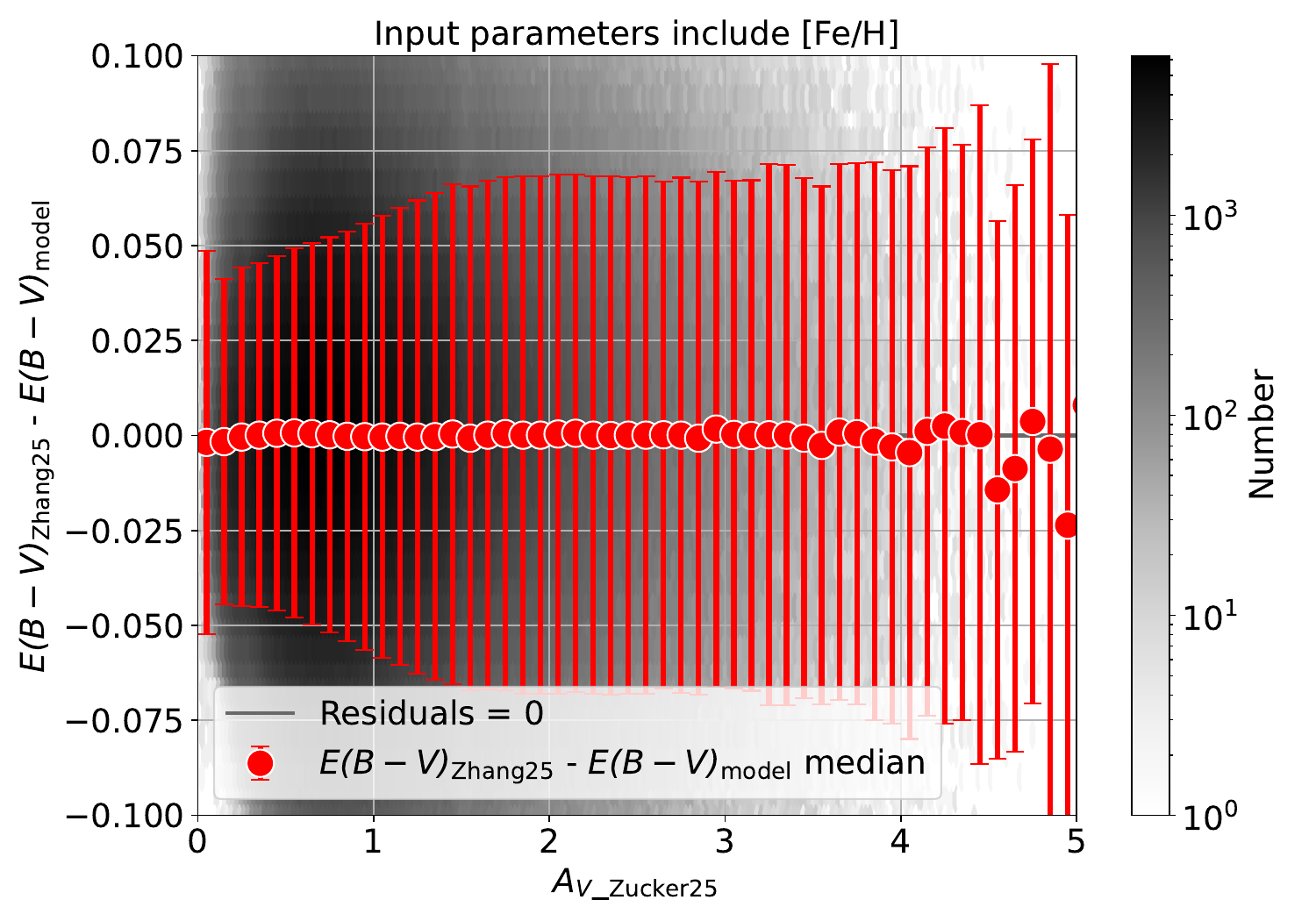}%
	\includegraphics[width=0.53\linewidth]{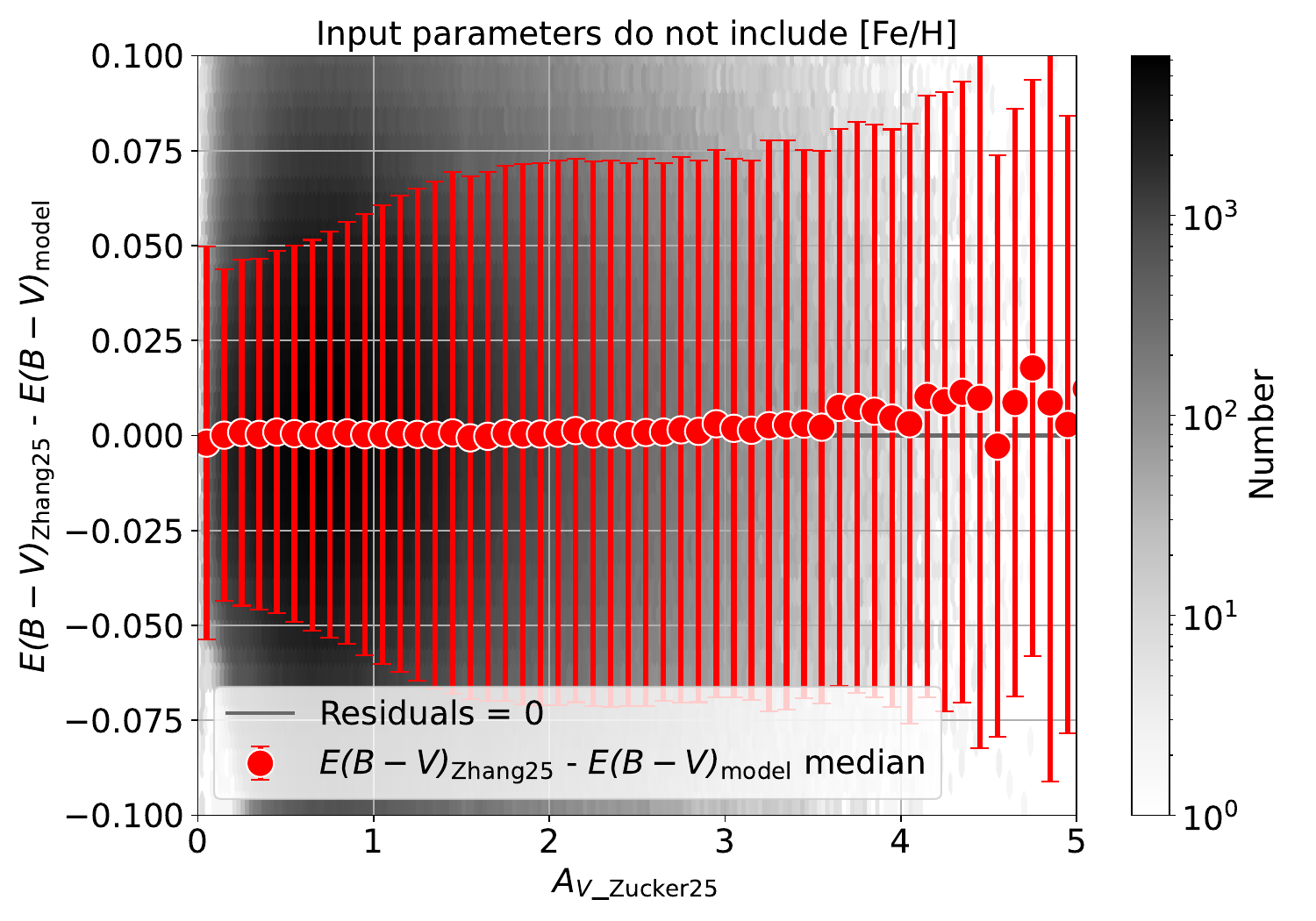}
	\caption{Median residuals and residual density of the target $E(B-V)_{\rm Zhang25}$ relative to the $E(B-V)_{\rm model}$ obtained with and without [Fe/H] among the inputs, as a function of the extinction $A_{V\_\mathrm{Zucker25}}$.}
	\label{feh_db}
\end{figure*}

Taking the Zucker25 data as an example, we test whether the input parameter [Fe/H] affects the result. When the inputs are \(A_{V\_\mathrm{Zucker25}}\), $T_{\rm eff}$, and [Fe/H], the best loss of the model is $0.0403$, and when the inputs are $E_{\rm Zhang25}$ and $T_{\rm eff}$, the best loss is $0.0416$. Fig.~\ref{feh_db} compares the residuals between $E(B-V)_{\rm model}$ and the target $E(B-V)_{\rm Zhang25}$ obtained with and without [Fe/H] among the inputs, and the results of the two models are similar, with the model that includes [Fe/H] performing slightly better.

\section{Sample Reliability Flags}\label{appendix_D}

We assign the sample reliability flags to the catalogues using two complementary criteria. The first is a per-dimension quantile interval, in which the core region $\mathcal{B}_{\mathrm{core}}$ and the extended region $\mathcal{B}_{\mathrm{ext}}$ are delimited by the $5\%$--$95\%$ and $0.5\%$--$99.5\%$ quantiles, respectively, computed solely from the training set with the test set excluded.
The second is a joint Mahalanobis distance  $d_{\mathrm{M}}=\sqrt{\mathbf{z}^{\top}\boldsymbol{\Sigma}^{-1}\mathbf{z}}$, where $\boldsymbol{\Sigma}$ is the covariance matrix of the training set. When $\mathbf{z}$ is approximately multivariate normal, $d_{\mathrm{M}}^{2}$ follows a chi-squared distribution with three degrees of freedom, $d_{\mathrm{M}}^{2}\sim\chi^{2}_{3}$; accordingly, the ellipsoidal radii
are set to $d_{\mathrm{M,core}}=\sqrt{\chi^{2}_{3;\,0.95}}\approx 2.80$ and  $d_{\mathrm{M,ext}}=\sqrt{\chi^{2}_{3;\,0.999}}\approx 4.03$. 
Samples are then labelled as follows: those satisfying both $\mathbf{z}\in\mathcal{B}_{\mathrm{core}}$ and $d_{\mathrm{M}}\le d_{\mathrm{M,core}}$ are labelled $2$ (supervised); those satisfying $\mathbf{z}\in\mathcal{B}_{\mathrm{ext}}$ and $d_{\mathrm{M}}\le d_{\mathrm{M,ext}}$ but not the core-region condition are labelled $1$ (weakly supervised); all remaining samples (including those with missing features or
outliers) are labelled $0$ (unsupervised). 
The per-dimension quantile criterion is robust to skewed distributions but ignores inter-feature correlations, whereas the Mahalanobis distance accounts for correlations but relies on the covariance estimate; requiring both to be satisfied yields a conservative high-confidence sample set and effectively reduces the risk of outliers being mislabelled as supervised samples. It is worth noting that the 'Flag' column contains two sub-columns, representing the supervision status for distance and extinction, respectively. Furthermore, since the Zhang25 catalogue serves as our baseline, we assign a flag of \(2\) (supervised) to all of its entries.

\section{Common Sources}\label{appendix_E}

\begin{figure*}
	\centering
	\includegraphics[width=0.53\linewidth]{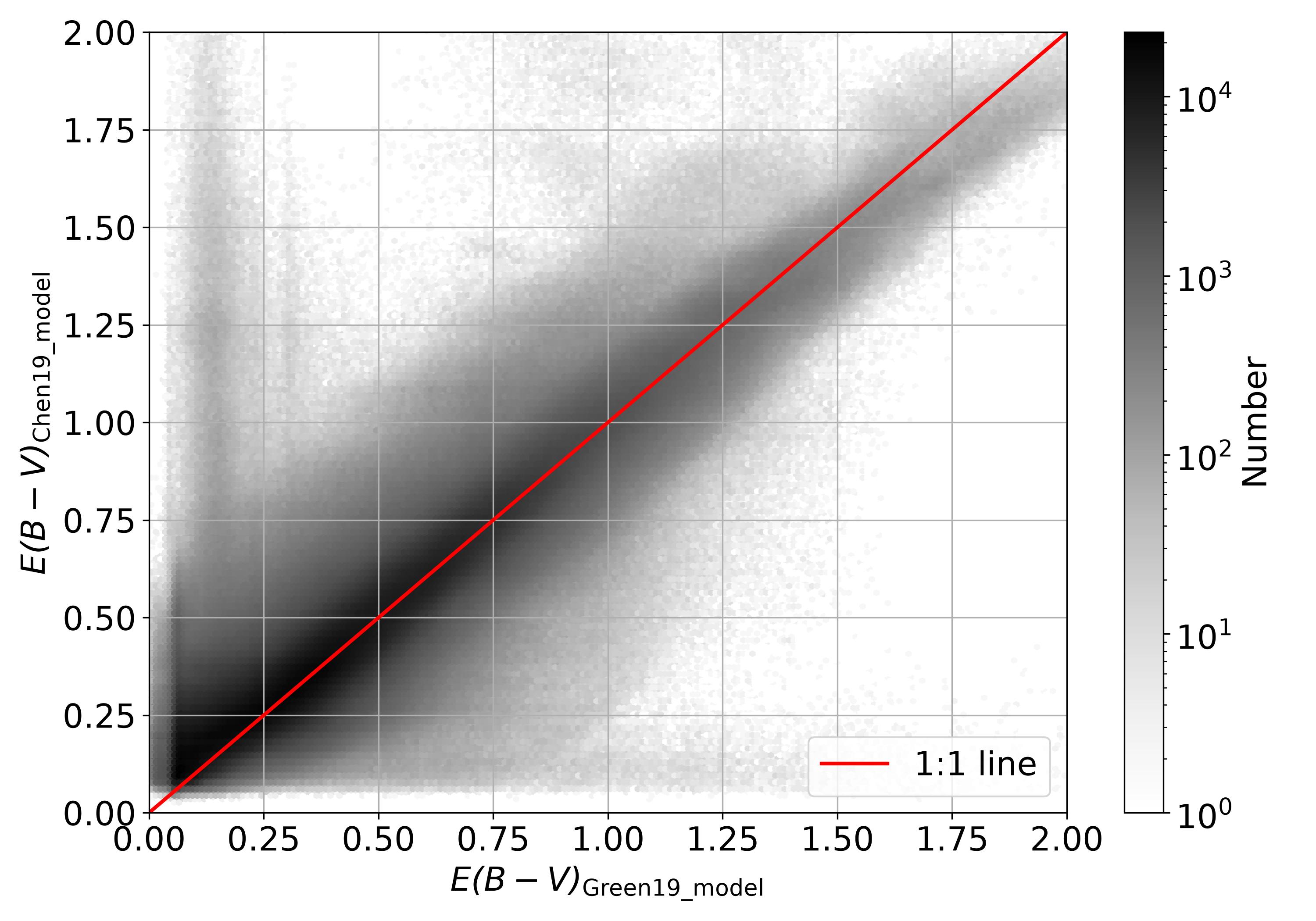}%
	\includegraphics[width=0.53\linewidth]{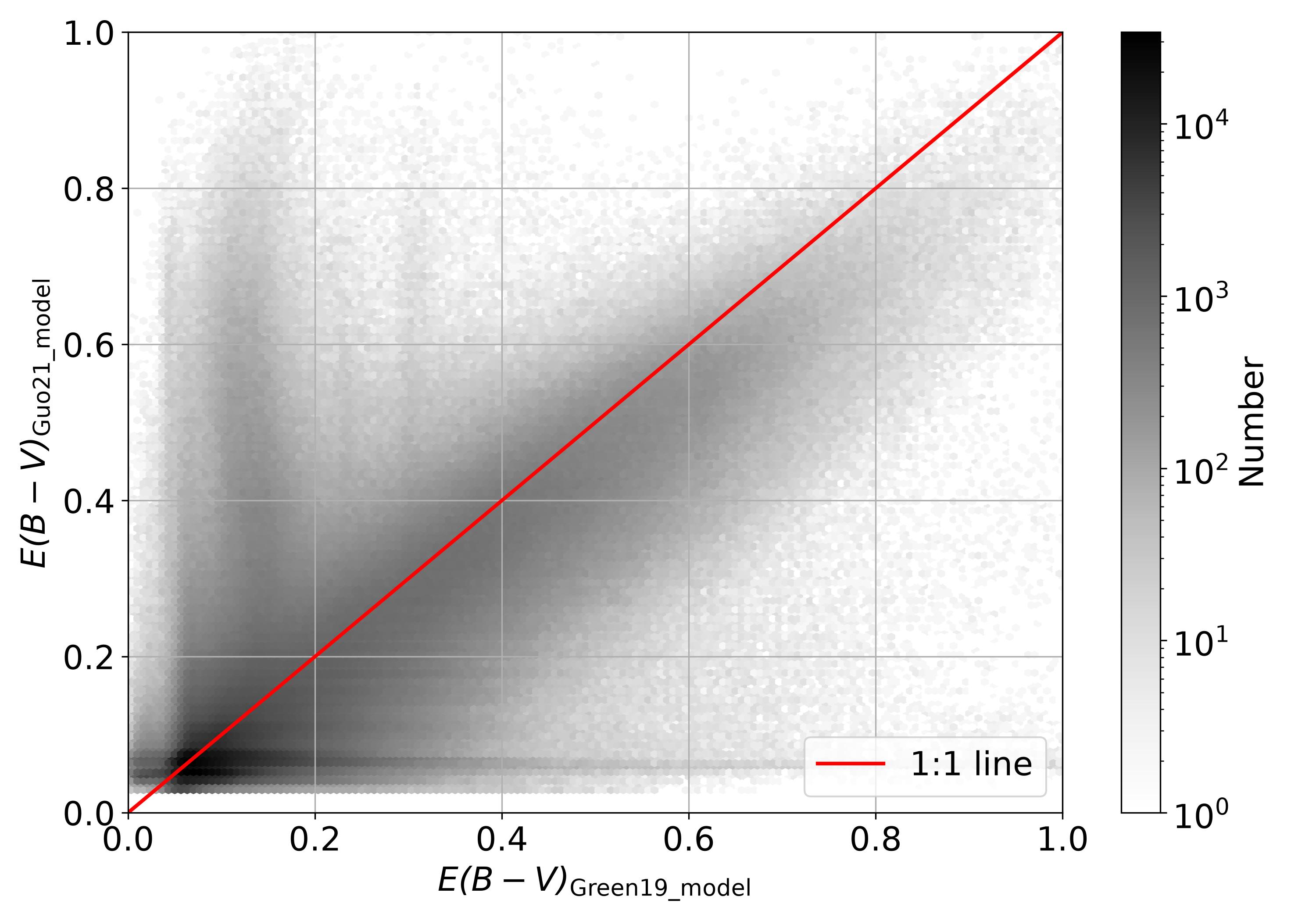}
	\\[0.2em]
	\includegraphics[width=0.53\linewidth]{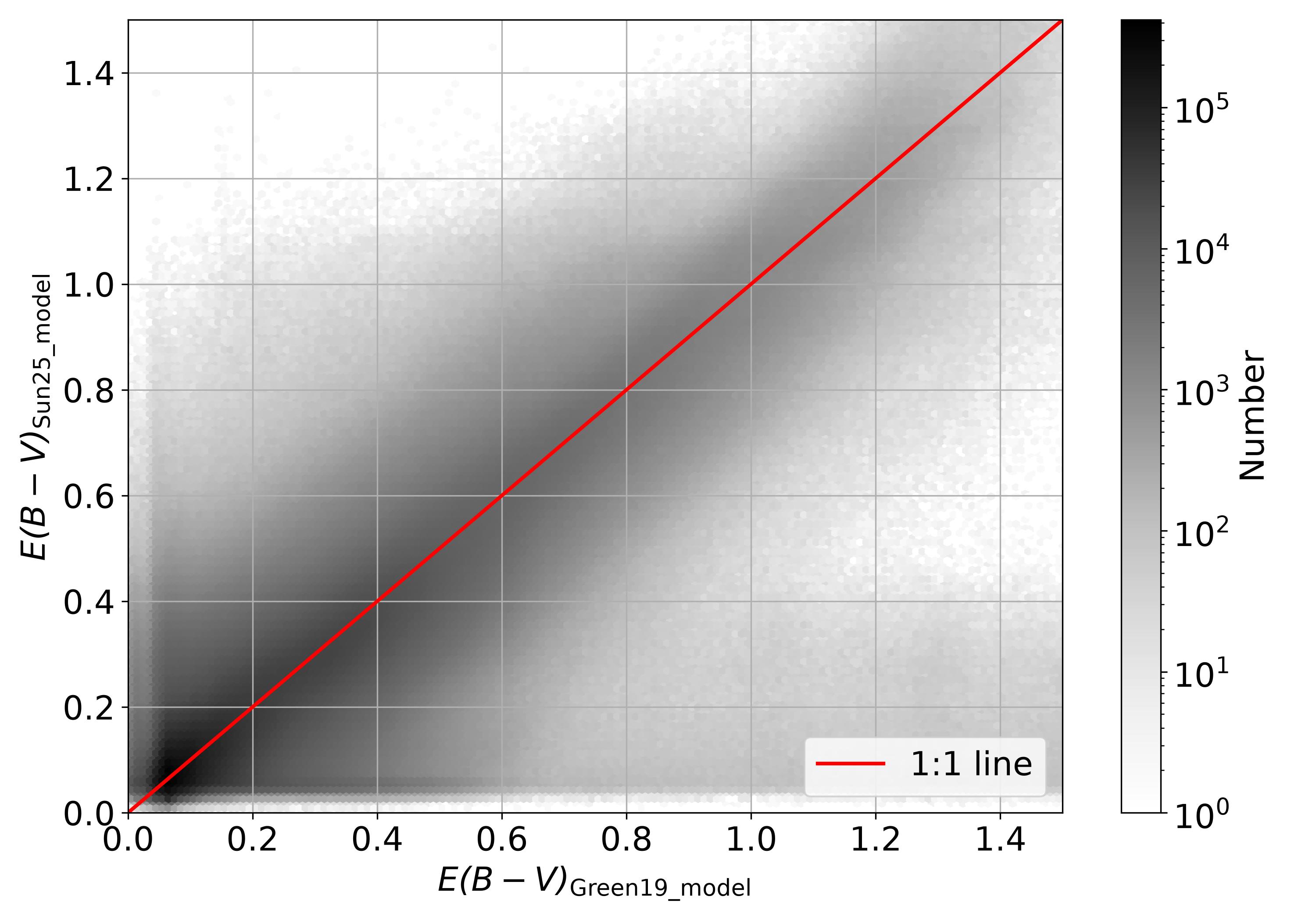}%
	\includegraphics[width=0.53\linewidth]{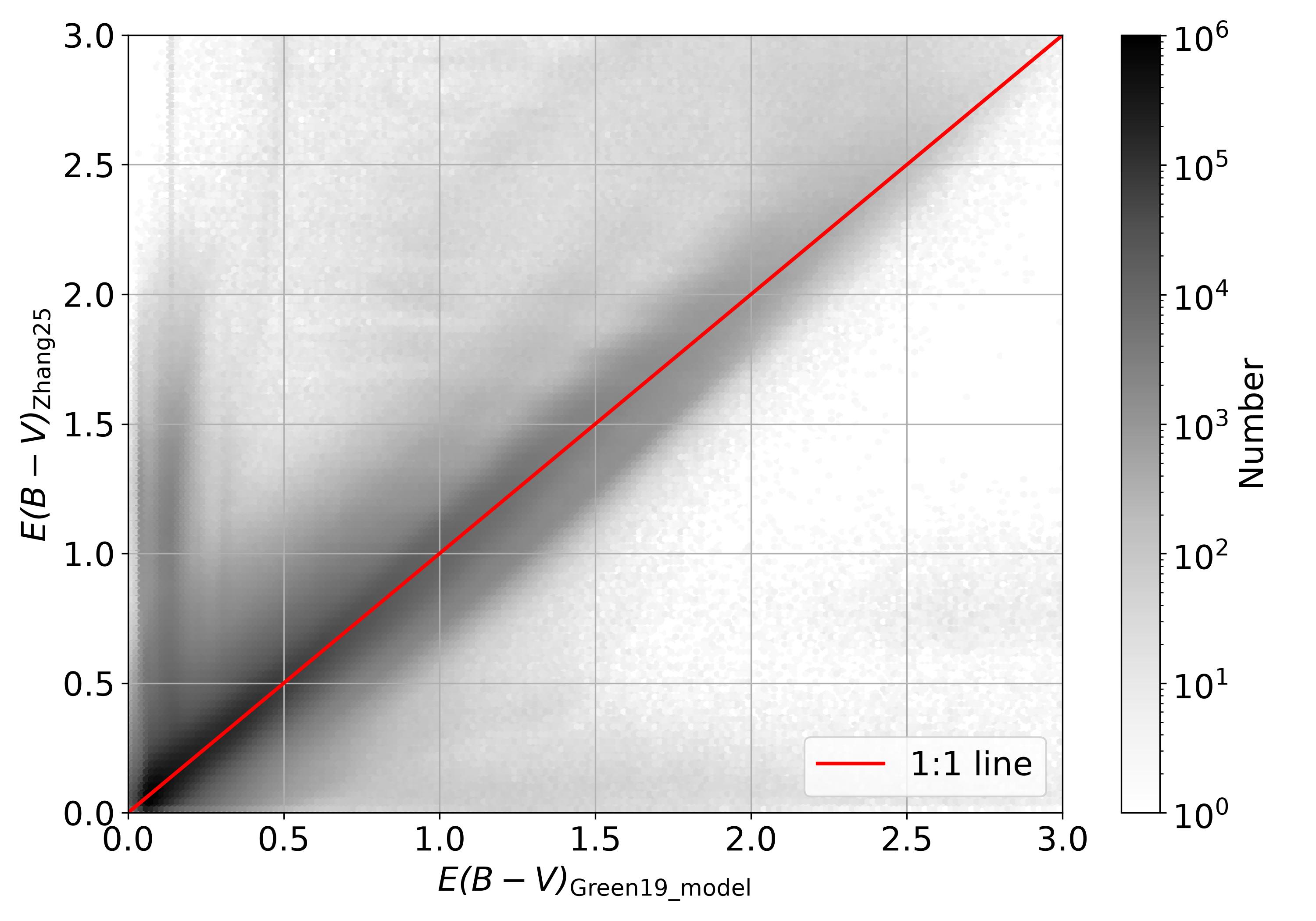}
	\\[0.2em]
	\includegraphics[width=0.53\linewidth]{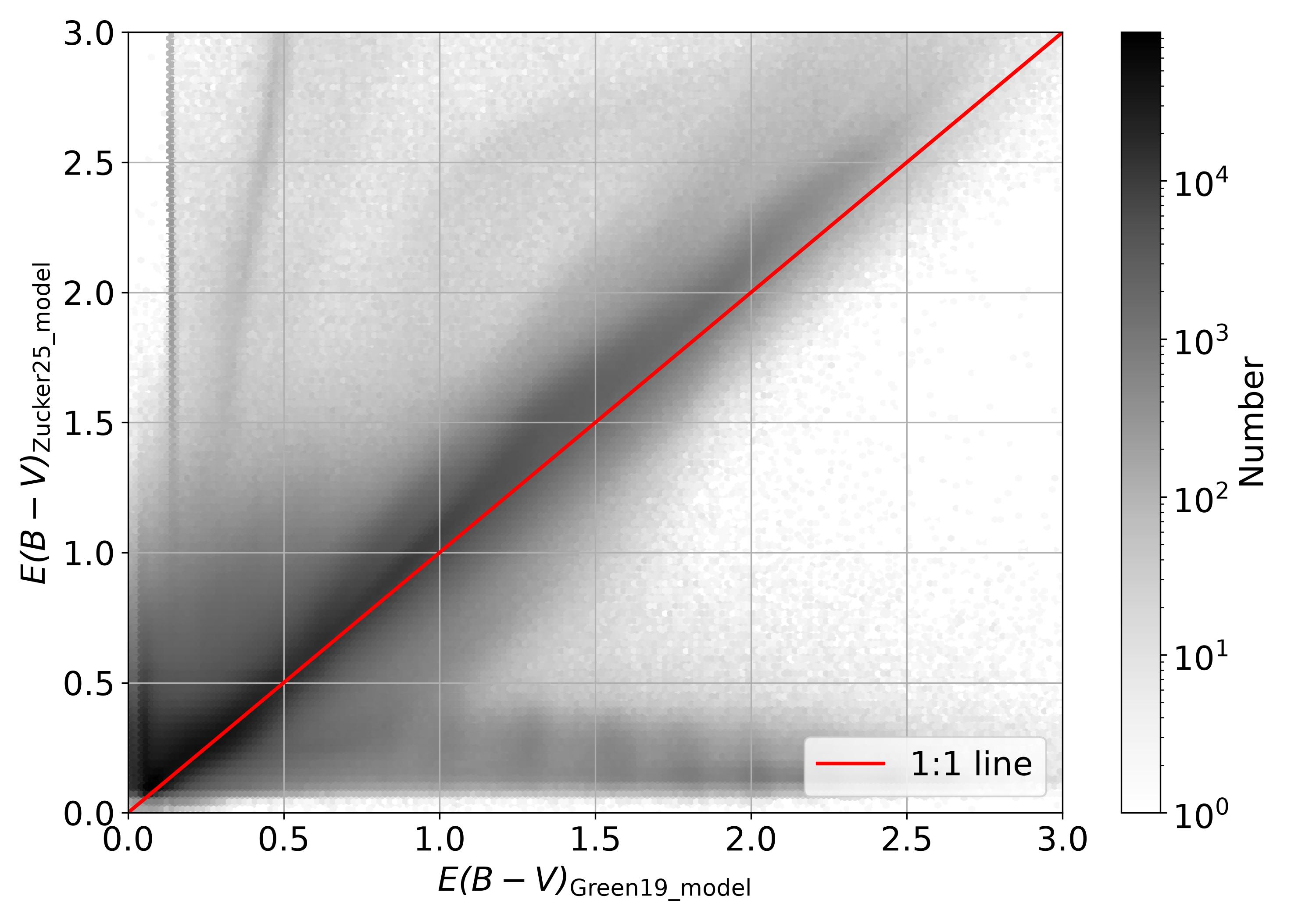}
	\caption{Comparison of \(E(B-V)\) for common sources cross-matched between different catalogues. The x-axis in all panels represents the corrected \(E(B-V)\) from Green19, while the y-axes correspond to the corrected \(E(B-V)\)  from Chen19 (top left), Guo21 (top right), Sun25 (middle left), Zhang25 (baseline, middle right), and Zucker25 (bottom). The red solid line represents the 1:1 reference line, illustrating the consistency and systematic deviations between Green19 and the other catalogues.}
	\label{common_green19ebv}
\end{figure*}

Fig.~\ref{common_green19ebv} compares the corrected \(E(B-V)\) values of sources in common between Green19 and five other catalogues. The abscissa is always \(E(B-V)_{\mathrm{Green19\_model}}\), while the ordinate is the corrected \(E(B-V)\) from Chen19 (top left panel), Guo21 (top right panel), Sun25 (middle left panel), Zhang25 (middle right panel, adopted as the baseline) and Zucker25 (bottom panel). 
The Green19 and Chen19 comparison spans 0 to 2 in \(E(B-V)\). While most sources follow the 1:1 line, the Chen19 values are systematically higher by \(0.05\)--\(0.1\) at low-to-intermediate reddening (\(E(B-V)_{\mathrm{Green19\_model}} \approx 0.3\)--\(0.5\)). The density ridge approaches the 1:1 line near unity but broadens into a diffuse tail above it at higher reddening.
The Green19 and Guo21 comparison spans 0 to 1 and shows the tightest relation and smallest scatter among the panels, with the density ridge closely tracking the 1:1 line. However, at low reddening (\(E(B-V)_{\mathrm{Green19\_model}} \lesssim 0.2\)), the Guo21 values exhibit a vertical dispersion and are systematically higher than the 1:1 line. A dense horizontal band accumulates near the lower bound of \(E(B-V)_{\mathrm{Guo21\_model}} \approx 0\).
For Sun25, the comparison range is 0 to 1.5. The distribution straddles the 1:1 line with moderate scatter, showing a mild excess at intermediate reddening. The scatter gradually increases toward the upper end of the range.
The Green19 and Zhang25 comparison spans from 0 to 3. The main density ridge closely follows the 1:1 line across the entire interval, indicating strong overall consistency between the two catalogues. A mild deficit in Green19 values at intermediate reddening (\(0.5 \lesssim E(B-V)_{\mathrm{Green19\_model}} \lesssim 1.5\)).
The Green19 and Zucker25 comparison also spans 0 to 3. The density ridge follows the 1:1 line overall. Furthermore, the Zucker25 values show a growing excess above the 1:1 line beyond \(E(B-V)_{\mathrm{Green19\_model}} \approx 1.5\).

\begin{figure*}
	\centering
	\includegraphics[width=0.53\linewidth]{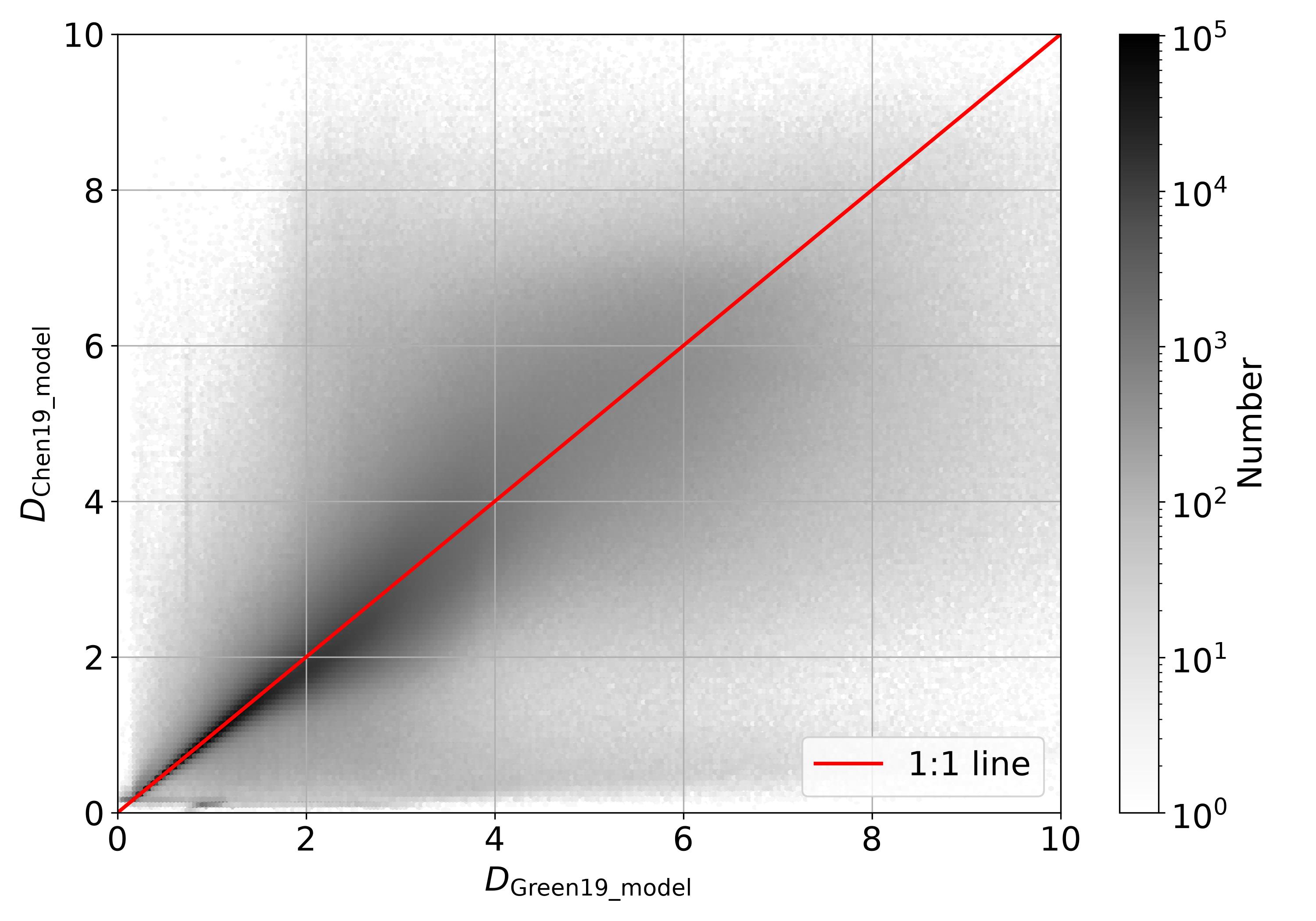}%
	\includegraphics[width=0.53\linewidth]{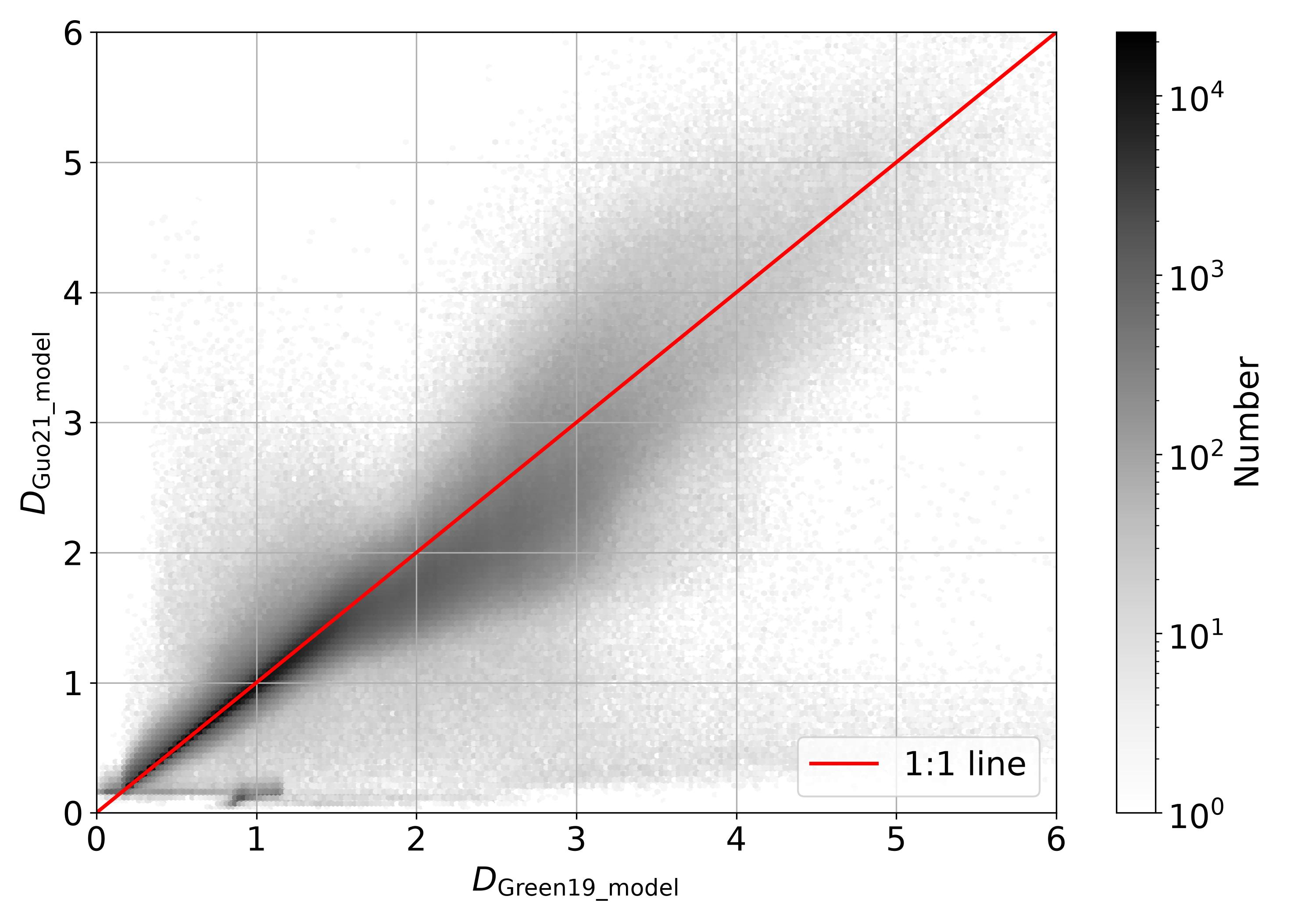}
	\\[0.2em]
	\includegraphics[width=0.53\linewidth]{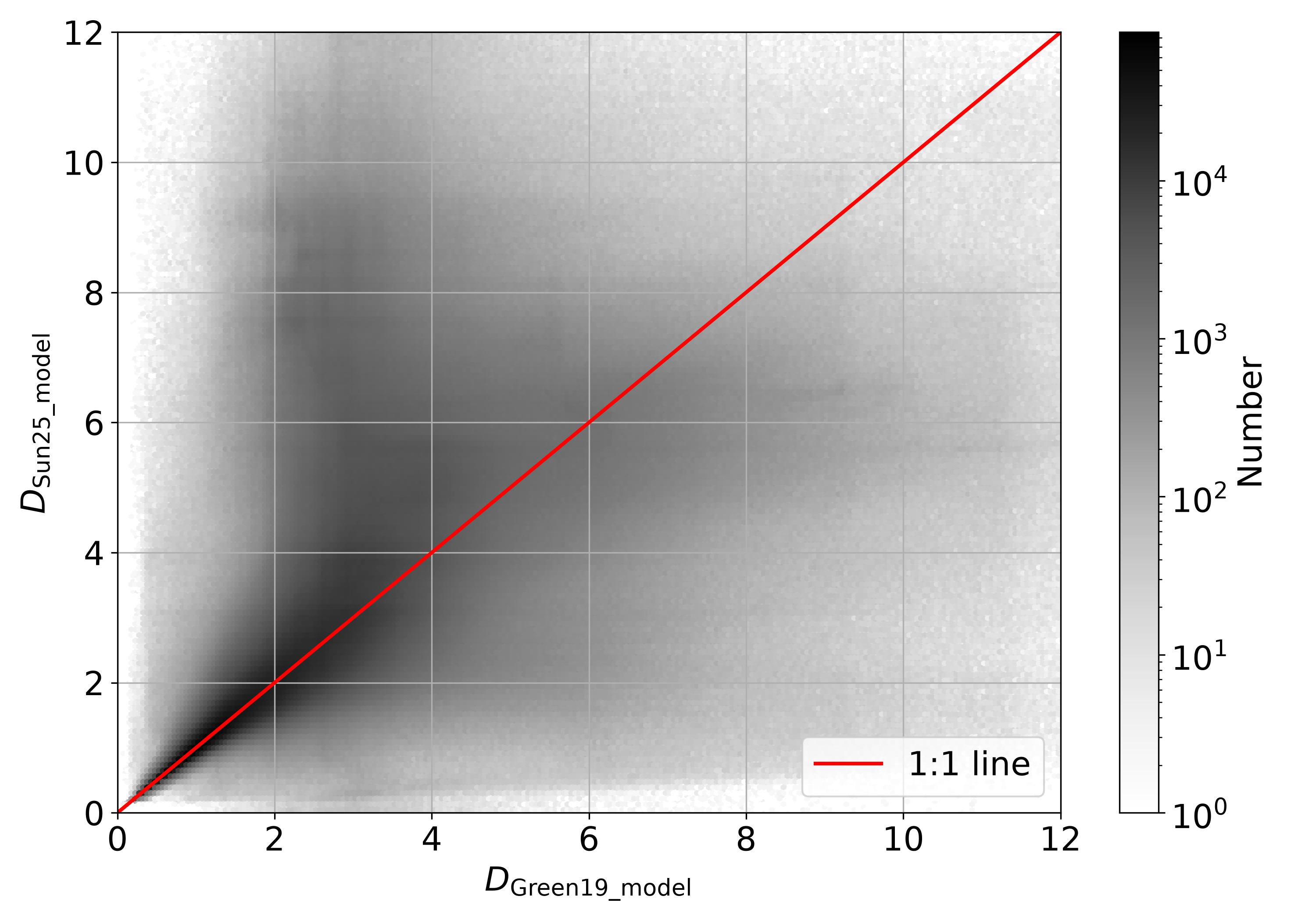}%
	\includegraphics[width=0.53\linewidth]{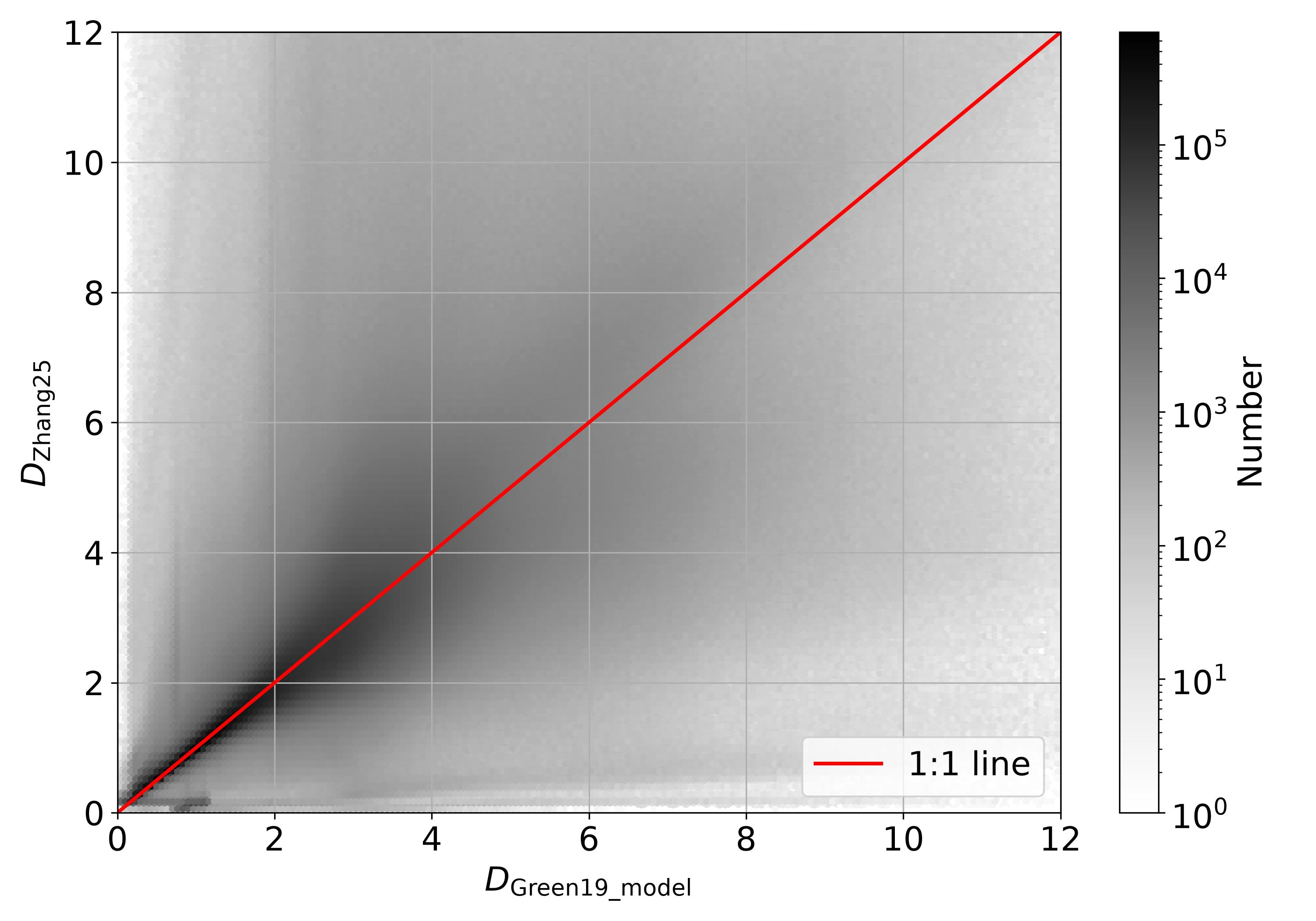}
	\\[0.2em]
	\includegraphics[width=0.53\linewidth]{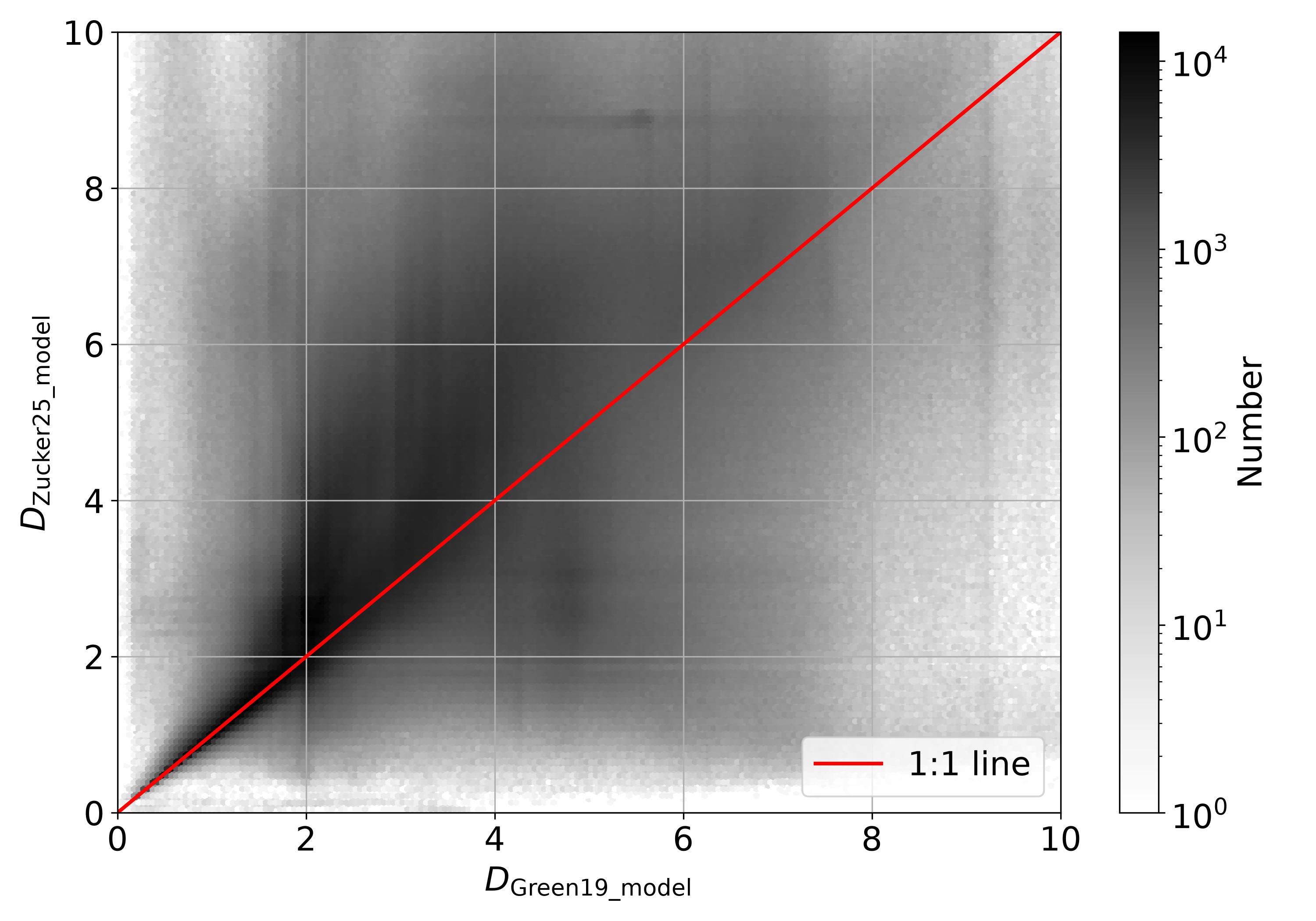}
	\caption{Same as Fig.~\ref{common_green19ebv}, but for the corrected \(D\).}
	\label{common_green19d}
\end{figure*}

Figure~\ref{common_green19d} compares the corrected \(D)\) values of sources in common between Green19 and five other catalogues. The Green19 and Chen19 comparison of \(D\) spans from 0 to 10. The main density ridge closely tracks the 1:1 line, demonstrating strong overall consistency between the two catalogues. However, the distribution broadens noticeably at higher values (\(D_{\mathrm{Green19}} \gtrsim 5\)). The Green19 and Guo21 comparison of \(D\) spans from 0 to 6. The main density ridge closely tracks the 1:1 line at low-to-intermediate distances. The Green19 and Sun25 comparison of \(D\) spans from 0 to 12. At low distances (\(D \lesssim 5\)), the main density ridge closely tracks the 1:1 line, indicating strong consistency. However, at larger distances, the distribution broadens significantly and shifts above the 1:1 line.
The Green19 and Zhang25 comparison of \(D\) spans from 0 to 12. The main density ridge closely tracks the 1:1 line across the entire interval, showing the overall consistency among the catalogues. The Green19 and Zucker25 comparison of \(D\) spans from 0 to 10. While the main density ridge tracks the 1:1 line at low distances, the distribution broadens significantly at higher values, with a diffuse component extending above the 1:1 line. 

The corrected \(E(B-V)\) and distance \(D\) values both exhibit a high degree of overall consistency, which validates the effectiveness and self-consistency of our normalization method. However, catalogue-specific residual biases remain. For \(E(B-V)\), these biases manifest as localized systematic offsets and varying degrees of scatter. For distance \(D\), the agreement is exceptionally high at close ranges but diverges significantly at larger distances.

\clearpage


\end{document}